\documentclass[12pt,preprint]{elsarticle}

\usepackage{amssymb,amsmath,amsfonts,graphicx,a4wide,array,bbm}
\usepackage[usenames,dvipsnames]{color} 
\usepackage{hyperref} 
\usepackage{paralist} 

\biboptions{sort}
\newcommand{\R}{\mathbb{R}}
\newcommand{\C}{\mathbb{C}}
\newcommand{\N}{\mathbb{N}}
\newcommand{\Z}{\mathbb{Z}}

\newcommand{\eps}{\varepsilon}
\newcommand{\BigOh}{\mathcal{O}}
\renewcommand{\phi}{\varphi}

\newcommand{\dd}{\mathrm{d}}
\renewcommand{\Im}{\mathrm{Im}}
\renewcommand{\Re}{\mathrm{Re}}

\newtheorem{theorem}{Theorem}[section]

\newtheorem{prop}[theorem]{Proposition}

\newtheorem{conjecture}{Conjecture}

\numberwithin{equation}{section}

\newcommand{\DEF}{{\,:=\,}}
\newcommand{\PT}[1]{\mathbf{#1}}
\DeclareMathOperator{\CAP}{cap}
\DeclareMathOperator{\digammafcn}{\psi}
\DeclareMathOperator{\TFDIAM}{diam}
\DeclareMathOperator{\extFIELD}{Q}
\DeclareMathOperator{\gammafcn}{\Gamma}
\DeclareMathOperator{\kKERNEL}{k}
\DeclareMathOperator{\KERNEL}{K}
\DeclareMathOperator{\METRIC}{m}
\DeclareMathOperator{\zetafcn}{\zeta}

\DeclareMathOperator{\supp}{supp}
\DeclareMathOperator{\WCE}{wce}

\begin{document}
\title{Distributing many points on spheres: minimal energy and designs}
\author[JSB,PJG]{Johann S. Brauchart\fnref{fn1,fn3}}
\ead{j.brauchart@tugraz.at}

\address[JSB]{School of Mathematics and Statistics, 
University of New South Wales, Sydney, NSW, 2052, Australia}

\fntext[fn1]{This research was supported under Australian Research Council's Discovery Projects funding scheme (project number DP120101816).}
\fntext[fn3]{This author is supported by the Austrian Science Fund FWF
  projects F5510 (part of the Special Research Program (SFB) ``Quasi-Monte
  Carlo Methods: Theory and Applications'').}

\author[PJG]{Peter J. Grabner\fnref{fn2}}
\ead{peter.grabner@tugraz.at}

\address[PJG]{Institut f\"ur Analysis und Computational Number Theory,
Technische Universit\"at Graz,
Steyrergasse 30,
8010 Graz,
Austria}

\fntext[fn2]{This author is supported by the Austrian Science Fund FWF
  projects F5503 (part of the Special Research Program (SFB) ``Quasi-Monte
  Carlo Methods: Theory and Applications'') and W1230 (Doctoral Program
  ``Discrete Mathematics'').}

\begin{abstract}
  This survey discusses recent developments in the context of spherical designs
  and minimal energy point configurations on spheres. The recent solution of
  the long standing problem of the existence of spherical $t$-designs on
  $\mathbb{S}^d$ with $\mathcal{O}(t^d)$ number of points by A.~Bondarenko,
  D.~Radchenko, and M.~Viazovska attracted new interest to this
  subject. Secondly, D.~P.~Hardin and E.~B.~Saff proved that point sets
  minimising the discrete Riesz energy on $\mathbb{S}^d$ in the hypersingular
  case are asymptotically uniformly distributed. Both results are of great
  relevance to the problem of describing the quality of point distributions on
  $\mathbb{S}^d$, as well as finding point sets, which exhibit good
  distribution behaviour with respect to various quality measures.
\end{abstract}

\maketitle

\centerline{Dedicated to Edward B. Saff on the occasion of his
  70\textsuperscript{th} birthday}

\section{Introduction}\label{sec:introduction}
The title of this survey is alluding to the fundamental paper by E.~B.~Saff and
A.~B.~J.~Kuijlaars \cite{Saff_Kuijlaars1997:distributing_many}, which discusses
in large generality how to construct point sets on the sphere
\begin{equation*}
  \mathbb{S}^d \DEF \left\{\PT{x}\in\R^{d+1}\mid\|\PT{x}\|=1\right\},
\end{equation*}
which have ``good distribution properties'' with respect to various
measures. That paper appeared in 1997 and motivated and initiated a fruitful
direction of research. Evenly distributed point sets have many
applications, the most prominent being numerical integration, approximation,
interpolation, and sampling; each of these applications needs a different
definition of what the word ``evenly'' should mean. 

The present survey on point distributions on the sphere is motivated by two
rather recent important contributions, which both shed new light on the
subject. The first was D.~P.~Hardin's and E.~B.~Saff's
\cite{Hardin_Saff2005:minimal_riesz} proof that minimal energy point
configurations on the sphere provide asymptotically uniformly distributed point
sets for hypersingular Riesz potentials. This result has become popular under
the name \emph{Poppy seed bagel theorem}.
This will be one subject of Section~\ref{sec:energy}. The second
breakthrough was A.~V.~Bondarenko's, D.~Radchenko's, and M.~S.~Viazovska's
\cite{Bondarenko_Radchenko_Viazovska2011:optimal_bounds_designs} proof that
spherical $t$-designs with $\mathcal{O}(t^d)$ number of points exist on
$\mathbb{S}^d$. The definition of spherical designs, their properties,
relevance, and occurrence in different contexts will be the subject of
Section~\ref{sec:designs}.


Throughout the paper we use the notation
$X_N=\{\PT{x}_1^{(N)},\PT{x}_2^{(N)},\ldots,\PT{x}_N^{(N)}\}$ for a set of $N$
points; $N$ will always denote the number of points, and the points
$\PT{x}_1^{(N)},\PT{x}_2^{(N)},\ldots,\PT{x}_N^{(N)}$ usually depend on $N$. In
order to keep the notation simple, we suppress this dependence.

Numerical computation of integrals over multidimensional domains is usually
done by weighted sums of point evaluations of the function $f$ to be
integrated. Since the sphere is a domain with very high symmetry, it is
preferable to use equal weight integration methods,
\begin{equation}\label{eq:num-int}
\int_{\mathbb{S}^d} f(\PT{x}) \, \dd \sigma_d(\PT{x}) \approx 
\frac{1}{N} \sum_{j=1}^N f( \PT{x}_j ).
\end{equation}
Here, and throughout the paper, we denote by $\sigma_d$ the \emph{normalised}
surface area measure on~$\mathbb{S}^d$.
For non-random collections of integration nodes, using the right-hand side of
\eqref{eq:num-int} as approximation for the integral is known as a
\emph{Quasi-Monte Carlo (QMC) method}.
In order to obtain a good approximation of the integral by the sum, the point
set $X_N$ should be \emph{well distributed} over the whole sphere
$\mathbb{S}^d$. The notion of good distribution of a point set is described in
an intrinsic way in this context: the distribution is \emph{good}, if the error
for numerical integration is small for a certain set of functions.
More precisely, sequence of node sets $(X_N)$ will be called
\emph{asymptotically uniformly distributed}, if the relation 
\begin{equation}\label{eq:caplim}
\lim_{N\to\infty}\frac1N\sum_{j=1}^N\mathbbm{1}_C(\PT{x}_j)=\sigma_d(C)
\end{equation}
holds for all spherical caps $C$ (here and throughout the paper $\mathbbm{1}_C$
denotes the indicator function of the set $C$). This is equivalent to
\begin{equation}\label{eq:Nflim}
  \lim_{N\to\infty}\frac1N\sum_{j=1}^Nf(\PT{x}_j)=
\int_{\mathbb{S}^d}f(\PT{x})\,\dd\sigma_d(\PT{x})
\end{equation}
for all continuous functions $f:\mathbb{S}^d\to\R$ (see
\cite{Kuipers_Niederreiter1974:uniform_distribution_sequences}).

Placing $N$ points on the sphere so that the least distance (or equivalently,
the smallest angle) between two points,
\begin{equation}\label{eq:least-distance}
\delta(X_N) \DEF \min_{1\leq i<j\leq N}\|\PT{x}_i-\PT{x}_j\|
\end{equation}
is maximised is a further classical approach to obtain evenly distributed
points, see~\cite{Habicht_Waerden1951:lagerung_von_punkten}. This is the
problem of best packing which has attracted attention for a long time in and
outside of mathematics. For instance, the best packing problem on the sphere is
attributed to P.~M.~L.~Tammes \cite{Tammes1930:pollen_grains}, a botanist, who
searched for an explanation of the surface structure of pollen grains.

A likewise geometric, but qualitatively different way of evenly distributing
$N$ points is to cover the sphere with $N$ spherical caps of radius $r$, and
to minimise $r$. The points are then chosen as the centres of these spherical
caps. Equivalently, the size of the largest cap, which does not contain a
point of $X_N$,
\begin{equation}\label{eq:mesh-norm}
\rho(X_N) \DEF \max_{\PT{y}\in\mathbb{S}^d}\min_{1\leq i\leq N}\|\PT{y}-\PT{x}_i\|,
\end{equation}
should be as small as possible. This quantity is called the \emph{covering
  radius}, and is also known as the \emph{mesh norm} or the \emph{fill radius}.
This is the problem of best covering which originates from the realm of
facility location problems where the farthest distance of a point on the sphere
to the nearest point of $X_N$ (service distance) is minimised.

Sampling function values on the sphere (\emph{e.g.}, for approximation or
interpolation by splines or radial basis functions) or exploring spatial
directions in an efficient way requires again \emph{well distributed} (but not
necessary uniformly distributed) point sets on $\mathbb{S}^d$. In this context,
the quality of a point set is measured differently: well distributed points are
required to be \emph{dense} in a quantifiable way.
As the mesh norm arises in the error of approximation and good separation is
generally associated with the ``stability'' of an approximation or
interpolation method, one would prefer a \emph{quasi-uniform} sequence of point
sets $(X_N)$ with uniformly bounded \emph{mesh-separation ratio}
\begin{equation}\label{eq:mesh-separation}
\gamma(X_N) \DEF \frac{\rho(X_N)}{\delta(X_N)}.  
\end{equation}
The mesh-separation ratio can be regarded as the ``condition number'' for the
point set. Quasi-uniformity is a crucial property for a number of methods
(see~\cite{FuWr2009,LeGSlWe2010,Sch1995}).


\section{Spherical Designs}\label{sec:designs}

Spherical designs were initially defined by P.~Delsarte, J.~M.~Goethals, and
J.~J.~Seidel \cite{Delsarte_Goethals_Seidel1977:spherical_designs} in the
context of algebraic combinatorics on spheres. Since then, spherical designs
have gained attraction in different areas of mathematics, ranging over number
theory, geometry, algebraic and geometric combinatorics, and numerical
analysis. We will give an account of these aspects in this section. For a
survey on further developments in the context of spherical designs mainly from
the point of view of algebraic combinatorics and number theory, we refer to
the survey \cite{Bannai_Bannai2009:survey_designs} by E.~Bannai and E.~Bannai.

\subsection{Definition}
A spherical $t$-design is a finite set of points $X_N\subset \mathbb{S}^d$, such
that
\begin{equation}\label{eq:design}
  \frac1{N}\sum_{\PT{x}\in X_N}f(\PT{x})= \int_{\mathbb{S}^d}f(\PT{x})\,\dd 
  \sigma_d(\PT{x})
\end{equation}
for all polynomials $f\in\R[x_1,\ldots,x_{d+1}]$ (restricted to the sphere
$\mathbb{S}^d$) of total degree $\leq t$. This definition is equivalent to
\begin{equation}\label{eq:design-gegenbauer}
\sum_{(\PT{x},\PT{y})\in X_N\times X_N}P_\ell^{(d)}(\langle \PT{x},\PT{y}\rangle)=0
\end{equation}
for $\ell=1,\ldots,t$; here $P_\ell^{(d)}$ denotes the Legendre polynomial of
degree $\ell$ for the sphere $\mathbb{S}^d$
(see~\cite{Mueller1966:spherical_harmonics}). These polynomials are multiples
of Gegenbauer polynomials $C_\ell^\alpha$ with $\alpha=\frac{d-1}2$ normalised
so that $P_\ell^{(d)}(1) = 1$.  The equivalence of \eqref{eq:design-gegenbauer}
and \eqref{eq:design} is then an immediate consequence of the fact that the
restrictions of polynomials to $\mathbb{S}^d$ are spanned by the harmonic
polynomials (\emph{i.e.}, polynomials $p$ satisfying $\triangle_{d+1}p=0$ for
the Laplace operator $\triangle_{d+1}$ in $\R^{d+1}$) and the addition theorem
for spherical harmonics (see~\cite{Mueller1966:spherical_harmonics})
\begin{equation}\label{eq:addition}
\sum_{k=1}^{Z(d,\ell)} Y_{\ell,k}(\PT{x}) Y_{\ell,k}(\PT{y}) = 
Z(d,\ell) \, P_\ell^{(d)}(\langle \PT{x},\PT{y} \rangle)
\text{ with } Z(d,\ell) \DEF\frac{2\ell+d-1}{d-1}\binom{\ell+d-2}{d-2};
\end{equation}
here and throughout this paper $Y_{\ell,k}$ ($k = 1, \ldots,Z(d,\ell)$) denotes
a real orthonormal basis of the space of spherical harmonics of total degree
$\ell$ with respect to the scalar product $\langle f, g\rangle \DEF
\int_{\mathbb{S}^d} f(\PT{x}) g(\PT{x}) \,\dd \sigma_d(\PT{x})$. Condition
\eqref{eq:design-gegenbauer} is then obtained by considering condition
\eqref{eq:design} for $f = Y_{\ell, k}$, squaring it, summing over $k$, and
using \eqref{eq:addition}.

There are two further characterisations of spherical $t$-designs,
which show the connection of this concept to other areas of mathematics. It was
observed in \cite{Lyubich_Vaserstein1993:isometric_banach} that a spherical
$2t$-design $X_N$ with $N$ points gives an isometric embedding of
$\ell_2^{d+1}$ into $\ell_{2t}^N$, which comes from the identity
\begin{equation}\label{eq:embed}
\frac1{N}\sum_{\PT{x}\in X_N}\langle \PT{x},\PT{a}\rangle^{2t}=
\frac{1\cdot 3\cdot 5\cdots (2t-1)}{(d+1)(d+2)\cdots (d+2t-1)}
\langle \PT{a},\PT{a}\rangle^t,
\end{equation}
which is valid for all $\PT{a}\in\R^{d+1}$. This identity is an immediate
consequence of \eqref{eq:design-gegenbauer} and the expansion of $x^{2t}$ as a
sum of Legendre polynomials.

A spherical $t$-design $X_N=\{\PT{x}_1,\PT{x}_2,\ldots,\PT{x}_N\}$ is called
\emph{rigid} (see~\cite{Bannai1987:rigid_designs}), if there exists an
$\eps>0$, such that for all $t$-designs
$X_N'=\{\PT{x}_1',\PT{x}_2',\ldots,\PT{x}_N'\}$ with
$\|\PT{x}_i-\PT{x}_i'\|<\eps$ (for $i=1,\ldots,N$) there exists a
rotation $\eta\in\mathrm{SO}(d+1)$ such that $X_N'=\eta X_N$.

It was observed in \cite{Grabner_Tichy1993:spherical_designs_discrepancy} in
passing and later rediscovered in
\cite{Sloan_Womersley2009:variational_designs} that spherical $t$-designs
can be characterised by a variational property. Let $p$ be a polynomial of
degree $t$ given by 
\begin{equation*}
p(z)=\sum_{\ell=1}^t a_\ell P_\ell^{(d)}(z)\quad\text{with }
a_\ell>0\text{ for }\ell=1,\ldots,t.
\end{equation*}
Then $X_N$ is a spherical $t$-design if and only if the energy functional
$E_p(X_N)$ defined by
\begin{equation}\label{eq:power-sum}
E_p(X_N) \DEF \sum_{(\PT{x},\PT{y})\in X_N\times X_N}
p(\langle \PT{x},\PT{y}\rangle)
\end{equation}
vanishes for $X_N$.  Since the sum \eqref{eq:power-sum} is non-negative for all
$X_N$, designs are minimisers of this sum.  This was used in
\cite{Sloan_Womersley2009:variational_designs} to characterise designs as
stationary points of $E_p(X_N)$.

In order to explain the connection to other questions of algebraic and
geometric combinatorics on $\mathbb{S}^d$, we need the following notion. For a
finite subset $X_N$ of $\mathbb{S}^d$, we define
\begin{equation}\label{eq:A(X)}
A(X_N) \DEF \{\langle \PT{x},\PT{y}\rangle\mid \PT{x},\PT{y}\in X_N, \PT{x}\neq 
\PT{y}\},
\end{equation}
the set of mutual inner products of distinct points. For a given
$A\subset[-1,1)$, $X_N$ is called an $A$-code, if $A(X_N)\subset A$. Then, for
instance, the problem of best packing of $N$ points on $\mathbb{S}^d$ can be
formulated as finding the minimal $\beta$, such that there exists an $A$-code
with $A=[-1,\beta]$. In particular, the determination of the \emph{kissing
  number}, that is the maximum number of non-overlapping unit spheres that can
touch another given unit sphere, is equivalent to finding the maximal
cardinality of an $A$-code for $A=[-1,\frac12]$ (see~\cite{Musin}).

\subsection{Relation to lattices}

Designs have a very interesting connection to the theory of lattices that we
want to explain here. A lattice in $\R^d$ is a $\Z$-module
\begin{equation}\label{eq:lattice}
\Lambda \DEF \Z\PT{v}_1\oplus\Z\PT{v}_2\oplus\cdots\oplus\Z\PT{v}_d,
\end{equation}
where $\PT{v}_1,\ldots,\PT{v}_d$ is a basis of $\R^d$. For more
detailed information on lattices, we refer to the book
\cite{Conway_Sloane1993:sphere_packings}. A lattice $\Lambda$ is called
\emph{even}, if all the squared norms $\|\PT{v}\|^2$ for
$\PT{v}\in\Lambda$ are even. Even lattices can only exist if $d$ is
divisible by $8$.

For any lattice $\Lambda$, the dual lattice is defined by
\begin{equation}\label{eq:dual}
\Lambda^\# \DEF \{\PT{x}\in\R^d\mid\forall\PT{v}\in\Lambda:
\langle\PT{v},\PT{x}\rangle\in\Z\}.
\end{equation}
The set $\Lambda^\#$ is again a lattice. A lattice $\Lambda$ is called
\emph{unimodular}, if $\Lambda^\#=\Lambda$. For an even unimodular lattice
$\Lambda$, the $\vartheta$-series
\begin{equation}\label{eq:theta}
  \vartheta_\Lambda(\tau)=\sum_{\PT{v}\in\Lambda}
e^{\pi i\|\PT{v}\|^2\tau}, \qquad \tau\in\C,\quad \Im(\tau)>0,
\end{equation}
is a modular form of weight $d/2$; \emph{i.e.}, the transformation formula
\begin{equation*}
\vartheta_\Lambda(-1/\tau)=\tau^{d/2} \, \vartheta_\Lambda(\tau)
\end{equation*}
holds for all $\tau$ with $\Im(\tau)>0$.  This follows from an application of
Poisson's summation formula.

The theory of modular forms (see~\cite{Apostol1990:modular_functions}) yields
that for a form $f$ of weight $d/2$, given by
\begin{equation*}
f(\tau)=\sum_{n=0}^\infty a_n \, e^{2\pi in\tau},
\end{equation*}
at least one of the coefficients $a_n$ for 
$n=1,\ldots,2+2\lfloor\frac d{24}\rfloor$ has to be non-zero. This implies that
every even unimodular lattice $\Lambda$ contains a non-zero vector $\PT{v}$ 
with $\|\PT{v}\|\leq2+2\lfloor\frac d{24}\rfloor$. If the shortest non-zero
vector $\PT{v}$ in $\Lambda$ satisfies
$\|\PT{v}\|^2=2+2\lfloor\frac d{24}\rfloor$, then the lattice is called
\emph{extremal}.

For a homogeneous polynomial $p$ of degree $j\geq1$, which is also harmonic
($\triangle_{d}\,p=0$), and an even unimodular lattice $\Lambda$, the series
\begin{equation}\label{eq:theta_p}
\vartheta_{\Lambda,p}(\tau)=\sum_{\PT{v}\in\Lambda} p(\PT{v}) \,
e^{\pi i\|\PT{v}\|^2\tau}, \qquad \Im(\tau)>0,
\end{equation}
is a modular form of weight $d/2+j$. It is immediate from the symmetry of
$\Lambda$ and the homogeneity of $p$ that $\vartheta_{\Lambda,p}=0$, if $j$ is
odd. If $j$ is even, the series $\vartheta_{\Lambda,p}$ is a cusp form;
\emph{i.e.},
\begin{equation*}
\lim_{\Im(\tau)\to\infty}\vartheta_{\Lambda,p}(\tau)=0,
\end{equation*}
since $p(\PT{0})=0$.

From the theory of modular forms (see~\cite{Apostol1990:modular_functions}) it
is known that there is a unique cusp form~$\Delta(\tau)$ (this standard
notation ``Delta'' is not to be confused with the Laplace operator $\triangle$
``triangle'') of weight $12$ having an expansion of the form
\begin{equation*}
\Delta(\tau)=e^{2\pi i\tau}+\sum_{n=2}^\infty a_n \, e^{2\pi in\tau}.
\end{equation*}
It follows from degree considerations that for an even, unimodular, extremal
lattice $\Lambda$ and a homogeneous, harmonic polynomial $p$ of degree $j\geq1$,
\begin{equation}
\label{eq:theta_p-Delta}
\vartheta_{\Lambda,p}(\tau)=\Delta(\tau)^{1+\lfloor\frac d{24}\rfloor}f(\tau),
\end{equation}
where $f(\tau)$ is a modular form of weight
\begin{equation*}
\frac d2-12\left\lfloor\frac d{24}\right\rfloor+j-12.
\end{equation*}
If this weight is negative, then $f$ has to vanish identically, since there are
no modular forms of negative weight. Consider $d=24m+k$ with $k\in\{0,8,16\}$
and insert this into the above equation. This yields $k/2+j-12$ for the weight
of $f$. For $k = 0$, $8$, and $16$, this yields negative weights for $j\leq
11$, $7$, and $3$, respectively, which shows that for all shells of the
corresponding lattices the sum
\begin{equation}\label{eq:lattice-shell}
\sum_{\|\PT{v}\|^2=n}p(\PT{v}) = n^{j/2}\sum_{\|\PT{v}\|^2=n}
p\left( \PT{v} / \sqrt n \right) = 0
\end{equation}
for all homogeneous harmonic polynomials of degree $1\leq j\leq11,7,3$,
respectively. Summing up, we have proved the following theorem.
\begin{theorem}[\cite{Venkov1984:even_unimodular_lattices}]\label{thm:design}
Let $\Lambda$ be an extremal even unimodular lattice in $\R^d$ with $d=24m+k$
for $k=0,8,16$. Then any non-empty shell of $\Lambda$ defines a $j$-design
$n^{-\frac12}\{\PT{v}\in\Lambda\mid\|\PT{v}\|^2=n\}$ for $j=11-k/2$.
\end{theorem}
For further developments in the context of designs, lattices, modular forms,
and algebraic codes, we refer to \cite{Bachoc2005:designs_lattices,
Bachoc_Venkov2001:modular_forms_designs,Nebe2013:venkov_lattices_designs,
NebeVenkov1,Bannai_Bannai2009:survey_designs,Coulangeon2006:spherical}.

\subsection{Lower bounds}
In \cite{Delsarte_Goethals_Seidel1977:spherical_designs} a linear programming
method has been developed, which allows to produce lower bounds for the number
of points $N$ of a spherical $t$-design. This method has been applied
successfully to related questions, such as find the minimal cardinality
of $A(X_N)$ for sets $X_N$ with $N$ points.

We give a short explanation of this powerful method, that has been applied with
great success to several problems of discrete geometry, the most prominent
being the solution of Kepler's conjecture by T.~Hales
\cite{Hales2005:kepler_conjecture}. Since we use the method only on the sphere,
we restrict the description to this case here.  Let $f:[-1,1]\to\R^+$ be a
continuous positive function given by its expansion in terms of Legendre
polynomials; \emph{i.e.},
\begin{equation}\label{eq:legendre}
f(x) = \sum_{n=0}^\infty \widehat{f}(n) \, Z(d,n) \, P_n^{(d)}(x).
\end{equation}
If the coefficients $\widehat{f}(n)$ are non-positive for $n>t$, then any
$t$-design $X_N$ has to have cardinality $N\geq f(1) / \widehat{f}(0)$. The
proof of this fact is given by the relations
\begin{align*}
f(1) N
&\leq \sum_{(\PT{x},\PT{y}) \in X_N\times X_N} f(\langle \PT{x},\PT{y}\rangle) \\
&= \widehat{f}(0) \left(N \right)^2 + \sum_{n>t} \widehat{f}(n) \, 
Z(d,n) \sum_{(\PT{x},\PT{y})\in X_N\times X_N} 
P_n^{(d)}(\langle \PT{x},\PT{y}\rangle) \\
&\leq \widehat{f}(0) N^2.
\end{align*}
The first inequality is a consequence of the positivity of $f$, the equality
uses the characterisation of $t$-designs by \eqref{eq:design-gegenbauer}, and
the second inequality uses that $\widehat{f}(n)\leq0$ for $n>t$ and the
non-negativity of the double sums $\sum_{(\PT{x},\PT{y})\in X_N\times X_N}
P_n^{(d)}(\langle\PT{x},\PT{y}\rangle)$. Equality can only occur, if $f(x)=0$
for all $x \in A(X_N)$ and $\widehat{f}(n) = 0$ for $n > t$; \emph{i.e.}, $f$
is a polynomial of degree $\leq t$.

In \cite{Delsarte_Goethals_Seidel1977:spherical_designs} a polynomial $p_t$ of
degree $t$ was constructed, which gives the lower bound
\begin{equation}\label{eq:delsarte}
 N\geq \begin{cases}
    \binom {d+t/2}{d}+\binom{d+t/2-1}{d}&\text{for }t\text{ even},\\
    2\binom{d+(t-1)/2}{d}&\text{for }t\text{ odd.}
  \end{cases}
\end{equation}
This polynomial actually provides the best possible lower bound that can be
obtained by polynomial functions $f$. Designs attaining this lower bound are
called \emph{tight}. It was shown in
\cite{Bannai_Damerell1979:tight_designs_i,Bannai_Damerell1980:tight_designs_ii}
that tight designs only exist for $d=1$ and all $t$, or finitely many values of
$t$ if $d\geq2$. The proof uses the fact that for a tight spherical $t$-design
$X_N$, the corresponding set $A(X_N)$ has to consist exactly of the zeros of the
polynomial $p_t$ constructed in
\cite{Delsarte_Goethals_Seidel1977:spherical_designs}. Galois theory is then
used to derive a contradiction, if $t$ is large enough. The same fact about
$A(X_N)$ shows that tight designs are rigid
(see~\cite{Bannai1987:rigid_designs}). Notice that the bound is of order
$t^d$. Later, V.~A.~Yudin \cite{Yudin1997:lower_design} considered a wider
class of functions than polynomials. He constructed a function $f$, which
allowed for a considerable improvement of the lower bound.  The bounds for
spherical designs he gives are again of order $t^d$, but they are larger by a
factor depending on $d$. This factor is exponential in $d$. For $d=2$, the gain
is asymptotically about $7\%$.

We give a short description of Yudin's construction. The function $f$ is
obtained as the spherical convolution of two positive functions $F$ and $G$,
which ensures the positivity of $f$. The spherical convolution of two functions
$F$ and $G$ is given by
\begin{equation*}
F \star G(\langle\PT{x},\PT{y}\rangle) \DEF
\int_{\mathbb{S}^d}F(\langle\PT{x},\PT{z}\rangle)G(\langle\PT{z},\PT{y}\rangle)
\, \dd\sigma_d(\PT{z}).
\end{equation*}
In order to obtain the required sign
change of the Laplace-Fourier coefficients of $f$, the function $G$ is chosen as
\begin{equation*}
  G(x) = (1-x^2)^{1-\frac d2}\left((1-x^2)^{\frac d2}F'(x)\right)'+(t+1)(t+d)F(x),
\end{equation*}
which is motivated by the fact that the Legendre polynomials are eigenfunctions
of the differential operator $L$ defined by
\begin{equation*}
LF(x) \DEF (1-x^2)^{1-\frac d2}\left((1-x^2)^{\frac d2}F'(x)\right)'.
\end{equation*}
If $F$ is expressed in terms of its Laplace-Fourier expansion
\begin{equation*}
F(x) = \sum_{n=0}^\infty \widehat{F}(n) \,  Z(d,n) \, P_n^{(d)}(x),
\end{equation*}
then the function $G$ is given by
\begin{equation*}
G(x) = \sum_{n=0}^\infty 
\left(t+1-n\right) \left(n+t+d\right)\widehat{F}(n) \, Z(d,n) \, P_n^{(d)}(x).
\end{equation*}
Hence the spherical convolution is given as
\begin{equation*}
f(x) = F \star G(x) = \sum_{n=0}^\infty 
\left(t+1-n\right) \left(n+t+d\right) \widehat{F}(n)^2 \, Z(d,n) \,P_n^{(d)}(x),
\end{equation*}
which has the required sign-change in its Laplace-Fourier coefficients. The
problem of maximising the quotient $f(1)/\widehat{f}(0)=f(1)/\widehat{F}(0)^2$
turns out to be a variational problem for $F$:
\begin{align*}
f(1)
&= \frac{\omega_{d-1}}{\omega_d} \int_{-1}^1 F(x) G(x) 
\left(1-x^2\right)^{d/2-1} \, \dd x \\
&= \frac{\omega_{d-1}}{\omega_d} \int_{-1}^1 
\Big( -\left(1-x^2\right) F^\prime(x)^2 + \left(t+1\right) \left(t+d \right) 
F(x)^2 \Big) \left(1-x^2\right)^{d/2-1} \, \dd x
\end{align*}
must be maximised subject to the condition
\begin{equation*}
\widehat{F}(0)=
\frac{\omega_{d-1}}{\omega_d}\int_{-1}^1 F(x)\left(1-x^2\right)^{d/2-1}\,\dd x=1.
\end{equation*}
The symbol $\omega_d$ denotes the
surface area of~$\mathbb{S}^d$ and it satisfies
\begin{equation*}
\int_{-1}^1 \left(1-x^2\right)^{d/2-1}\dd x = \frac{\omega_d}{\omega_{d-1}}=
\frac{\sqrt{\pi} \, \gammafcn( d/2 )}{\gammafcn( (d+1)/2 )}.
\end{equation*}
Furthermore, $f$ has to be non-negative on $[-1,1]$. This is
achieved by assuming that $F$ and $G$ are both non-negative. The solution of
this variational problem with the additional condition on the sign of $F$ is
then given by
\begin{equation*}
F(x)=
\begin{cases}
  P_{t+1}^{(d)}(x) - P_{t+1}^{(d)}(\alpha_t)&\text{for }\alpha_t\leq x\leq1, \\
  0&\text{for }-1\leq x\leq\alpha_t,
\end{cases}
\end{equation*}
where $\alpha_t$ is the largest zero of $\frac{d}{dx}P_{t+1}^{(d)}(x)$. In this
case $G$ is a piecewise constant function. Putting everything together yields
the lower bound
\begin{equation}\label{eq:yudin}
N \geq \frac{\int_{-1}^1 \left( 1-x^2 \right)^{d/2-1}\, \dd x}
{\int_{\alpha_t}^1 \left( 1-x^2 \right)^{d/2-1}\, \dd x} = 
\frac{1}{\sigma_d( \{ \PT{y} \in \mathbb{S}^d \mid 
\langle \PT{y}, \PT{x} \rangle \geq \alpha_t \} )} \gg_d t^d
\end{equation}
for a $t$-design $X_N$.  For a more detailed exposition, we refer to
\cite{Yudin1997:lower_design}.  Even if the obtained function $f$ gives a
better value for the lower bound than the polynomial given in
\cite{Delsarte_Goethals_Seidel1977:spherical_designs}, the technical
requirement of non-negativity of $F$ seems to leave room for further
improvement.

A similar construction for a function $f$ is used in
\cite{Cohn_Elkies2003:bounds_sphere_packings_i} to obtain linear programming
bounds for the packing density of spheres in $\R^d$. The function obtained
there has similar features; in particular, it is supported on a short
interval. In \cite[Section~5]{Cohn_Elkies2003:bounds_sphere_packings_i} it is
mentioned that the function constructed by this convolution method does
\textbf{not} produce optimal bounds
(see~\cite{Cohn2002:bounds_sphere_packings_ii}). The reason seems to be exactly
the non-negativity requirement on the corresponding function $F$. Thus there is
reason to believe that Yudin's lower bound for the cardinality of $t$-designs
can still be improved.

\subsection{Existence}

On the other hand, the question of the existence of spherical $t$-designs has
been answered affirmatively. First, a rather general result obtained in
\cite{Seymour_Zaslavsky1984:averaging_designs} shows that, for given $t$ and
large enough $N$, there exists a $t$-design with $N$ points. Actually, the
result given in \cite{Seymour_Zaslavsky1984:averaging_designs} is more general:
given a path connected topological space $\Omega$ and a finite measure $\mu$
that charges every non-empty open set, then for any finite set of continuous
real valued functions $f_1,\ldots,f_t$ there exists an $N_0$, such that for
every $N>N_0$ there exists a set $X_N\subset\Omega$, such that
\begin{equation*}
\frac1N\sum_{x\in X_N}f_j(x)=\frac1{\mu(\Omega)}\int_\Omega f_j(x)\,\dd\mu(x)
\end{equation*}
for $j=1,\ldots,t$. The result gives a bound for the number of points needed by
geometric quantities defined in terms of $\Omega$ and the functions
$f_1,\ldots,f_t$. These quantities are very difficult to compute even in the
special case of the sphere.

In order to discuss the question of existence of spherical designs further, we
introduce two quantities:
\begin{align*}
  N(t,d)&\DEF \min\left\{N\mid \exists X_N\subset\mathbb{S}^d: X_N
\text{ is a }t\text{-design}\right\}, \\
  N^*(t,d)&\DEF \min\left\{K\mid \forall N\geq K:\exists 
X_N\subset\mathbb{S}^d:  X_N\text{ is a }t\text{-design}\right\}.
\end{align*}
By definition it is clear that $N^*(t,d)\geq N(t,d)$.

Only very recently, the long standing problem of the existence of spherical
$t$-designs with $\mathcal{O}(t^d)$ points was answered affirmatively by
A.~V.~Bondarenko, D.~Radchenko, and M.~S.~Viazovska
\cite{Bondarenko_Radchenko_Viazovska2011:optimal_bounds_designs}. They proved
that $N^*(t,d)=\mathcal{O}(t^d)$ with an explicit, but large implied constant.
The proof puts the existence of spherical $t$-designs with $N=\mathcal{O}(t^d)$
in the context of Brouwer's degree theorem: Let $\mathcal{P}_t$ denote the
Hilbert space of polynomials $P$ of total degree $\leq t$ with
\begin{equation*}
\int_{\mathbb{S}^d}P(\PT{x})\,\dd\sigma_d(\PT{x})=0
\end{equation*}
equipped with the scalar product
\begin{equation*}
\langle P,Q\rangle=\int_{\mathbb{S}^d}P(\PT{x})Q(\PT{x})\,\dd\sigma_d(\PT{x}).
\end{equation*}
Then for every $\PT{x}\in\mathbb{S}^d$ there exists a polynomial $G_{\PT{x}}$
such that
\begin{equation*}
  \langle G_{\PT{x}},Q\rangle=Q(\PT{x})
\end{equation*}
for all $Q\in\mathcal{P}_t$. A set of points $X_N$ then is a spherical
$t$-design, if
\begin{equation*}
G_{\PT{x}_1}+\cdots+G_{\PT{x}_N}\equiv0.
\end{equation*}
Then continuous maps $\PT{x}_i:\mathcal{P}_t\to\mathbb{S}^d$ are constructed,
which in turn define the map $f:\mathcal{P}_t\to\mathcal{P}_t$ by
\begin{equation*}
f(P)\DEF G_{\PT{x}_1(P)}+\cdots+G_{\PT{x}_N(P)}.
\end{equation*}
Then for any $P\in\mathcal{P}_t$ the identity
\begin{equation*}
\langle P,f(P)\rangle=\sum_{i=1}^NP(\PT{x}_i(P))
\end{equation*}
holds. A polynomial $\tilde P$ with $f(\tilde P)=0$ then gives a spherical
$t$-design $\{\PT{x}_1(\tilde P),\ldots,\PT{x}_N(\tilde P)\}$. The construction
of the maps $\PT{x}_i$ is the crucial part of the proof. Starting with points
$\{\PT{x}_1(0),\ldots,\PT{x}_N(0)\}$ in the parts of an equal area partition of
the sphere, the maps $\PT{x}_i$ are defined by an intricate geometric
procedure. The proof finishes by considering the set
\begin{equation*}
\Omega=\left\{P\in\mathcal{P}_t\mid 
\int_{\mathbb{S}^d}\|\nabla P(\PT{x})\|\,\dd\sigma_d(\PT{x})<1\right\}
\end{equation*}
and observing that the construction of $\PT{x}_i$ yields
\begin{equation*}
  \langle P,f(P)\rangle>0
\end{equation*}
for $P\in\partial\Omega$. This last inequality is verified by an application of
a spherical version of the Marcinkiewicz--Zygmund inequality. In this step the
precise choice of the number $N$ of points in relation to $t$ is
significant. The Brouwer degree theorem gives the existence of a point $\tilde
P$ with $f(\tilde P)=0$, which the yields the desired spherical $t$-design
$\{\PT{x}_1(\tilde P),\ldots,\PT{x}_N(\tilde P)\}$.

In a recent paper \cite{Bondarenko_Radchenko_Viazovska2014:well_separated}, the
same authors showed that there exist well separated $t$-designs with optimal
order of the number of points:
\begin{theorem}\label{thm:BRV2}
Let $d\geq 2$. Then there exist positive constants $C_d$ and $\beta_d$ such
that for every $N\geq C_d t^d$ there exists a $t$-design $X_N$ and
\begin{equation*}
\left\| \PT{x} - \PT{y} \right\| \geq \beta_d \, N^{-1/d} \qquad 
\text{for all $\PT{x}, \PT{y} \in X_N$ with $\PT{x} \neq \PT{y}$.}
\end{equation*}
\end{theorem}
The proof is a refinement of their original proof by keeping control on the
distance of distinct points.

\subsection{Numerical results}
Besides the theoretical investigation of the existence of spherical designs,
several attempts were made to compute lists of $t$-designs for moderately large
values of $t$ and $N$. Since most of these computations have been done for
$d=2$, we will restrict to that case in this section.

A first list of $t$-designs for $t\leq21$ was provided by R.~H.~Hardin and
N.~J.~A.~Sloane (see~\cite{Hardin_Sloane1996:mclarens_snub_cube,
Hardin_Sloane1995:codes_designs,Hardin_Sloane1993:optimal_designs}). Their list
is still available on the web \cite{Hardin_Sloane2002:table}. These numerical
computations, as well as those performed in
\cite{Sloan_Womersley2009:variational_designs}, seem to suggest that
$N(t,2)$ is close to $\frac12(t+1)^2$.
Furthermore, X.~Chen and R.~S.~Womersley found spherical $t$-designs by
numerical computations for $t\leq100$
(see~\cite{Chen_Womersley2006:existence_designs}). Their computations seem to
indicate that there exist $t$-designs with less than $(t+1)^2$ points. Later
X.~Chen, A.~Frommer, and B.~Lang
\cite{Chen_Frommer_Lang2011:computational_designs} used interval arithmetic to
prove that there exist spherical $t$-designs with $(t+1)^2$ points for $1\leq
t\leq100$.
Recently, M.~Gr\"af and D.~Potts
\cite{Graef_Potts2011:fourier,Graef2013:efficient_algorithms_computation}
derived a new method based on fast Fourier transform, which allows to find
$t$-designs numerically with high precision for values of $t$ up to
$1000$. They also provide their results on the web
\cite{Graef_Potts2013:table}. Again these numerical results show that $N(t,2)$
is close to $\frac12(t+1)^2$ for small values of $t$.

\begin{table}[h]\setlength\extrarowheight{4pt}\tiny
  \centering

  \caption{\label{tab:design} The lower bounds for the number of points of a 
    $t$-design derived in
    \cite{Delsarte_Goethals_Seidel1977:spherical_designs} ($\mathrm{DGS}(t)$)
    and in 
    \cite{Yudin1997:lower_design} ($\mathrm{Y}(t)$) compared to the number of
    points of $t$-designs $N(t,2)$ obtained in \cite{Graef_Potts2011:fourier} 
    (as provided on the website \cite{Graef_Potts2013:table}).}

\medskip

  \begin{tabular}{|l|c|c|c|c|c|c|c|c|c|c|c|c|c|c|c|}
\hline
$t$& 5 & 7 & 9 & 10 & 20 & 30 & 40 & 50 & 60 & 70 & 80 & 90 & 100 & 114 & 124 
\\[2mm]
\hline\hline
$\mathrm{DGS}(t)$& 12 & 20 & 30 & 36 & 121 & 256 & 441 & 676 & 961 & 1296 & 
1681 & 2116 & 2601 & 3364 & 3969 \\[2mm]
\hline
$\mathrm{Y}(t)$& 12 & 20 & 31 & 37 & 127 & 271 & 470 & 723 & 1031 & 1394 & 
1810 & 2282 & 2808 & 3635 & 4292 \\[2mm]
\hline
$N(t,2)$& 12 & 24 & 48 & 60 & 216 & 480 & 840 & 1296 & 1860 & 2520 & 3276 
& 4140 & 5100 & 6612 & 7812 \\[2mm]
\hline
$\lfloor\frac{(t+1)^2}2\rfloor$&18&32&50&60&220&480&840&1300&1860&2520&3280&
4140&5100&6612
&7812\\[2mm]
\hline
\end{tabular}
\end{table}

\subsection{Designs and uniform distribution}
\label{sec:designs-unif-distr}
It was observed independently in
\cite{Grabner_Tichy1993:spherical_designs_discrepancy} and in
\cite{Korevaar_Meyers1993:spherical_faraday_chebyshev} that spherical designs
provide well-distributed point sets on the sphere. 

The \emph{spherical cap discrepancy} of a point set $X_N$ of $N$ points is
given by
\begin{equation}\label{eq:discrepancy}
D(X_N) \DEF \sup_{\substack{\PT{y}\in\mathbb{S}^d, \\ \phi\in[0,\pi]}}
\left|\frac{1}{N}\sum_{j=1}^N\mathbbm{1}_{C(\PT{y},\phi)}(\PT{x}_j)
-\sigma_d(C(\PT{y},\phi))\right|;
\end{equation}
the supremum is extended over all spherical caps
\begin{equation*}
C(\PT{y},\phi)\DEF\left\{\PT{x}\in\mathbb{S}^d\mid 
\langle\PT{x},\PT{y}\rangle>\cos(\phi)\right\},
\end{equation*}
and measures the maximum
deviation between the empirical distribution of the point set~$X_N$ from
uniform distribution. In \cite{Grabner1991:erdoes_turan_type} the estimate
\begin{equation}\label{eq:erdoes-turan}
D(X_N)\leq\frac{C_1(d)}{M} + \sum_{\ell=1}^M\frac{C_2(d)}{\ell} 
\sum_{k=1}^{Z(d,\ell)}
\left| \frac{1}{N} \sum_{j=1}^N Y_{\ell,k}(\PT{x}_j) \right|
\end{equation}
was proved; here $M$ is an arbitrary positive integer, and $C_1(d)$ and
$C_2(d)$ are explicit constants depending only on the dimension $d$. A similar
inequality was later given in \cite{Li_Vaaler1999:trigonometric_extremal}. The
inequality \eqref{eq:erdoes-turan} resembles the classical
Erd\H{o}s-Tur\'an-Koksma inequality estimating the Euclidean discrepancy of a
point set in $[0,1]^d$ in terms of trigonometric sums
(see~\cite{Kuipers_Niederreiter1974:uniform_distribution_sequences}). In
\cite{Korobov1959:approximate_integrals}, N.~M.~Korobov introduced \emph{good
  lattice points} $(g_1,\ldots,g_d)\in\mathbb{Z}^d$ by the requirement that the
point set $\{ (\{\frac{jg_1}N\},\ldots,\{\frac{jg_d}N\}) \mid j=0,\ldots,N-1
\}$ has small discrepancy.

In \cite{Grabner_Tichy1993:spherical_designs_discrepancy} spherical $t$-designs
were regarded as spherical analogues of good lattice points, in the sense that
the estimate \eqref{eq:erdoes-turan} becomes particularly simple when applied
to a $t$-design and choosing the parameter $M$ to be $t$: the estimate then
reduces to
\begin{equation*}
D(X_N)\leq\frac{C_1(d)}{t}.
\end{equation*}
Similarly, for a continuous function $f:\mathbb{S}^d\to\R$ satisfying the
Lipschitz-condition $|f(\PT{x})-f(\PT{y})|\leq
C_f\arccos(\langle\PT{x},\PT{y}\rangle)$, the estimate
\begin{equation}\label{eq:integration_error}
\left|\frac{1}{N}\sum_{\PT{x}\in X_N}f(\PT{x})-
\int_{\mathbb{S}^d}f(\PT{x})\,\dd\sigma_d(\PT{x})\right|
\leq C_f\left(6\frac dM+\pi\sum_{\ell=1}^{2M}\sum_{k=1}^{Z(d,\ell)}
\left|\frac{1}{N}\sum_{\PT{x}\in X_N} Y_{\ell,k}(\PT{x})\right|\right)
\end{equation}
was shown in the same paper. Again, taking $X_N$ to be a $2t$-design and $M=t$
gives an estimate $6dC_f/t$ for the integration error.

J.~Korevaar and J.~L.~H.~Meyers
\cite{Korevaar_Meyers1993:spherical_faraday_chebyshev} take a potential
theoretic point of view.  In their papers
\cite{Korevaar_Meyers1993:spherical_faraday_chebyshev,
  Korevaar_Meyers1994:chebyshev_quadrature} they conjecture that
$N(t,d)=\mathcal{O}(t^d)$, which was finally proved in
\cite{Bondarenko_Radchenko_Viazovska2011:optimal_bounds_designs}.  Let $\mu_N
\DEF \frac{1}{N}\sum_{\PT{x}\in X_N}\delta_{\PT{x}}$ be the discrete equal
weight distribution supported on $X_N\subset\mathbb{S}^d$. Then the deviation
of $\mu_N$ and the equilibrium measure $\sigma_d$ is measured by the deviation
of the potential
\begin{equation}\label{eq:potential}
U_{d-1}^{\mu_N}(\PT{x}) = \frac{1}{N}\sum_{\PT{y}\in X_N}
\frac1{\|\PT{x}-\PT{y}\|^{d-1}}=
\int_{\mathbb{S}^d}\frac1{\|\PT{x}-\PT{y}\|^{d-1}}\,
\dd\mu_N(\PT{y})
\end{equation}
from the equilibrium potential
\begin{equation*}
U_{d-1}^{\sigma_d}(\PT{x})=
\int_{\mathbb{S}^d}\frac1{\|\PT{x}-\PT{y}\|^{d-1}}\,
\dd\sigma_d(\PT{y})=1
\end{equation*}
for $\|\PT{x}\|\leq r<1$.

Taking a spherical $t$-design $X_N$ and using the identity
(see~\cite{Magnus_Oberhettinger_Soni1966:formulas_theorems})
\begin{equation*}
\frac1{\|r\PT{x}-\PT{y}\|^{d-1}}=
\sum_{n=0}^\infty\binom{n+d-2}{d-2}
P_n^{(d)}(\langle\PT{x},\PT{y}\rangle) \, r^n,
\end{equation*}
we obtain
\begin{equation}\label{eq:potential_mun}
U_{d-1}^{\mu_N}(r\PT{x})-1=
\sum_{n=t+1}^\infty\binom{n+d-2}{d-2}r^n\frac{1}{N}\sum_{\PT{y}\in X_N}
P_n^{(d)}(\langle\PT{x},\PT{y}\rangle).
\end{equation}
Estimating the right hand side by $|P_n^{(d)}(\cdot)|\leq1$, we obtain
\begin{equation}\label{eq:potential_est}
\left|U_{d-1}^{\mu_N}(r\PT{x})-1\right|\leq
\sum_{n=t+1}^\infty \binom{n+d-2}{d-2}r^n.
\end{equation}
The sum on the right hand side can be expressed in closed form.
\begin{align*}
&\sum_{n=t+1}^\infty \binom{n+d-2}{d-2}r^n\\
&\phantom{equals}=(d-1)\binom{t+d-1}{d-1}\frac1{(1-r)^{d-1}}
\int_0^r(1-\rho)^{d-2}\rho^{t}\,\dd\rho\\
&\phantom{equals}=\frac{r^{t+1}}{(1-r)^{d-1}}(d-1)\binom{t+d-1}{d-1}
\sum_{\ell=0}^{d-2}\binom{d-2}\ell(-1)^\ell\frac{r^\ell}{t+\ell+1},
\end{align*}
which can be proved by multiplying with $(1-r)^{d-1}$ and differentiating.
Estimating the integral in the second line by $r^{t+1}/(t+1)$ we get the
estimate 
\begin{equation}\label{eq:potential_design}
\left|U^{\mu_N}_{d-1}(r\PT{x})-1\right|\leq
\binom{t+d-1}{d-2}\frac{r^{t+1}}{(1-r)^{d-1}},
\end{equation}
valid for any $t$-design $X_N$, $\PT{x}\in\mathbb{S}^d$, and $0\leq
r<1$. This, together with the fact that $N(t,d)=\mathcal{O}(t^d)$, generalises
\cite[Theorem~2.2]{Korevaar_Meyers1993:spherical_faraday_chebyshev} to
arbitrary dimension.

\subsection{Applications to numerical integration on 
$\mathbb{S}^d$}

Equal weight quadrature formulas like spherical design QMC methods are
especially useful for integrating functions taken from suitably defined Sobolev
spaces on $\mathbb{S}^d$. These spaces are \emph{reproducing kernel Hilbert
  spaces}, which makes the study of the \emph{worst case error} in integration
particularly simple and transparent
(see~\cite{Novak_Wozniakowski2008:tractability_vol1}).

In order to describe the results on numerical integration in more detail, we
give a precise definition of the function spaces. The negative Laplace-Beltrami
operator $-\triangle_d^*$ on $\mathbb{S}^d$ has the eigenvalues $\lambda_\ell
\DEF \ell (\ell+d-1)$, $\ell\in\N_0$. The space of eigenfunctions for the
eigenvalue $\lambda_\ell$ is spanned by the spherical harmonics $Y_{\ell,k}$
for $k=1,\ldots,Z(d,\ell)$. Since the eigenfunctions of $-\triangle_d^*$ form a
complete function system, every function $f\in L^2(\mathbb{S}^d)$ can be
represented by its Laplace-Fourier series expansion
\begin{equation}\label{eq:fourier}
f(\PT{x}) = \sum_{\ell =0}^\infty \sum_{k=1}^{Z(d,\ell)} \widehat{f}_{\ell,k} \, 
Y_{\ell,k}(\PT{x}),
\end{equation}
where the Laplace-Fourier coefficients are given by 
\begin{equation*}
\widehat{f}_{\ell,k} = \int_{\mathbb{S}^d}f(\PT{x}) \, Y_{\ell,k}(\PT{x}) \, \dd 
\sigma_d(\PT{x}).
\end{equation*}
The series \eqref{eq:fourier} has to be interpreted in the $L^2$-sense.
Furthermore, Parseval's identity 
\begin{equation}\label{eq:parseval}
\sum_{\ell=0}^\infty\sum_{k=1}^{Z(d,\ell)}\left|\widehat{f}_{\ell,k}\right|^2 =
\int_{\mathbb{S}^d} \left| f(\PT{x}) \right|^2\,\dd\sigma_d(\PT{x}) = \|f\|_2^2 
\end{equation}
and the Funk-Hecke formula
\begin{equation}\label{eq:funk-hecke}
  \sum_{k=1}^{Z(d,\ell)}\widehat{f}_{\ell,k} \, Y_{\ell,k}(\PT{x}) =
Z(d,\ell) \int_{\mathbb{S}^d} f(\PT{y}) \, 
P_\ell^{(d)}\big(\langle \PT{x}, \PT{y} \rangle\big) \, \dd\sigma_d(\PT{y})
\end{equation}
hold. For more details on the harmonic analysis on $\mathbb{S}^d$, we refer to
\cite{Mueller1966:spherical_harmonics}.

In the following we adopt the notation of
\cite{Brauchart_Saff_Sloan+2014:qmc-designs}.  For $s\geq0$, we define the
space
\begin{equation}\label{eq:Hs}
H^s(\mathbb{S}^d) \DEF \left\{ f\in L^2(\mathbb{S}^d) \,\,\, \Bigg| \,\,\,
\sum_{\ell=0}^\infty(1+\ell)^{2s} \sum_{k=1}^{Z(d,\ell)}
\left|\widehat{f}_{\ell,k}\right|^2 < \infty \right\}.
\end{equation}
From this definition, it is clear that the ``sequence of weights''
$((1+\ell)^{2s})$ can be replaced by any other \emph{comparable} sequence
$(w_\ell)$, in the sense of
\begin{equation*}
\exists C_1,C_2>0:\forall \ell\in\N_0: C_1(1+\ell)^{2s}\leq w_\ell
\leq C_2(1+\ell)^{2s}.
\end{equation*}
We define a norm on $H^s(\mathbb{S}^d)$ by
\begin{equation}\label{eq:norm}
\|f\|_{H^s} \DEF \left( \sum_{\ell=0}^\infty
(1+\ell)^{2s}\sum_{k=1}^{Z(d,\ell)}
\left|\widehat{f}_{\ell,k}\right|^2\right)^{1/2}.
\end{equation}
From the definition it is clear that the spaces $H^s(\mathbb{S}^d)$ are getting
smaller as the index of smoothness $s$ increases.  Furthermore, the Sobolev
embedding theorem ensures that $H^s(\mathbb{S}^d)$ embeds continuously into
$C^k(\mathbb{S}^d)$ when $s>k+d/2$; in particular, $H^s(\mathbb{S}^d)$ embeds
continuously into $C(\mathbb{S}^d)$ when $s>d/2$.

As a consequence of the embedding of $H^s(\mathbb{S}^d)$ into $C(\mathbb{S}^d)$
for $s>d/2$, point evaluation of a function is a continuous
functional, which can be represented as a scalar product by the Riesz
representation theorem. This ensures the existence of a reproducing kernel
given by
\begin{equation}\label{eq:kernel}
\KERNEL^{(s)}\big(\langle \PT{x},\PT{y} \rangle\big) \DEF
\sum_{\ell=0}^\infty \left(1+\ell\right)^{-2s} Z(d,\ell) \,
P_\ell^{(d)}\big(\langle\PT{x},\PT{y}\rangle\big).
\end{equation}
With this definition the reproducing kernel properties
\begin{equation}\label{eq:reproducing}
\forall \PT{y}\in\mathbb{S}^d: \KERNEL^{(s)}(\langle\cdot,\PT{y} \rangle) \in 
H^s(\mathbb{S}^d) \text{ and }
\forall f\in H^s(\mathbb{S}^d)\,\,\forall \PT{x}\in\mathbb{S}^d:
f(\PT{x})=\langle f,\KERNEL^{(s)}(\langle\cdot,\PT{x}\rangle)\rangle_{H^s}
\end{equation}
can be immediately verified.

For $f\in H^s(\mathbb{S}^d)$, $s>d/2$, the integration error of the
QMC method with node set $X_N$ is given by
\begin{equation}\label{eq:int-error}
Q(X_N)(f) \DEF \frac{1}{N} \sum_{\PT{x}\in X_N}f(\PT{x}) -
\int_{\mathbb{S}^d} f(\PT{x})\, \dd\sigma_d(\PT{x}) =
\left\langle f, R(X_N) \right\rangle_{H^s},
\end{equation}
where the function
\begin{equation}\label{eq:representer}
R(X_N)(\PT{y}) \DEF \frac{1}{N} \sum_{\PT{x}\in X_N} 
\KERNEL^{(s)}(\langle\PT{y},\PT{x}\rangle)-1
\end{equation}
is called the \emph{representer of the integration error}. It is now a
consequence of elementary Hilbert space theory that the worst case error takes
the form
\begin{equation}\label{eq:worst-case}
\WCE_{H^s}(X_N) \DEF \sup_{\|f\|_{H^s}=1} \left| Q(X_N)(f) \right| = 
\left\| R(X_N) \right\|_{H^s}.
\end{equation}
The squared norm $\|R(X_N)\|^2 = \langle R(X_N), R(X_N)\rangle$ can be
expressed in terms of the kernel function $\KERNEL^{(s)}$ by means of
\begin{equation}\label{eq:worst-case-K}
\left\|R(X_N)\right\|_{H^s}^2=\frac{1}{N^2} \sum_{\PT{x},\PT{y}\in X_N}
\KERNEL^{(s)}(\langle\PT{x},\PT{y}\rangle)-1.
\end{equation}
For a more precise explanation of this formalism we refer to
\cite{Brauchart_Saff_Sloan+2014:qmc-designs}.  The expression
\eqref{eq:worst-case-K} is a special case of an energy functional as discussed
further in Section~\ref{sec:energy}.

Taking a spherical $t$-design for the set $X_N$ in \eqref{eq:worst-case-K}, the
first $t$ terms in the Laplace-Fourier expansion of $\KERNEL^{(s)}$ are
annihilated:
\begin{equation}\label{eq:error-design}
\WCE_{H^s}(X_N)^2 = \sum_{\ell=t+1}^\infty 
\left(1+\ell\right)^{-2s} Z(d,\ell) \, \frac{1}{N^2}
\sum_{\PT{x},\PT{y}\in X_N} P_\ell^{(d)}(\langle\PT{x},\PT{y}\rangle).
\end{equation}
Trivially estimating this sum and taking square roots would give a bound
$\mathcal{O}(t^{d/2-s})$ for the worst case error.

A method is introduced in \cite{Hesse_Sloan2005:worst_case_sobolev} and
\cite{Hesse_Sloan2006:cubature_s_sobolev}, which provides much better estimates
for the worst case error in $H^s(\mathbb{S}^2)$. This method was extended and
generalised to higher dimensions in
\cite{Brauchart_Hesse2007:numerical_integration}. This is now a standard
technique in this context. While too technical to give a precise description
here, we explain the two main steps. For a full description of the method we
refer to \cite{Brauchart_Hesse2007:numerical_integration}.

First, the tail of the series defining the kernel
\begin{equation*}
\KERNEL_t^{(s)}(x) \DEF \sum_{\ell=t+1}^\infty 
\left(1+\ell\right)^{-2s} Z(d,\ell) \, P_\ell^{(d)}(x)
\end{equation*}
is rewritten in terms of Jacobi polynomials instead of Legendre
polynomials. This is done using the fact that the Legendre polynomials are
special cases of Jacobi polynomials and that there exist connection formulas
involving hypergeometric expressions
(see~\cite{Magnus_Oberhettinger_Soni1966:formulas_theorems}). After this
transformation, the function $K_t^{(s)}$ is expressed as a polynomial of degree
$t$ plus a series $\widetilde{K}_t^{(s)}$ involving higher order polynomials
(``kernel splitting method''). The polynomial part is integrated exactly by the
QMC method supported in $X_N$; the remaining part
\begin{equation*}
\sum_{\PT{x},\PT{y}\in X_N}\widetilde{\KERNEL}_t^{(s)}
(\langle\PT{x},\PT{y}\rangle)
\end{equation*}
is estimated using bounds for Jacobi polynomials. This part is rather
delicate, since different estimates have to be used for
$\langle\PT{x},\PT{y}\rangle\in[-1+c/t,1-c/t]$ and
$|\langle\PT{x},\PT{y}\rangle|>1-c/t$. Here, a coarse equidistribution property
of spherical $t$-designs is used to obtain optimal order estimates: there exist
constants $c_1$ and $c_2$, independent of $t$ and $N$, such that
\begin{equation}\label{eq:reimer}
\#\left(X_N\cap C\left(\PT{x},\frac{c_1}{t}\right)\right) \leq 
c_2N\sigma_d \left( C\left(\PT{x},\frac{c_1}{t}\right) \right).
\end{equation}
It was proved in \cite{Reimer2000:hyperinterpolation_sphere} (see also
\cite{Reimer2001:geometry_nodes}) that this is a general property of positive
weight quadrature formulas, which integrate polynomials of degree $\leq t$
exactly; thus \eqref{eq:reimer} holds \emph{a forteriori} for spherical
$t$-designs.

Using the technique described above, it was proved in
\cite{Brauchart_Hesse2007:numerical_integration} that the worst case
integration error on $H^s(\mathbb{S}^d)$ (for $s>d/2$) is
\begin{equation}\label{eq:wce}
\WCE_{H^s}(X_N) = \BigOh_{s,d}\big(t^{-s}\big).
\end{equation}
Using the recent progress on the existence of spherical designs with optimal
growth order
(see~\cite{Bondarenko_Radchenko_Viazovska2011:optimal_bounds_designs}), this
gives
\begin{equation}\label{eq:wce2}
\WCE_{H^s}(X_N) = \BigOh_{s,d}\big( N^{-s/d}\big)
\end{equation}
if $X_N$ is a spherical $t$-design with $N=\mathcal{O}(t^d)$. This is known to
be the optimal order for the worst case error in $H^s(\mathbb{S}^d)$
(see~\cite{Hesse2006:lower_worst-case,Hesse_Sloan2005:optimal_lower}).

The estimate \eqref{eq:wce2} led to the following definition in
\cite{Brauchart_Saff_Sloan+2014:qmc-designs}. Let $s>d/2$, then a sequence of
point sets $(X_N)$ on $\mathbb{S}^d$ is called a \emph{sequence of
  QMC-designs} for $H^s(\mathbb{S}^d)$, if \eqref{eq:wce2} holds. The supremum
over all $s$ for which \eqref{eq:wce2} holds, is called the \emph{strength} of
the sequence $(X_N)$. By the above description, a sequence of spherical
$t$-designs of optimal growth order $\mathcal{O}(t^d)$ has strength
$\infty$. For a more detailed exposition of QMC-designs and their properties,
we refer to the original paper
\cite{Brauchart_Saff_Sloan+2014:qmc-designs}. These investigations were
extended to the $L^p$-setting in \cite{BrSaSl2014arXiv}.

\subsection{Applications in approximation and interpolation}
\label{sec:appl-appr-interp}

Well distributed point sets can also be used as sample points for interpolation
formulas. Let $f:\mathbb{S}^d\to\R$ be a continuous function and $X_N$ a set of
points with $N\geq Z(d+1,t)$. Observe that
\begin{equation*}
Z(d+1,t)=\sum_{k=0}^t Z(d,k)
\end{equation*}
is the dimension of the space of all polynomials of total degree $\leq t$ on
$\mathbb{S}^d$. Then we seek a polynomial $p$ of degree $\leq t$, such that
\begin{equation}\label{eq:interp}
p(\PT{x}_i)=f(\PT{x}_i)\qquad\text{ holds for }i=1,\ldots,N.
\end{equation}
Choosing $Y_{\ell,k}$ ($\ell=0,\ldots,t$, $k=1,\ldots,Z(d,\ell)$) as a basis
for the polynomials and $N=Z(d+1,t)$, we have to solve a system of $N$ linear
equations; the points must be chosen so that this system has full rank. One
possible method for choosing the set $X_N$ is to maximise the modulus of the
determinant of the system, which optimises the numerical stability of solving
the system. This approach is in the spirit of finding the extremal value of a
functional depending on the points, such as the energy functionals discussed in
Section~\ref{sec:energy}. This has been done in
\cite{Sloan_Womersley2004:extremal_sphere} for $d=2$.

In \cite{An_Chen_Sloan+2010:spherical_designs} it is proposed to use spherical
$t$-designs as interpolation points; for $N\geq Z(d+1,t)$ the determinant of
the matrix
\begin{equation*}
 H_t \DEF \left(\sum_{j=1}^N
    Y_{\ell,k}(\PT{x}_j) Y_{\ell',k'}(\PT{x}_j)\right)_{(\ell,k),(\ell',k')}
\!\!\!\! 1 \leq \ell,\ell'\leq t,\quad 1\leq k \leq Z(d,\ell),\quad 
1\leq k'\leq Z(d,\ell'),
\end{equation*}
is maximised under the constraint that the point set $X_N$ is a spherical
$t$-design. This leads to the definition of \emph{well conditioned} designs.
In \cite{An_Chen_Sloan+2010:spherical_designs} this is worked out for $d=2$.

It should be mentioned that, based on the numerical experiments in
\cite{Hardin_Sloane2002:table,Graef_Potts2013:table}, there exist $t$-designs
with $\lfloor(t+1)^2/2\rfloor$ points for $t\leq100$. In the context
of interpolation, the number of points has to be chosen $\geq(t+1)^2$, which
gives further freedom for choosing the points that can be used to maximise
the determinant of $H_t$.

In \cite{An_Chen_Sloan+2012:regularized_approximations}, the application of
spherical designs in the context of approximation of functions
$f:\mathbb{S}^d\to\R$ is proposed. Usually, the approximation is computed by
integrating $f$ against an approximation kernel $G_L$, which is a polynomial of
degree $L$ in $\langle\PT{x},\PT{y}\rangle$
(see~\cite{Reimer2003:multivariate_approximation,
  Berens_Butzer_Pawelke1969:limitierungsverfahren_reihen_kugelfunktionen}).  In
\cite{An_Chen_Sloan+2012:regularized_approximations} the integral is replaced
by the equal weight quadrature rule given by a $2L$-design $X_N$. This approach
is worked out for $d=2$ and various classical approximation kernels. In order
to make this procedure numerically more stable, a regularisation procedure is
investigated: the polynomial $p$ is chosen to minimise
\begin{equation}\label{eq:regul}
\sum_{j=1}^N\left(p(\PT{x}_j)-f(\PT{x}_j)\right)^2+
\lambda\sum_{j=1}^N\left(\mathcal{R}p(\PT{x}_j)\right)^2
\end{equation}
amongst all polynomials of degree $L$; here $\mathcal{R}$ is an operator on the
space of polynomials and $\lambda>0$ is the regularisation parameter. The fact
that $X_N$ is chosen as a $2L$-design simplifies the linear algebra behind the
least square approximation.


\section{Energy}\label{sec:energy}

\subsection{Minimal energy in applications}
\label{sec:minim-energy-appl}

A surprising number of diverse applications can be formulated as a discrete or continuous minimal energy problem or a mixture of both. In the discrete setting this means finding a collection of $N$ distinct points in a subset $\Omega \subset \R^p$, $p \geq 1$, that minimises a \emph{discrete $\KERNEL$-energy functional}
\begin{equation*}
E_{\KERNEL,\extFIELD}( \PT{x}_1, \dots, \PT{x}_N ) \DEF \mathop{\sum_{i = 1}^N \sum_{j = 1}^N}_{i \neq j} \Big[ \KERNEL( \PT{x}_i, \PT{x}_j ) + \extFIELD( \PT{x}_i ) + \extFIELD( \PT{x}_j ) \Big]
\end{equation*}
among all sets of $N$ points from $\Omega$. The diagonal self-interaction terms are removed to allow singular kernels~$\KERNEL$. The external field $\extFIELD$ is often taken to be zero. Evidently, further requirements on the set~$\Omega$, the kernel $\KERNEL$, and the external field $\extFIELD$ are needed to ensure existence of a solution. A suitable $\extFIELD$ compatible with $\KERNEL$ introduces soft boundaries and thus prevents points from escaping to infinity in the case of unbounded sets $\Omega$. On the other hand, 
the restriction of the points to a finite set, fractal, torus or sphere of Hausdorff dimension $d \leq p$ introduces fractal or topological aspects. 
A standard assumption is that $\KERNEL$ is symmetric and lower semi-continuous on $\Omega \times \Omega$ and that $\extFIELD$ is also lower semi-continuous on $\Omega$. This implies that the minimal $\KERNEL$-energy problem with external field $\extFIELD$ has a solution for every $N \geq 2$ when solved for an infinite compact set $\Omega$. A solution is called \emph{$N$-point minimal $\KERNEL$-energy configuration associated with $\extFIELD$} and its $\KERNEL$-energy is equal to the \emph{minimum $N$-point $\KERNEL$-energy of $\Omega$ associated with $\extFIELD$} given by
%
%
%
\begin{equation*}
\mathcal{E}_{\KERNEL,\extFIELD}( \Omega; N ) \DEF \inf\Big\{ E_{\KERNEL,\extFIELD}( \PT{x}_1, \dots, \PT{x}_N ) : \PT{x}_1, \dots, \PT{x}_N \in \Omega \Big\}. 
\end{equation*}
%
Furthermore, if $\mathcal{E}_{\KERNEL,\extFIELD}( \Omega; N ) > 0$ for all $N$, then the quantities $N ( N - 1 ) / \mathcal{E}_{\KERNEL,\extFIELD}( \Omega; N )$ form a non-increasing sequence that is bounded from below by lower semi-continuity.\footnote{The reciprocals $\mathcal{E}_{\KERNEL,\extFIELD}( \Omega; N ) / [ N ( N - 1 ) ]$ always form a non-decreasing sequence that may be unbounded.} Thus the limit exists and it discriminates between two types of sets $\Omega$ depending on whether or not this limit vanishes. This gives rise to two different regimes characterised by a complete change in the nature of the minimization problem with regard to properties of the solution and methods that are used to study it. 
In the field-free setting ($\extFIELD \equiv 0$), the normalised \emph{discrete minimal $N$-point energy of $\Omega$}, given by $\mathcal{E}_{\KERNEL,0}( \Omega; N ) / [ N ( N - 1 ) ]$, is also known as the $N$th diameter of $\Omega$. The limit in the extended sense (as $N \to \infty$) is called the \emph{transfinite diameter of $\Omega$}.

%
%
%
%
%
%
A fundamental question concerns the ``limit distribution'' (if such exists) of a sequence $(X_N^*)$ of minimal energy configurations $X_N^*$ on $\Omega$ as $N \to \infty$; \emph{i.e.}, is there a (unique) Borel probability measure $\mu_\Omega^*$ supported on $\Omega$ that is the weak limit of the sequence formed by the discrete equal weight distribution supported on $X_N^*$,
\begin{equation*}
\mu_{X_N^*} \DEF \frac{1}{N} \sum_{j = 1}^N \delta_{\PT{x}_j}.
\end{equation*}
%
%

Let $\mathcal{M}( \Omega )$ denote the collection of Borel probability measures supported on $\Omega$. 
The analogous continuous energy problem is to find a measure in $\mathcal{M}( \Omega )$ that minimises the \emph{weighted $\KERNEL$-energy associated with $\extFIELD$} given by
\begin{equation*}
\mathcal{I}_{\KERNEL, \extFIELD}( \mu ) \DEF \mathcal{I}_{\KERNEL}( \mu ) + 2 \int_\Omega \extFIELD( \PT{x} ) \, \dd \mu( \PT{x} )  
\end{equation*}
among all measures in $\mathcal{M}( \Omega )$, where 
\begin{equation*}
\mathcal{I}_{\KERNEL}( \mu ) \DEF \int_\Omega \int_\Omega \KERNEL( \PT{x}, \PT{y} ) \, \dd \mu( \PT{x} ) \dd \mu( \PT{y} )
\end{equation*}
is the \emph{$\KERNEL$-energy of $\mu \in \mathcal{M}( \Omega )$}.
A minimising measure ${\mu_{\KERNEL,\extFIELD; \Omega} \in \mathcal{M}( \Omega )}$ with
\begin{equation*}
\mathcal{I}_{\KERNEL, \extFIELD}( \mu_{\KERNEL,\extFIELD; \Omega} ) = W_{\KERNEL, \extFIELD}( \Omega ) \DEF \inf\Big\{ \mathcal{I}_{\KERNEL, \extFIELD}( \mu ) \  \Big| \  \mu \in \mathcal{M}( \Omega ) \Big\},
\end{equation*}
is called a \emph{$\KERNEL$-extremal (or positive equilibrium) measure on $\Omega$ associated with $\extFIELD$}. In the field free setting $\extFIELD \equiv 0$, a minimising measure $\mu_{K;\Omega}$ is called \emph{$\KERNEL$-equilibrium measure on $\Omega$}. In this case, the $\KERNEL$-energy $\mathcal{I}_{\KERNEL}( \mu_{K;\Omega} )$ is equal to the \emph{Wiener energy of $\Omega$},
\begin{equation*}
W_{\KERNEL}( \Omega ) \DEF \inf\Big\{ \mathcal{I}_{\KERNEL}( \mu ) \  \Big| \  \mu \in \mathcal{M}( \Omega ) \Big\}.
\end{equation*}
A fundamental question concerns the relationship between the Wiener energy of $\Omega$, the transfinite diameter of $\Omega$, and the Chebyshev constant of $\Omega$ which is the limit as $N \to \infty$ of the $N$th Chebyshev constant of $\Omega$ defined as
\begin{equation*}
M_{\KERNEL}( \Omega; N ) \DEF \sup_{\PT{x}_1, \dots, \PT{x}_N \in \Omega} \inf_{\PT{x} \in \Omega} \frac{1}{N} \sum_{j=1}^N \KERNEL( \PT{x}, \PT{x}_j );
\end{equation*}
see~\cite{FaNa2008} and references cited therein for further details.


%
%

We conclude this section by a discussion of applications that make use of the minimal energy problem or are related to it. 

\begin{description}
\item[The Thomson problem and its generalizations.] A classical problem in electrostatics is to find the distribution of $N$ unit point charges on a conductor in the most stable equilibrium (the charges interact according to the Coulomb potential $1/r$, where $r$ is the Euclidean distance between two interacting charges). This leads to a minimization problem for the potential energy of the discrete charge system named after J.~J.~Thomson who posed it for the sphere~\cite{thomson:1904}.\footnote{Recently, T.~LaFave~Jr.~\cite{LaF2013} investigated correspondences between the Thomson problem and atomic electronic structure, and he applied discrete transformations to the Thomson problem in \cite{LaF2014} to study the minimal Coulomb energy.}
Generalizations of Thomson problem that utilise a Riesz $s$-potential $1/r^s$ are used, \emph{e.g.}, to model multi-electron bubbles and arrangements of protein subunits which form the shells (capsids) of viruses; see \cite{Bowick_Cacciuto_Nelson+2002:crystalline_order_sphere} (also \cite{BoCaNeTr2006, Bowick_Giomi2009}) for a discussion. 
In material physics (see \cite{BrUr2008}) minimal energy points on the sphere have been used to model ``spacer particles'' in powders which ensure that spherical host-particles do not touch. 
%

%

%
\item[Polarization.] A problem related to finding the minimal $\KERNEL$-energy configurations on a compact set $\Omega \subset \R^p$ is to find optimal \emph{$N$-point $K$-polarization configurations on~$\Omega$}, which are configurations on $\Omega$ that maximise the minimal value over $\Omega$ of the potential
\begin{equation*}
\frac{1}{N} \sum_{j=1}^N \KERNEL( \PT{x}, \PT{x}_j ).
\end{equation*}
An optimal configuration realises the $N$th Chebyshev constant of $\Omega$. 
%
%
For a Riesz kernel $\KERNEL_s( \PT{x}, \PT{y} ) = 1 / [ \METRIC( \PT{x}, \PT{y} ) ]^{s}$ with metric $\METRIC$ and $s > 0$, one has the following duality: the minimal $s$-energy configurations tend to \emph{best-packing configurations} on~$\Omega$ as $s \to \infty$, whereas the optimal $s$-polarization configurations on $\Omega$ tend to \emph{best-covering configurations} on $\Omega$. 
Questions concerning optimal polarization configurations, their limit distributions, polarization inequalities, asymptotic behaviour of the $N$th Chebyshev constant are addressed in \cite{AmBaEr2013,HaKeSa2013,ErSa2013,BoBo2014,PrSaWi2014}. The paper~\cite{ErSa2013} proposes a conjecture for the dual of the Poppy-seed Bagel theorem\footnote{Cf.~\url{http://news.vanderbilt.edu/2004/11/the-poppy-seed-bagel-theorem-59497/}} for optimal polarization points (which is proven for the boundary case $s = \dim( \Omega )$ in~\cite{BoBo2014}). 



\item[Smale's 7th Problem.] In the seminal work \cite{ShSm1993III}, M.~Shub and
  S.~Smale define a condition number of a polynomial at a point in $\C$, which
  connects the problem of solving a polynomial equation with the discrete
  minimal logarithmic energy problem on $\mathbb{S}^2$ (see
  Section~\ref{sec:cont-minim-energy}). They show that a monic polynomial
  whose zeros are the stereographic projection of minimal logarithmic energy
  points on the \emph{Riemann sphere}, called ``elliptic Fekete polynomial'',
  is ``well conditioned'' and thus gives a good starting polynomial for a
  homotopy method for solving a polynomial equation or system  (see \cite{ShSm1993I}).
This is the background for \emph{Smale's 7th problem}~\cite{Sm1998}: \emph{Find an algorithm which, on input~$N$, outputs distinct points $\PT{x}_1, \dots, \PT{x}_N$ on the Riemann sphere such that} 
\begin{equation} \label{eq:Smale.s.7.th.problem}
\mathop{\sum_{i = 1}^N \sum_{j = 1}^N}_{i \neq j} \log \frac{1}{\| \PT{x}_i - \PT{x}_j \|} - V_N \leq c \, \log N, \qquad \text{\emph{where $c$ is  a universal constant.}}
\end{equation}
Here $V_N$ is the minimal logarithmic energy of $N$ points on the Riemann sphere. 
Smale further specifies that the algorithm is a real number algorithm in the sense of Blum, Cucker, Shub, and Smale (see~\cite{BCSS1997}) with halting time polynomial in $N$. In~\cite{Be2013}, C.~Beltr{\'a}n showed that there are $N$-point sets with logarithmic energy that differs from $V_N$ by at most $1/9$. These points are rational with coordinates of order $\log_2 N$ bit length. Thus, there exists an exponential running time algorithm.
%
%
%
Mean value considerations yield that the \emph{typical} logarithmic energy of $N$ i.i.d. uniformly distributed random points on the Riemann sphere is of the form $\frac{1}{2} N^2 - \frac{1}{2} N$. 
%
One the other hand, D.~Armentano, C.~Beltr{\'a}n, and M.~Shub~\cite{ArBeSh2011} \footnote{The paper~\cite{FeZe2013} gives a generalization to higher dimensions and other manifolds.} observe that the typical logarithmic energy of the zeros of certain \emph{random polynomials}, $\frac{1}{2} N^2 - \frac{1}{2} N \log N - \frac{1}{2} N$, is surprisingly small in the sense that the first two terms in the asymptotics of the minimal $N$-point logarithmic energy for the Riemann sphere are recovered. A recent account of the state of the art regarding Smale's 7th problem can be found in \cite{Be2013b} (see also \cite{Beltran2015:facility_location_formulation}).

%

\item[Log gases, Coulomb gases, and random matrices.] A log or Coulomb gas is a system of interacting particles in which the repelling interaction is governed by a logarithmic or Coulomb potential. An external field confines the particles to a finite volume of the space. Typically, the \emph{mean-field regime} is considered. In this setting the number $N$ of particles is large and the pair-interaction strength (coupling parameter) scales as the inverse of~$N$. The Hamiltonian 
\begin{equation*}
H( \PT{x}_1, \dots, \PT{x}_N ) \DEF \mathop{\sum_{i=1}^N \sum_{j=1}^N}_{i \neq j} k( \PT{x}_i - \PT{x}_j ) + N \sum_{i=1}^N V( \PT{x}_i ), \qquad \PT{x}_1, \dots, \PT{x}_N \in \mathbb{R}^d,
\end{equation*}
where $k( \PT{x} ) = - \log \| \PT{x} \|$ if $d = 2$ or $k( \PT{x} ) = \| \PT{x} \|^{2-d}$ if $d \geq 3$, is minimised.\footnote{Instead of the harmonic potentials one can also consider Riesz $s$-potentials $\| \PT{\cdot} \|^{-s}$. For $s \geq d$ one needs to adjust the coupling parameter to ensure comparability between the pair-interaction part and external field part.} 
Very recent progress provides a deep connection with the discrete minimum energy problem on the sphere (see discussion at the end of Section~\ref{sec:asympt-expans-minim.log}). We refer further to S.~R.~Nodari and S.~Serfaty~\cite{NoSe2014} and N.~Rougerie and S.~Serfaty~\cite{RoSe2013arXiv}. The concept of renormalised energy  can be also successfully applied to random matrices; see A.~Borodin and S.~Serfaty~\cite{BoSe2013}. 
For the connection between log gases and random matrix applications and theory we refer to the book of P.~Forrester~\cite{Fo2010} and, \emph{e.g.}, T.~Claeys, A.~B.~J.~Kuijlaars and M.~Vanlessen~\cite{ClKuVa2008} and A.~Mays~\cite{Mays2013}. 
A related problem is the discrete energy of periodic point sets in the Euclidean space; see D.~P.~Hardin, E.~B.~Saff, and B.~Simanek~\cite{HaSaSi2014arXiv}.

\item[Half-toning.] Loosely speaking, half-toning is a way of creating an illusion of a grey-value image by appropriately distributing black dots. 
In \cite{GrPoSt2012}, M.~Gr{\"a}f, D.~Potts,  and G.~Steidl show how the process of half-toning can be seen as a numerical integration process with the aim to minimise a worst-case error which can be interpreted as an external field problem where the picture drives the external field which guides the interacting points; see also \cite{TeStGw2011} and \cite{Graef2013:efficient_algorithms_computation}.
\item[Maximizing Determinants.] 
%
%
Points that maximise a Vandermonde-like determinant are well-suited for
interpolation and numerical integration. They are called Fekete points due to
the paper \cite{Fe1923} by M.~Fekete. Given a compact set in the complex plane,
Fekete points are, indeed, minimal logarithmic energy points. However, in
higher dimensions, Fekete's optimisation problem is different from minimising
the logarithmic energy. I.~H.~Sloan and R.~S.~Womersley~\cite{Sloan_Womersley2004:extremal_sphere} used the logarithm of the determinant of an interpolation matrix to calculate extremal systems which yield interpolatory cubature rules with positive weights on the sphere. J.~Marzo and J.~Ortega-Cerd{\'a}~\cite{Marzo_Ortega-Cerda2010:equidistribution_fekete} established that Fekete (or extremal) points are asymptotically uniformly distributed.


%
%

\item[Diffusion on a sphere with localised traps.] As an application in cellular signal transport, D.~Coombs, R.~Straube, and M.~Ward \cite{CoStWa2009} calculate the principal eigenvalue for the Laplacian on the unit sphere in the presence of $N$ traps on the surface of the sphere of asymptotically small radii. The positions of the traps are chosen to minimise the discrete logarithmic energy given in \eqref{eq:logarithmic.energy.sphere} below.


\end{description}

%






\subsection[The discrete and continuous minimal logarithmic and Riesz energy]{The discrete and continuous minimal logarithmic and Riesz energy problem}
\label{sec:cont-minim-energy}

The discrete logarithmic energy problem on $\mathbb{S}^d$ is concerned with the properties of $N$-point configurations $\{ \PT{x}_1^*, \dots, \PT{x}_N^* \} \subset \mathbb{S}^d$ that maximise the product of all mutual pairwise Euclidean distances 
\begin{equation} \label{eq:product.mutual.distances}
\mathop{\prod_{i=1}^N \prod_{j=1}^N}_{i \neq j} \left\| \PT{x}_i - \PT{x}_j \right\|,
\end{equation}
or equivalently, minimise the discrete logarithmic energy 
\begin{equation} \label{eq:logarithmic.energy.sphere}
E_{\log}( \PT{x}_1, \dots, \PT{x}_N ) \DEF \mathop{\sum_{i=1}^N \sum_{j=1}^N}_{i \neq j} \log \frac{1}{\left\| \PT{x}_i - \PT{x}_j \right\|}
\end{equation}
over all $N$-point configurations $\{ \PT{x}_1, \dots, \PT{x}_N \}$ on $\mathbb{S}^d$. 
%
The discrete logarithmic energy can be understood as a limiting case (as $s \to 0$) of the Riesz $s$-energy
\begin{equation*}
E_s( \PT{x}_1, \dots, \PT{x}_N ) \DEF \mathop{\sum_{i=1}^N \sum_{j=1}^N}_{i \neq j} \frac{1}{\left\| \PT{x}_i - \PT{x}_j \right\|^s};
\end{equation*}
\emph{i.e.}, $E_s( \PT{x}_1, \dots, \PT{x}_N ) = N \left( N - 1 \right) + s \, E_{\log}( \PT{x}_1, \dots, \PT{x}_N ) + o(s)$ as $s \to 0$.

The discrete Riesz $s$-energy problem for $s > 0$ is concerned with the properties of $N$-point configurations $\{ \PT{x}_1^*, \dots, \PT{x}_N^* \} \subset \mathbb{S}^d$ that minimise the Riesz $s$-energy over all $N$-point configurations $\{ \PT{x}_1, \dots, \PT{x}_N \}$ on $\mathbb{S}^d$. For convenience we set $\kKERNEL_{s}( \PT{x}, \PT{y} ) \DEF - \log \| \PT{x} - \PT{y} \|$ for $s = \log$ and $\kKERNEL_{s}( \PT{x}, \PT{y} ) \DEF 1 / \| \PT{x} - \PT{y} \|^s$ for $s \in \R$. Then we are interested in the \emph{optimal $N$-point $s$-energy} of an infinite compact set $\Omega \subset \mathbb{S}^d$ defined by
\begin{equation*}
\mathcal{E}_s( \Omega; N ) \DEF 
\begin{cases}
\min\big\{ E_s( \PT{x}_1, \dots, \PT{x}_N ) \, \mid \, \PT{x}_1, \dots, \PT{x}_N \in \Omega \big\} & \text{for $s = \log$ or $s \geq 0$,} \\
\max\big\{ E_s( \PT{x}_1, \dots, \PT{x}_N ) \, \mid \, \PT{x}_1, \dots, \PT{x}_N \in \Omega \big\} & \text{for $s < 0$.}
\end{cases}
\end{equation*}
Observe that $\mathcal{E}_0( \Omega; N ) = N^2 - N$ (which is attained by any $N$-point set on $\Omega$). The Riesz $s$-kernel is conditionally positive definite of order $1$ for $-2 < s < 0$. Alternatively, as in the setting of numerical integration on $\mathbb{S}^d$, one can minimise
\begin{equation*}
2 \int_\Omega \int_\Omega \left\| \PT{x} - \PT{y} \right\|^{-s} \dd \sigma_d( \PT{x} ) \dd \sigma_d( \PT{y} ) - \mathop{\sum_{i=1}^N \sum_{j=1}^N} \left\| \PT{x}_i - \PT{x}_j \right\|^{-s}
\end{equation*}
in this case.  The papers \cite{Saff_Kuijlaars1997:distributing_many,
  Hardin_Saff2004:discretizing_manifolds} are standard references for the
discrete logarithmic and Riesz $s$-energy problem.

The \emph{$s$-potential} and the \emph{$s$-energy} of a measure $\mu$ in the class $\mathcal{M}(\Omega)$ of Borel probability measures supported on $\Omega$ are given, respectively, by 
\begin{equation} \label{eq:potential.energy.integral}
U_s^{\mu}( \PT{x} ) \DEF \int_\Omega \kKERNEL_s( \PT{x}, \PT{y} ) \, \dd \mu( \PT{y} ), \quad \PT{x} \in \R^{d+1}, \qquad \mathcal{I}_s( \mu ) \DEF \int_\Omega \int_\Omega \kKERNEL_s( \PT{x}, \PT{y} ) \, \dd \mu( \PT{x} ) \dd \mu( \PT{y} ).
\end{equation}
For $s > 0$, the \emph{$s$-capacity} of $\Omega$ is the reciprocal of the Wiener energy ${\inf\{ \mathcal{I}_s( \mu ) \mid \mu \in \mathcal{M}( \Omega ) \}}$. When the Wiener energy is finite, it will be denoted by~$W_s( \Omega )$. Because of the possibility of negative logarithmic energy integral, we set ${\CAP_{\log}( \Omega ) \DEF \exp( {-\inf\{ \mathcal{I}_{\log}( \mu ) \mid \mu \in \mathcal{M}( \Omega ) \} })}$. When finite, the infimum is denoted by $W_{\log}( \Omega )$.
The lower semi-continuous logarithmic kernel is bounded from below and thus the kernel $\kKERNEL_s$ (plus a constant) is strictly positive definite for $s = \log$ and $0 < s < d$. Consequently, the $s$-equilibrium measure $\mu_{\Omega,s}$ on $\Omega$ is unique for every compact set $\Omega \subset \R^p$ with finite $s$-energy; see \cite{Landkof1972:potential_theory,BoHaSaBook}. 
For the range $-2 < s < 0$, one also has a unique $s$-equilibrium measure on $\Omega$; see~\cite{Bj1956} for the potential theoretic quantities and variational inequalities. 

In the remaining section we consider the sphere $\mathbb{S}^d$ and subsets of $\mathbb{S}^d$.
An \emph{external field} is a lower semi-continuous function $\extFIELD:\mathbb{S}^d \to (-\infty,\infty]$ such that ${\extFIELD(\PT{x})<\infty}$ on a set of positive Lebesgue surface measure. We note that the lower semi-continuity implies the existence of a finite $c_{\extFIELD}$ such that $\extFIELD(\PT{x})\geq c_{\extFIELD}$ for all $\PT{x}\in
\mathbb{S}^d$. The {\em weighted energy associated with $\extFIELD(\PT{x})$} is then given by
\begin{equation} \label{energy}
\mathcal{I}_{\extFIELD,s}(\mu) \DEF \mathcal{I}_s(\mu) + 2 \int \extFIELD(\PT{x}) \dd \mu(\PT{x}), \qquad \mu \in \mathcal{M}( \mathbb{S}^d ).
\end{equation}
We recall from~\cite{Dragnev_Saff2007:riesz_potential_separation} the following Frostman-type result that deals with existence and uniqueness of the $s$-extremal measure on $\Omega$ associated with $\extFIELD$ and its characterization in terms of weighted potentials. 
The result is stated for the Riesz case but a similar result holds also for the logarithmic case.
The potential theory used in the context of this survey is formulated by
G.~Bj{\"o}rck~\cite{Bj1956} (dealing with Riesz potential with negative
exponent), by E.~B.~Saff, and
V.~Totik~\cite{Saff_Totik1997:logarithmic_potentials} (logarithmic external
field problem in the plane), by
N.~S.~Landkof~\cite{Landkof1972:potential_theory} (Riesz and logarithmic
potential and general reference) and, in particular, for Riesz external field
problems by N.~V.~Zori{\u\i}~\cite{Zo2003, Zo2004}.
\begin{prop} \label{prop:1}
Let $0< s < d$.
For the minimal energy problem on $\mathbb{S}^d$
with external field $\extFIELD$ the following properties hold:
\begin{itemize}
\item[\rm (a)] $W_{\extFIELD,s} \DEF \inf\big\{ \mathcal{I}_{\extFIELD,s}(\mu) \big| \mu \in \mathcal{M}( \mathbb{S}^d ) \big\}$ is finite.
\item[\rm (b)] There is a unique $s$-extremal measure
$\mu_{\extFIELD,s}\in \mathcal{M}(\mathbb{S}^d)$ associated with $\extFIELD$.
Moreover, the support $S_{\extFIELD,s} \DEF \supp( \mu_{\extFIELD,s} )$ of this measure is contained in the
compact set $E_M \DEF \{ \PT{x} \in \mathbb{S}^d : \extFIELD(\PT{x}) \leq M
\}$ for some $M>0$.
\item[\rm (c)] The measure $\mu_{\extFIELD,s}$ satisfies the variational inequalities
\begin{align}
U_s^{\mu_{\extFIELD,s}}(\PT{x}) + \extFIELD(\PT{x}) &\geq F_{\extFIELD,s} \quad \text{q.e. on $\mathbb{S}^d$,} \label{VarEq1} \\
U_s^{\mu_{\extFIELD,s}}(\PT{x}) + \extFIELD(\PT{x}) &\leq F_{\extFIELD,s} \quad
\text{everywhere on $S_{\extFIELD,s}$,} \label{VarEq2}
\end{align}
where
\begin{equation}
F_{\extFIELD,s} \DEF W_{\extFIELD,s} - \int \extFIELD(\PT{x}) \dd \mu_{\extFIELD,s}(\PT{x}).
\label{VarConst}
\end{equation}
\item[\rm (d)] Inequalities \eqref{VarEq1} and
\eqref{VarEq2} completely characterise the $s$-extremal measure
$\mu_{\extFIELD}$ in the sense that if $\nu \in \mathcal{M}(\mathbb{S}^d)$ is
a measure with finite $s$-energy such that for some constant~$C$ we
have
\begin{align}
U_s^{\nu}(\PT{x}) + \extFIELD(\PT{x}) &\geq C \quad \text{q.e. on $\mathbb{S}^d$,} \label{VarEq3} \\
U_s^{\nu}(\PT{x}) + \extFIELD(\PT{x}) &\leq C \quad \text{everywhere on $\supp (\nu)$,} \label{VarEq4}
\end{align}
then $\nu=\mu_{\extFIELD,s}$ and $C=F_{\extFIELD,s}$.
\end{itemize}
\end{prop}

A property holds \emph{quasi-everywhere} if the exceptional set has $s$-capacity zero.
If $\extFIELD$ is continuous on $\mathbb{S}^d$, then the inequalities \eqref{VarEq1} and \eqref{VarEq3} hold everywhere on~$\mathbb{S}^d$.

In principle, once $\supp( \mu_{\extFIELD} )$ is known, then the measure
$\mu_{\extFIELD}$ can be recovered by solving an integral equation for the
weighted $s$-potential arising from \eqref{VarEq3} and \eqref{VarEq4}. Finding
$\supp( \mu_{\extFIELD} )$ when it is a proper subset of $\mathbb{S}^d$ can be
a very difficult problem. It is a substantially easier task to find a signed measure that has constant weighted $s$-potential everywhere on~$\mathbb{S}^d$. Given a compact subset~$\Omega \subset \mathbb{S}^d$ with $\CAP_s( \Omega ) > 0$ and an external field~$\extFIELD$, there is a unique (if it exists, see~\cite[Lemma~23]{BrDrSa2009}) \emph{signed $s$-equilibrium} measure $\eta_{\KERNEL, \extFIELD}$ supported on~$\Omega$ of total charge one associated with $\extFIELD$  with constant weighted $s$-potential; \emph{i.e.}, 
\begin{equation*}
U_s^{\eta_{\KERNEL, \extFIELD}}(\PT{x}) + \extFIELD(\PT{x}) = G_{\KERNEL,\extFIELD,s} \qquad \text{everywhere on $\Omega$}
\end{equation*}
for some constant $G_{\KERNEL,\extFIELD,s}$. 
A remarkable connection exists to the analogue of the \emph{Mhaskar-Saff functional} from classical planar potential theory (\cite{MhSa1985} and \cite[Chapter~IV, p.~194]{Saff_Totik1997:logarithmic_potentials}) given by
\begin{equation*}
\mathcal{F}_s(\Omega^\prime) \DEF W_s(\Omega^\prime) + \int \extFIELD(\PT{x}) \, \dd \mu_{\Omega^\prime}(\PT{x}), \qquad \text{$\Omega^\prime \subset \mathbb{S}^d$ compact with $\CAP_s( \Omega^\prime ) > 0$,}
\end{equation*}
where $W_s(\Omega^\prime)$ is the $s$-energy of $\Omega^\prime$ and $\mu_{\Omega^\prime}$ is the $s$-equilibrium measure (without external field) on $\Omega^\prime$. 
%
%
Namely, if the signed $s$-equilibrium on a compact set $\Omega^\prime$ associated with~$\extFIELD$ exists, then $\mathcal{F}_s(\Omega^\prime) = G_{\KERNEL,\extFIELD,s}$. 
The essential property of the $\mathcal{F}_s$-functional is that it is minimised for the support of the $s$-extremal measure (see \cite[Proposition~8]{BrDrSa2014} for a precise statement).
The papers \cite{Dragnev_Saff2007:riesz_potential_separation,BrDrSa2009,BrDrSa2014,BrDrSa2012} determine the signed $s$-equilibrium on the full sphere and on spherical caps associated with logarithmic and Riesz $s$-external fields due to a single point charge (or a axis-supported superposition of such fields) to derive separation results for minimal Riesz $s$-energy points and to characterise the $s$-extremal measure on~$\mathbb{S}^d$. 
The use of balayage techniques (the signed equilibrium can be expressed as the difference of two balayage measures) together with a restricted principle of domination and a restricted maximum principle yields that the Riesz parameter $s$ is restricted to the interval $[d-2,d)$. New phenomena occur when $s = d - 2$. For example, a boundary charge appears in the signed $(d-2)$-equilibrium for a spherical cap which vanishes when the cap coincides with the support of the $s$-extremal measure on~$\mathbb{S}^d$. Numerical methods for external field problems for Riesz potentials have been developed and applied in \cite{OfWeZo2010,HaWeZo2012,HaWeZo2014,Harbrecht_Wendland_Zorii2014:rapid}.

\subsection{The distribution of minimal logarithmic and Riesz energy points}
\label{sec:distribution.min.energy}

Let $\Omega$ be an infinite compact set $\Omega \subset \R^p$ with Hausdorff dimension $d$ having finite logarithmic or Riesz $s$-energy ($s > 0$) and $\extFIELD \equiv 0$. Then classical potential theory implies that the minimal energy configurations $X_N^*$ on~$\Omega$ are distributed according to the unique equilibrium measure $\mu_\Omega$ on $\Omega$ and the discrete measures $\mu_{X_N^*}$ have $\mu_\Omega$ as a weak limit.
%
%
The $s$-equilibrium measure on $\mathbb{S}^d$ is the uniform measure $\sigma_d$. In general, the measure $\mu_\Omega$ will not be uniform as 
the example of a circular torus shows (see \cite{Hardin_Saff_Stahl2007:support_logarithmic} and \cite{Brauchart_Hardin_Saff2007:riesz_energy_revolution, Brauchart_Hardin_Saff2009:riesz_energy_revolution}).\footnote{It seems to be unresolved for which $\Omega$ the $s$-equilibrium measure on $\Omega$ is the uniform measure on $\Omega$.} 
%
%
%
%
%
%
In this potential theoretic regime ($s = \log$ or $0 < s < d$), the support of $\mu_\Omega$ 
depends on whether the kernel is superharmonic, harmonic or subharmonic (see~\cite{Landkof1972:potential_theory}). In the superharmonic case ($s = \log$ or $0 < s < p - 2$), the measure $\mu_\Omega$ is supported on the \emph{outer boundary} (\emph{i.e.}, the boundary of $\Omega$ shared with the unbounded component of the ambient space $\mathbb{R}^p$), whereas in the strictly subharmonic case, the measure $\mu_\Omega$ can be supported on all of $\Omega$.
%
%
%

The intuition is that in the regime $s = \log$ or $0 < s < d$ global effects dominate (points interact as if they are subject to long-range forces and the range increases as $s$ becomes smaller). 
In the hypersingular case $s > d$, local effects dominate (points interact as if they are responding to a short range force). Both kinds of interactions intermingle when $s = d$.

In the hypersingular case $s \geq d$, the energy integral attains $+\infty$ for every $\mu \in \mathcal{M}( \Omega )$. Geometric measure theory yields that the limiting distribution of a sequence $( X_N^*)$ of minimal Riesz $s$-energy $N$-point sets on $\Omega$ (even asymptotically $s$-energy minimising would suffice) is uniformly distributed with respect to the $d$-dimensional Hausdorff measure $\mathcal{H}_d$,
\begin{equation*}
\mu_{X_N^*} \substack{*\\\longrightarrow} \frac{\mathcal{H}_d\big|_\Omega}{\mathcal{H}_d( \Omega )} \qquad \text{as $N \to \infty$,} 
\end{equation*}
for a large class of sets $\Omega$ with $\mathcal{H}_d( \Omega ) > 0$ (see~\cite{Borodachov_Hardin_Saff2008:asymptotics_riesz} and earlier 
work~\cite{Borodachov_Hardin_Saff2007:asymptotics_best-packing,Hardin_Saff2005:minimal_riesz,Hardin_Saff2004:discretizing_manifolds}). 
It should be noted that in the case $s = d$ an additional regularity assumption on $\Omega$ is required. 
In the limit $s \to \infty$ only the nearest-neighbour interaction matters and the optimal solutions are best-packing configurations which solve the Tammes problem \cite{Tammes1930:pollen_grains}. 
The paper~\cite{Borodachov_Hardin_Saff2008:asymptotics_riesz} also shows analogous results for weighted Riesz $s$-energy 
\begin{equation*}
\mathop{\sum_{i=1}^N \sum_{j=1}^N}_{i \neq j} \frac{w( \PT{x}_i, \PT{x}_j)}{\left\| \PT{x}_i - \PT{x}_j \right\|^s},
\end{equation*}
where $w$ is (almost everywhere) {\bf c}ontinuous and {\bf p}ositive on the {\bf d}iagonal (called CPD weight function).
We remark that the case $s = d$ for $\mathbb{S}^d$ has already been dealt with in \cite{GoSa2001} using results from  \cite{Kuijlaars_Saff1998:asymptotics_minimal_energy}. Furthermore, in \cite{Calef_Hardin2009:riesz_equilibrium} it was shown that the $s$-equilibrium measures on $\Omega$ converge to the normalised $d$-dimensional Hausdorff measure restricted to $\Omega$ as $s \to d^-$ under rather general assumptions on $\Omega$.
We remark that I.~Pritzker~\cite{Pr2011} studied the discrete approximation of the equilibrium measure on a compact set $\Omega \subset \R^p$, $p \geq 2$, with positive $s$-capacity by means of points which do not need to lie inside $\Omega$. He also obtained discrepancy estimates in the harmonic case. The properties of so-called \emph{
greedy $\KERNEL$-energy points} were studied by A.~L{\'o}pez Garc{\'{\i}}a and E.~B.~Saff~\cite{Lopez_Saff2010:asymptotics_greedy}.

Summarizing, for $-2 < s < 0$, $s = \log$ and $s > 0$, the (asymptotically) $s$-energy minimising $N$-point configuration on $\mathbb{S}^d$ are uniformly distributed with respect to the surface area measure $\sigma_d$. 

In certain applications one prefers to generate well distributed $N$-point sets on a compact $d$-rectifiable set in $\R^p$ which have a prescribed non-uniform asymptotic distribution~$\rho$ with respect to $\mathcal{H}_d$ as $N \to \infty$. It is shown in~\cite{HaSaWh2012} that such points can be obtained by minimising the energy of $N$ points on $\Omega$ interacting via a weighted power law potential  $w( \PT{x}, \PT{y} ) / \| \PT{x} - \PT{y} \|^s$, where $s > d$ and $w( \PT{x}, \PT{y} ) \DEF [ \rho( \PT{x} ) \rho( \PT{y} ) ]^{-s/(2d)}$. Furthermore, such point sets are ``quasi-uniform'' in the sense that the ratio of the covering radius to the separation distance is uniformly bounded in $N$. As mentioned in the introduction, quasi-uniformity is crucial for a number of numerical methods (see~\cite{FuWr2009,LeGSlWe2010,Sch1995}).
S.~V.~Borodachov, D.~P.~Hardin, and E.~B.~Saff show in \cite{BoHaSa2014b} that it suffices to use a varying truncated weight $w( \PT{x}, \PT{y}) \, \Phi( \| \PT{x} - \PT{y} \| / r_N )$. Thus only those pairs of points that are located at a distance of at most $r_N = C_N \, N^{-1/d}$ from each other contribute to the energy sum. (The positive sequence $(C_N)$ can be taken to tend to $\infty$ as slowly as one wishes.) 
In this way, under suitable assumptions, the complexity of the energy computation can be greatly reduced, leading to order $N \, C_N^d$ computations for generating ``low energy'' $N$-point approximations.\footnote{We note that in the astronomical community a hierarchical equal area iso-latitude pixelisation (HEALPix~\cite{HEALPix2005}) is used to generate large numbers of uniformly distributed points on $\mathbb{S}^2$; cf. also~\cite{Le2006}. A.~Holho{\c{s}} and D.~Ro{\c{s}}ca~\cite{HoRo2014} study Riesz energy of points derived from an area-preserving map from the $2$-sphere to the 
octahedron.
}

A point charge approaching the sphere subject to the same law of interaction as
the points on the sphere affects the charge distribution on the
sphere. Sufficiently close to the sphere, it will generate a spherical cap with
zero charge. The papers \cite{BrDrSa2009,BrDrSa2012,BrDrSa2014}, in particular,
provide explicit representations of the charge distributions due to a single
external charge. They also address a question attributed to A.~A.~Gonchar,
namely to find a critical distance from the sphere surface of a point charge
generating the external field so that the support of the $s$-extremal measure
on $\mathbb{S}^d$ only just becomes the whole sphere. In the harmonic case this
distance is characterised by the largest zero of polynomials dubbed Gonchar
polynomials.\footnote{For $d = 2$ and $d = 4$, the critical distance is the
  golden ratio and the plastic number.}

For a small $N$, numerical optimization methods can be used to find putative minimal Riesz $s$-energy configurations. T.~Erber and G. M.~Hockney~\cite{Erber_Hockney1997:complex_systems} noticed that the number of local minimal energy configurations (most of which are not global ones) seems to grow exponentially with $N$. 
Most of the numerical data is for the $2$-sphere and for the Coulomb case and a few other values of $s$. Regarding data, we refer to online resources~\cite{Hardin_Sloane_Smith1997:table,CCD2014,bowick_etal:_thomson_applet} and a more recent study of the energy landscape in~\cite{Ca2009,CaEtal2013}, whereas~\cite{BeCaEnGe2009} provides a complexity analysis for the logarithmic case. Higher dimensional configurations have also been investigated (see~\cite{Ballinger_Blekherman_Cohn+2009:experimental_spheres}). 
In \cite{NeBrKie2013} monotonicity properties of the second discrete derivative were considered and led to new putative low-energy configurations in two cases.
Numerical results in \cite{MeKnSm1977} suggest that a minimal $s$-energy $N$-point set may transit through one or more (depending on $N$) basic configuration as $s$ grows.\footnote{The phenomenon of transiting through several basic types of configurations can also be observed in the external field setting when $s$ is fixed but the distance of the external field source varies (see~\cite[Figure~4]{BrDrSa2014}).} 
The smallest only partly resolved problem is the \emph{five point problem on~$\mathbb{S}^2$}. Five points cannot form a universally optimal system~\cite{Cohn_Kumar2007:universally_optimal_sphere}.\footnote{A \emph{universally optimal} point set minimises all energy functionals with kernels of the form $\KERNEL( \PT{x}, \PT{y} ) = f( \| \PT{x} - \PT{y} \|^2 )$, where $f$ is a completely monotonic $C^\infty$ function like the Riesz potential $1/r^s$ for $s > 0$, meaning that $(-1)^k \, f^{(k)}(x) \geq 0$ for all $k$; see~\cite{Cohn_Kumar2007:universally_optimal_sphere}.}
Melnyk \emph{et al.}~\cite{MeKnSm1977} identified two basic configurations: triangular bi-pyramid and quadratic pyramid. According to numerical results, the regular triangular bi-pyramid is the putative energy-minimising configuration for $−2 \leq  s \leq 15.048077392\dots$, whereas for higher values of $s$ it seems to be the square pyramid (with adjusted height); see also \cite{NeBrKie2013} for a finer analysis. 
Moreover, it is shown in~\cite{BoHaSa2014} that there are sequences of $s$-energy minimising configurations that tend to a square pyramid best packing configuration as $s \to \infty$.
In general, the five point problem is a difficult problem to analyse rigorously. Recently, the papers \cite{Sch2013} (for the Coulomb case $s = 1$ and for $s = 2$) and \cite{HouSh2011} (for sum of distances, $s = -1$) provided computer-assisted proofs that the triangular bi-pyramid is optimal, whereas in the logarithmic case a conventional proof was given in \cite{DrLeTo2002}. In~\cite{Tu2013} a bi-quadratic energy functional is considered. 
%
%
Other rigorously proved minimising configurations are rare and are often universally optimal.  Proved minimising configurations on $\mathbb{S}^2$ are the antipodal and equilateral configuration and the Platonic solids with $N = 4$, $6$ and $12$ vertices. For higher dimensions, we refer to \cite[Table~1]{Cohn_Kumar2007:universally_optimal_sphere}, \cite{BachocVallentin1} and \cite[Section~5.3]{CoWo2012}.



\subsection{Asymptotic expansion of minimal Riesz energy}
\label{sec:asympt-expans-minim}
Let $s > 0$. As sets we shall consider the unit sphere $\mathbb{S}^d$ and, more generally, infinite compact sets $\Omega \subset \R^p$ with Hausdorff dimension $0 < d \leq p$.
The leading term of the asymptotic expansion of the $N$-point Riesz $s$-energy of $\Omega$ is 
well-understood if $\Omega$ has positive $s$-capacity (\emph{i.e.}, finite Riesz
$s$-energy). This is the \emph{potential-theoretic regime}. A standard argument
from classical potential theory yields that the positive quantities ${N ( N - 1
  ) / \mathcal{E}_s( \Omega; N )}$ form a monotonically decreasing sequence. The
limit $\TFDIAM_s( \Omega )$, called the \emph{generalised transfinite diameter of
  $\Omega$}, is equal to the $s$-capacity of $\Omega$
(cf.~\cite{Polya_Szego:transfiniten_durchmesser}). Thus, the leading term of
$\mathcal{E}_s( \Omega; N )$ grows like $N^2$ as $N \to \infty$ and the leading
coefficient is given by the Riesz $s$-energy of $\Omega$, or equivalently, by the
reciprocals of the $s$-capacity and transfinite diameter of $\Omega$:
\begin{equation}
\lim_{N \to \infty} \frac{\mathcal{E}_s( \Omega; N )}{N^2} = W_s( \Omega ) =
\frac{1}{\CAP_s( \Omega )} = \frac{1}{\TFDIAM_s( \Omega )}. 
\end{equation}
For $0 < s < d$, the Riesz $s$-energy of the sphere $\mathbb{S}^d$ has the explicit form
\begin{equation} \label{eq:W.s.sphere}
W_s( \mathbb{S}^d ) = \mathcal{I}_s[ \sigma_d ] = 
2^{d-1-s} \frac{\gammafcn( ( d + 1 ) / 2 ) \, 
\gammafcn( ( d - s ) / 2 )}{\sqrt{\pi} \, \gammafcn( d - s / 2 )},
\end{equation}
expressed in terms of the gamma function $\gammafcn$. By identifying $W_s(
\mathbb{S}^d )$ with the analytic continuation of the right-hand side above to
the complex $s$-plane\footnote{The meromorphic function $W_s( \mathbb{S}^d )$,
  which appears in the conjecture for the asymptotics in the hypersingular
  case, has simple poles (finitely many if $d$ is even and infinitely many if
  $d$ is odd). The effect of this dichotomy on the asymptotic expansion of the
  minimal Riesz $s$-energy is completely open for $d \geq 2$ and leads to $\log
  N$ terms for the unit circle,
  cf.~\cite{Brauchart_Hardin_Saff2009:riesz_energy_roots}.}, we can define the
Riesz $s$-energy of $\mathbb{S}^d$ for Riesz parameter $s$ for which the
$s$-energy integral \eqref{eq:potential.energy.integral} is~$+\infty$ for every Borel
probability measure on $\mathbb{S}^d$.
%
%
%
%
%
The combined effort of \cite{Wagner1992:means_distances_upper,
  Rakhmanov_Saff_Zhou1994:minimal_discrete_energy,
  Kuijlaars_Saff1998:asymptotics_minimal_energy,
  Wagner1990:means_distances_lower, Brauchart2006:riesz_energy} resulted in the
following bounds for the second term of the minimal energy
asymptotics:\footnote{Similar estimates but with negative constants $c, C$
  hold for the sum of generalised distances (\emph{i.e.},~${-2 < s < 0}$);
  see~\cite{Al1972, Al1977, Stolarsky1973:sums_distances_ii, Ha1982, Be1984,
    Stolarsky1972:sums_distances_i, AlSt1974} and culminating
  in~\cite{Wagner1990:means_distances_lower,
    Wagner1992:means_distances_upper}.} there exist constants $c, C > 0$
depending only on $d \geq 2$ and $0 < s < d$ such that
\begin{equation*}
c \, N^{1+s/d} \leq \mathcal{E}_s( \mathbb{S}^d; N ) - 
W_s( \mathbb{S}^d ) \, N^2 \leq C \, N^{1+s/d}, \qquad N \geq 2.
\end{equation*}
These estimates give the correct order of growth and sign for the second-order
term. It is an open problem if the sequence $\big( \mathcal{E}_s( \mathbb{S}^d;
N ) - W_s( \mathbb{S}^d ) \, N^2 \big) / N^{1+s/d}$ has a limit as $N \to
\infty$. A.~A.~Berezin~\cite{Be1986} used a semi-continuum approach (a classical
method from solid state physics, cf.~\cite{GoAd1960}) to derive the plausible
asymptotics
\begin{equation*}
\begin{split}
\mathcal{E}_s( \mathbb{S}^2; N ) 
&\approx N^2 \frac{2^{1-s}}{2-s} \left[ 1 - \left( n / N \right)^{1-s/2} \right] + N^{1+s/2} \left( \frac{\sqrt{3}}{8 \pi} \right)^{s/2} \\
&\phantom{=\pm\times}\times \left\{ \frac{6}{1^s} + \frac{6}{( \sqrt{3} )^s} + \frac{6}{2^s} + \frac{12}{( \sqrt{7} )^s} + \frac{6}{3^s} + \frac{6}{( 2 \sqrt{3} )^s} + \frac{12}{( \sqrt{13} )^s} + \cdots \right\}
\end{split}
\end{equation*}
based on the assumptions that a typical point (and most of its immediate
neighbours) in a minimal Riesz $s$-energy $N$-point configuration on
$\mathbb{S}^2$ gives rise to a hexagonal Voronoi cell (sixfold symmetry)
whereas the defects according to the curved surface of the sphere will have no
significant influence on the second term in the asymptotics. Thus the
contribution to the Riesz $s$-energy due to a typical point can be split into a
local part which uses $n$ nearest neighbour points from a suitably adjusted flat hexagonal lattice and a distant part
where the $N-n$ points are replaced by the continuous uniform distribution. The
expression in curly braces gives the formal series expansion of the Epstein
zeta function of the hexagonal lattices truncated to include only the $n-1$
shortest distances in the lattice\footnote{Indeed, the first few most frequent
  distances in a putative minimal energy configuration emulate remarkably well
  the first few distances in a hexagonal lattice (see, in particular,
  \cite[Figure~1]{ Brauchart_Hardin_Saff2012:riesz}).}.
(In order to get a non-trivial expansion, $n$ has to grow slowly in terms of $N$ to
infinity. The paper~\cite{Be1986} mentions some numerical experiments for
slowly growing $n$ but a rigorous investigation has not been undertaken.) The
discussion leading to Conjecture~\ref{conj:d.2.4.8.24} below suggests that
the semi-continuum approach would also work for $d = 4$, $8$, and $24$. In
general, it is not clear which local approximation should be used.

For $\Omega$ in $\R^p$ with vanishing $s$-capacity, the leading term is
rather well-understood. In the \emph{strictly hypersingular regime $s > d$},
Hardin and Saff~\cite{Hardin_Saff2005:minimal_riesz} (for rectifiable $d$-dimensional
manifolds including the sphere $\mathbb{S}^d$) and Borodachov, Hardin, and
Saff~\cite{Borodachov_Hardin_Saff2008:asymptotics_riesz} (for infinite compact
$d$-rectifiable sets\footnote{A $d$-rectifiable set is the Lipschitz image of a
  bounded set in $\R^d$.}) established the existence of a constant $C_{s,d}$ such that for a large class of sets $\Omega$ \footnote{The boundedness of
  $\mathcal{E}_s( \mathbb{S}^d; N ) / N^{1+s/d}$ has already been shown
  in~\cite{Kuijlaars_Saff1998:asymptotics_minimal_energy}.}
\begin{equation} \label{eq:Poppy.seed.Bagel.Theorem}
\lim_{N \to \infty} \frac{\mathcal{E}_s( \Omega; N )}{N^{1+s/d}} = \frac{C_{s,d}}{\left[ \mathcal{H}_d( \Omega ) \right]^{s/d}},
\end{equation}
where $\mathcal{H}_d$ denotes the $d$-dimensional Hausdorff
measure in $\R^p$ normalised such that the $d$-dimensional unit cube has
$\mathcal{H}_d$-measure~$1$.
This result is referred to as the \emph{Poppy-seed Bagel Theorem} because
of its interpretation for distributing points on a
torus.
%
%
Except for one-dimensional sets (when $C_{s,1}$ is
twice the Riemann zeta function at $s$, see~\cite[Thm.~3.1]{MaMaRaSa2004}), the
precise value of $C_{s,d}$ is not known. Its determination is a challenging
open problem. The significance and difficulty of obtaining $C_{s,d}$ is due to
the deep connection to densest
packings. In~\cite{Borodachov_Hardin_Saff2007:asymptotics_best-packing} it is
shown that $C_{s,d}$ is tied to the largest sphere packing density $\Delta_d$
in~$\R^d$ and the best-packing distance $\delta_N^*$ of $N$-points on
$\mathbb{S}^d$ by means of the limit relations\footnote{Indeed, one can recast
  this relation as $\Delta_d = \lim_{s \to \infty} \lim_{N \to \infty} \big[
  \mathcal{E}_s( \frac{1}{2} \mathbb{B}^d; N ) / N^{1+s/d} \big]^{-d/s}$.}
\begin{equation} \label{eq:density.limit.relations}
\lim_{s \to \infty} \left[ C_{s,d} \right]^{-1/s} = 
2 \left[ \frac{\Delta_d}{\mathcal{H}_d( \mathbb{B}^d )} \right]^{1/d} 
= \lim_{N \to \infty} N^{1/d} \delta_N^*.
\end{equation}
Here, $\mathcal{H}_d( \mathbb{B}^d )$ is the volume of the unit ball in
$\R^d$. We recall that $\Delta_d$ is only known for three cases: $\Delta_1=1$,
$\Delta_2 = \pi / \sqrt{12}$ (Thue~\cite{Thue1910} and L. Fejes T{\'o}th~\cite{Fe1972})
and $\Delta_3=\pi/\sqrt{18}$ (Kepler conjecture proved by
Hales~\cite{Hales2005:kepler_conjecture}). The connection to (regular) lattices
is evident in the upper estimate of the constant $C_{s,d}$ in terms of the
Epstein zeta function $\zetafcn_\Lambda( s )$ of a lattice $\Lambda$ in
$\mathbb{R}^d$ which can be obtained by considering the Riesz $s$-energy of the
$N = n^d$ points of the rescaled lattice $\frac{1}{n} \Lambda$ lying in the
fundamental parallelotope $\Omega$ of $\Lambda$ and which implies that
$\mathcal{E}_s( \Omega; N ) \leq n^{d+s} \zetafcn_\Lambda( s ) = N^{1 + s / d}
\zetafcn_\Lambda( s )$ and thus for $s > d$
(cf.~\cite[Prop.~1]{Brauchart_Hardin_Saff2009:riesz_energy_roots}),
\begin{equation} \label{eq:C.s.d..Epstein.zeta.fcn}
C_{s,d} \leq \min_\Lambda \left| \Lambda \right|^{s/d} \zetafcn_\Lambda( s ),
\end{equation}
where the minimum is extended over all lattices $\Lambda$ in $\R^d$ with
positive co-volume $| \Lambda |$.
Because of \eqref{eq:density.limit.relations}, the sharpness of this inequality
touches on questions regarding densest lattice sphere packings. 
%
For ${1 \leq d \leq 8}$ and $d = 24$, the unique densest lattice in $\R^d$ up
to scaling and isometries is the root lattice $\mathsf{A}_1$, $\mathsf{A}_2$,
$\mathsf{A}_3$, $\mathsf{D}_4$, $\mathsf{D}_5$, $\mathsf{E}_6$, $\mathsf{E}_7$,
$\mathsf{E}_8$, and the Leech lattice $\Lambda_{24}$, respectively
(cf.~\cite{Cohn_Kumar2009:optimality_leech}). Among those the hexagonal lattice
$\mathsf{A}_2$ in $\R^2$, the $\mathsf{E}_8$ root lattice in $\R^8$ and the
Leech lattice $\Lambda_{24}$ in~$\R^{24}$ are conjectured to be universally
optimal, whereas the remaining lattices are provably not universally optimal
(cf.~\cite{Cohn_Kumar2007:universally_optimal_sphere, CohnKumarSchurmann}). See
\cite{Sarnak_Stroembergsson2006:minima_epsteins} for local optimality results
and \cite{Coulangeon_Schuermann2012:energy_minimization} for improvements.
Montgomery~\cite{Montgomery1988:minimal_theta} proved that the hexagonal
lattice is universally optimal amongst all lattices in $\R^2$ (which is weaker
than universal optimality amongst all periodic point configurations). H.~Cohn and
N.~Elkies~\cite{Cohn_Elkies2003:bounds_sphere_packings_i} conjectured that
$\mathsf{E}_8$ and the Leech lattice $\Lambda_{24}$ solve the sphere packing
problem in their dimension.
It is generally expected that for sufficiently large $d$, lattice packings are
not densest packings and \cite{ToSt2006} suggests that best-packings are highly
``disordered'' as $d \to \infty$. This motivates the following
conjecture\footnote{The conjecture for $d = 2$ appeared
  in~\cite{Kuijlaars_Saff1998:asymptotics_minimal_energy}.}.
%
%
\begin{conjecture}[\cite{Brauchart_Hardin_Saff2009:riesz_energy_roots}] \label{conj:d.2.4.8.24}
  For $d = 2$, $4$, $8$, and $24$, one has $C_{s,d} = \left| \Lambda_d
  \right|^{s/d} \zetafcn_{\Lambda_d}( s )$, where $\Lambda_d$ denotes,
  respectively, the hexagonal lattice $\mathsf{A}_2$, the root lattices
  $\mathsf{D}_4$ and $\mathsf{E}_8$, and the Leech lattice $\Lambda_{24}$.
\end{conjecture}

We remark that in \cite{Brauchart_Hardin_Saff2012:riesz} very coarse lower and
upper bounds are obtained for $\mathcal{E}_s( \mathbb{S}^d; N )$ which imply for $s > d \geq 2$ and $( s - d ) /
2$ not an integer, the estimates  
\begin{equation*}
\frac{d}{s-d} \left[ \frac{1}{2} \frac{\gammafcn((d+1)/2) \gammafcn(1+(s-d)/2)}{\sqrt{\pi}\gammafcn(1+s/2)} \right]^{s/d} \leq \frac{C_{s,d} }{ \left[ \mathcal{H}_d(\mathbb{S}^d) \right]^{s/d}} \leq \left[ \frac{\mathcal{H}_d(\mathbb{B}^d)}{\mathcal{H}_d(\mathbb{S}^d)( 1-d/s )} \right]^{s/d}.
\end{equation*}
%
%

In the \emph{hypersingular case $s = d$}, 
more can be said. 
It has been known from~\cite{Kuijlaars_Saff1998:asymptotics_minimal_energy}
that the leading term of $\mathcal{E}_d( \mathbb{S}^d; N )$ grows like $N^2
\log N$ and
\begin{equation}
\lim_{N \to \infty} \frac{\mathcal{E}_d( \mathbb{S}^d; N)}{N^2 \log N} = \frac{\mathcal{H}_d(\mathbb{B}^d)}{\mathcal{H}_d(\mathbb{S}^d)} = \frac{1}{d} \frac{\omega_{d-1}}{\omega_{d}} = \frac{1}{d} \frac{\gammafcn((d+1)/2)}{\sqrt{\pi} \gammafcn(d/2)}.
\end{equation}
The best estimates so far for the second-order term has been obtained recently
in~\cite{Brauchart_Hardin_Saff2012:riesz},
%
\begin{equation*} 
- c(d) \, N^2 + \mathcal{O}(N^{2-2/d} \log N)\le \mathcal{E}_d( \mathbb{S}^d; N )-\frac{\mathcal{H}_d(\mathbb{B}^d)}{\mathcal{H}_d(\mathbb{S}^d)} \, N^2 \log N  \leq   \frac{\mathcal{H}_d(\mathbb{B}^d)}{\mathcal{H}_d(\mathbb{S}^d)} \, N^2 \log \log N + \mathcal{O}(N^2)  
\end{equation*}
as $N \to \infty$, where the constant $c(d)$ is given by 
\begin{equation*}
  c(2) = 1/2, \qquad c(d) \DEF \frac{\mathcal{H}_d(\mathbb{B}^d)}{\mathcal{H}_d(\mathbb{S}^d)} \left\{ 1 - \log \frac{\mathcal{H}_d(\mathbb{B}^d)}{\mathcal{H}_d(\mathbb{S}^d)} + d \left[ \digammafcn(d/2) - \digammafcn(1) - \log 2 \right] \right\} > 0.
\end{equation*}
(Recall, that $\digammafcn=\digammafcn'/\digammafcn$ denotes the digamma function.) Based on a limiting process $s \to d$ in Conjecture~\ref{conj:fundamental.conj} given below, it is conjectured in \cite{Brauchart_Hardin_Saff2012:riesz} that the correct order of the second term is $N^2$. Furthermore, in the case $d = 2$ a conjecture is posed for the constant of the $N^2$-term. 
%
%

\subsection{Higher Order Terms -- Complete Asymptotic 
Expansions --  Fundamental Conjecture}

Very little is known about higher-order terms of the asymptotics of the
minimal Riesz $s$-energy except for the unit circle. As the $N$th roots of
unity are universally optimal,\footnote{For $s \geq -1$ (and $s \neq 0$) a
convexity argument can be applied to get optimality for Riesz $s$-energy
(see~\cite{AlSt1974,Fe1964,Go2003}). The much more general
result~\cite[Theorem~1.2]{Cohn_Kumar2007:universally_optimal_sphere} provides
optimality for $s > -2$.}
the complete asymptotic expansion can be obtained by direct computation of the
Riesz $s$-energy $\mathcal{L}_s( N )$ of the $N$th roots of unity,
see~\cite{Brauchart_Hardin_Saff2009:riesz_energy_roots} for Euclidean and
\cite{BrHaSa2012} for geodesic metric. Indeed, for $s \in \C$ with $s
\neq 0, 1, 3, 5, \dots$ and fixed $p = 1, 2, 3, \dots$, one has\footnote{The
  precise formulas for finite $N \geq 2$ are obtained in~\cite{Br2011arXiv}.}
\begin{equation} \label{eq:circle.asympt}
\begin{split}
\mathcal{L}_s(N) 
&= W_s(\mathbb{S}^1) \, N^2 + \frac{2\zetafcn(s)}{(2\pi)^s} \, N^{1+s} + \sum_{n=1}^p \alpha_n(s) \frac{2\zetafcn(s-2n)}{(2\pi)^s} \, N^{1+s-2n} \\
&\phantom{=\pm}+ \BigOh_{s,p}(N^{-1+\Re( s )-2p}) \qquad \text{as $N \to \infty$,}
\end{split}
\end{equation}
where the coefficients $\alpha_n(s)$, $n\geq0$, are defined by the generating function
relation
\begin{equation*} \label{sinc.power.0}
\left( \frac{\sin \pi z}{\pi z} \right)^{-s} = \sum_{n=0}^\infty \alpha_n(s) \, z^{2n}, \quad |z|<1, \ s\in \C.
\end{equation*}
Explicit formulas for $\alpha_n(s)$ in terms of generalised Bernoulli polynomials
$B_n^{(\alpha)}(x)$ are given in~\cite{BrHaSa2012}.
The asymptotics~\eqref{eq:circle.asympt} has two noteworthy features: this expansion is valid for complex $s$ and the Riemann zeta function plays an essential role.
%
The coefficients of the terms in the asymptotics are best understood as functions in the complex $s$-plane. This is called the \emph{principle of analytic continuation}.\footnote{This principle breaks
  down when the perfectly symmetric unit circle is replaced by some other
  smooth closed curve $\Gamma$. Then the $s$-equilibrium measure on $\Gamma$ is
  not the normalised arc-length measure for each $0 < s < 1$ which plays a role in
  the characterization of the coefficient of $N^2$ in the hypersingular regime
  $1 < s < 3$; see~\cite{Borodachov2012:lower_riesz}.} 
The interplay between the simple poles of the coefficient $W_s( \mathbb{S}^1 )$ of the $N^2$-term in \eqref{eq:W.s.sphere} and the simple poles of the shifted Riemann zeta functions then gives rise to a logarithmic term whenever $s$ tends to one of the exceptional cases $s = 0, 1, 3, 5, \dots$.
%


%
%
%

By combining the results for the potential theoretic and the hypersingular
regime, the principle of analytic continuation motivates the following
fundamental conjecture.

\begin{conjecture}[{see~\cite{Brauchart_Hardin_Saff2012:riesz}}] \label{conj:fundamental.conj}
Let $d \geq 2$. Then for $0 < s < d + 2$ with $s \neq d$, 
\begin{equation*}
\mathcal{E}_s( \mathbb{S}^d; N ) = W_s( \mathbb{S}^d ) \, N^2 + \frac{C_d( s )}{\left[ \mathcal{H}_d( \mathbb{S}^d ) \right]^{s/d}} \, N^{1+s/d} + o( N^{1 + s/d} ) \qquad \text{as $N \to \infty$,}
\end{equation*}
where $W_s( \mathbb{S}^d )$ is the analytic continuation of the right-hand side
of~\eqref{eq:W.s.sphere} and $C_d( s )$ is the analytic continuation of
$C_{s,d}$ in \eqref{eq:Poppy.seed.Bagel.Theorem}. Furthermore, for $d = 2, 4,
8$, and $24$, the constant $C_d( s )$ is the analytic continuation of $\left| \Lambda_d
\right|^{s/d} \zetafcn_{\Lambda_d}( s )$, where $\Lambda_d$ is given in
Conjecture~\ref{conj:d.2.4.8.24}.
\end{conjecture}

It should be discussed briefly that the asymptotic expansion of the minimal
$s$-energy can also be studied from a geometrical point of view by identifying
which features of the Voronoi cell decomposition (or its dual, the Delaunay
triangulation) induced by minimal $s$-energy configurations contribute in which
way to the asymptotics. For the $2$-sphere and large $N$, the typical picture
is a vast sea of hexagonal Voronoi cells --- thus the local approximation of
the neighbourhood of a typical point is done by a suitably scaled hexagonal
lattice to get the second term of the asymptotics. The topology of the sphere
gives rise to \emph{geometric frustration} (cf.~\cite{SaMo1999}) where certain
points pick up a \emph{topological charge} that measures the discrepancy from
the ideal coordination number (six) of the planar triangular lattice. Euler's
celebrated Polyhedral formula yields that the total topological charge on
$\mathbb{S}^2$ is always $12$.
Numerically, one observes ``scars'' for large $N$ emerging from $12$ pentagonal
centres. These scars attract pentagon-heptagon pairs which have total
topological charge zero. It is an unresolved question if there are Voronoi cells
with more than $7$ sides in minimising configurations.
It is not well understood how (\emph{i.e.}, on which level of the asymptotic
scale) scars, the type of Voronoi cells, and the variation in their sizes
affect higher-order terms of the asymptotics; see ~\cite{CaEtal2013} for
numerical background and \cite{Bowick_Giomi2009} for an approach using elastic
continuum formalism.


\subsection{Asymptotic expansion of logarithmic energy}
\label{sec:asympt-expans-minim.log}

The leading term of the asymptotic expansion for a compact set $\Omega$ in $\R^p$ with positive logarithmic capacity (\emph{i.e.}, finite logarithmic energy) follows from classical potential theory. 
It should be noted that (see~\cite{BoHaSaBook})
\begin{equation} \label{eq:log.energy.diff}
\frac{\dd}{\dd s} \mathcal{E}_s( \Omega; N ) \Big|_{s = 0^+} = \mathcal{E}_{\log}( \Omega; N ),  \qquad N \geq 2.
\end{equation}
For the unit sphere $\mathbb{S}^d$, one has 
\begin{equation*}
\lim_{N \to \infty} \frac{\mathcal{E}_{\log}( \mathbb{S}^d; N )}{N^2} = W_{\log}( \mathbb{S}^d ) = \log \frac{1}{\CAP_{\log}( \mathbb{S}^d )},
\end{equation*}
where the \emph{logarithmic energy of $\mathbb{S}^d$} is given by
\begin{equation*}
W_{\log}( \mathbb{S}^d ) = \frac{\dd W_s( \mathbb{S}^d )}{\dd s} \Big|_{s = 0^+} = \log \frac{1}{2} + \frac{1}{2} \left[ \digammafcn( d ) - \digammafcn( d/2 ) \right].
\end{equation*}
Here, $\digammafcn( z )$ is the digamma function. An averaging argument that uses an equal-area partition of $\mathbb{S}^d$ and bounds of G.~Wagner~\cite{Wagner1992:means_distances_upper} and \cite{Brauchart2008:optimal_logarithmic} yields
\begin{equation*}
\mathcal{E}_{\log}( \mathbb{S}^d; N ) = W_{\log}( \mathbb{S}^d ) \, N^2 - \frac{1}{d} \, N \log N + \BigOh( N ), \qquad N \to \infty.
\end{equation*}
Relation~\eqref{eq:log.energy.diff} and Conjecture~\ref{conj:fundamental.conj}
provide the basis for the following conjecture posed
in~\cite{Brauchart_Hardin_Saff2012:riesz}.

\begin{conjecture}
For $d = 2$, $4$, $8$, and $24$, 
\begin{equation*}
\mathcal{E}_{\log}( \mathbb{S}^d; N ) = W_{\log}( \mathbb{S}^d ) \, N^2 - \frac{1}{d} \, N \log N + C_{\log,d} \, N + o(N) \qquad \text{as $N \to \infty$,}
\end{equation*}
where 
\begin{equation*}
C_{\log,d} = \frac{1}{d} \log \frac{\mathcal{H}_d( \mathbb{S}^d )}{| \Lambda_d |} + \zetafcn_{\Lambda_d}^\prime( 0 ).
\end{equation*}
For $d = 2$ one has
\begin{equation*}
C_{\mathrm{log},2} = 2 \log 2 + \frac{1}{2} \log \frac{2}{3} + 3 \log \frac{\sqrt{\pi}}{\gammafcn(1/3)} = -0.05560530494339251850\dots.
\end{equation*}
\end{conjecture}

For more details see~\cite{Brauchart_Hardin_Saff2012:riesz}. Very recently,
L.~B{\'e}termin~\cite{Be_arXiv:1404.4485v2} found a surprising connection
between the problem of minimising a planar ``Coulombian renormalised energy''
derived from the Ginzburg-Landau model of superconductivity introduced by
E.~Sandier and S.~Serfaty in~\cite{SaSe2012} (also see the survey~\cite{Se2014}) and
the discrete logarithmic energy problem on~$\mathbb{S}^2$. The preliminary
results are: (i) the asymptotics of the minimal logarithmic energy on
$\mathbb{S}^2$ has a term of order $N$, (ii) whose constant is bounded from
above by $C_{\mathrm{log},2}$ given above, and (iii) this constant equals
$C_{\mathrm{log},2}$ if and only if a certain triangular lattice of density $1$
(called ``Abrikosov'' triangular lattice $\Z + e^{i\pi/3} \, \Z$, properly
scaled, see \cite{Se2014}) is the minimiser of the Coulombian renormalised
energy. 
%

\subsection{Numerical integration and discrepancy from the energy point of 
view}\label{sec:numer-integr-from}

The reproducing kernel Hilbert space approach (see Section~\ref{sec:designs}) enables us to write the squared worst-case error as
\begin{equation*}
\frac{1}{N^2} \sum_{i=1}^N \sum_{j=1}^N \KERNEL( \PT{x}_i, \PT{x}_j ) - \int_{\mathbb{S}^d} \int_{\mathbb{S}^d} \KERNEL( \PT{x}, \PT{y} ) \, \dd \sigma_d( \PT{x} ) \dd \sigma_d( \PT{y} ),
\end{equation*}
which can be interpreted as
$\KERNEL$-energy of the node set $\{ \PT{x}_1, \dots, \PT{x}_N \}$ of the QMC
method. Here, the energy kernel is a reproducing kernel for $H^s( \mathbb{S}^d
)$. Optimal node sets are solutions of the minimal energy problem for
this kernel. In \cite{Brauchart_Dick2013:characterization_sobolev_spaces} and
\cite{Brauchart_Dick2013:stolarskys_invariance} it is shown how the distance
kernel
\begin{equation} \label{eq:K.gd.s}
\KERNEL_{\mathrm{gd}}^{(s)}( \PT{x}, \PT{y} ) \DEF 2 W_{d-2s}( \mathbb{S}^d ) - \left\| \PT{x} - \PT{y} \right\|^{2s-d},\qquad \PT{x},\PT{y}\in\mathbb{S}^d,
\end{equation}
arises in a natural way as a reproducing kernel for $H^s( \mathbb{S}^d )$ for $s$ in $(d/2, d/2+1)$.
Evidently, there is a close connection between finding optimal QMC nodes and the problem of maximizing the sum of all pairwise distances taken to the power $2s-d$. 
Wagner's bounds (see~\cite{Wagner1990:means_distances_lower,Wagner1992:means_distances_upper})
yield that a sequence of $N$-point maximisers of such a generalised sum of distances is a QMC design sequence for
$H^s( \mathbb{S}^d )$, $d/2 < s < d/2 + 1$;
see~\cite{Brauchart_Saff_Sloan+2014:qmc-designs}.
Explicit constructions for QMC design sequences are not known. 
Based on numerical evidence it is conjectured in
\cite{Aistleitner_Brauchart_Dick2012:point_sets_sphere,Brauchart_Dick2012:QMC_rules}
that a $(0,2)$-sequence (a special digital net sequence in the sense of
\cite{DiPi2010}) or a Fibonacci lattice (also see~\cite{DiPi2010}) in the square
$[0,1)^2$ mapped to the $2$-sphere via an
\emph{area-preserving} map (\emph{Lambert azimuthal equal-area projection})
will be a QMC design sequence for $H^{3/2}( \mathbb{S}^2 )$.
For $s \geq d/2 + 1$, the space $H^s( \mathbb{S}^d )$ can also be provided with a
reproducing kernel which is essentially a distance kernel with power
$2s-d$. In order to ensure that the reproducing kernel is positive definite (in the sense of
Schoenberg~\cite{Schoenberg1938:positive_definite}) a polynomial correction term is needed when $s-d/2$ is not a positive integer.\footnote{A logarithm of the distance appears in the reproducing kernel when $s - d/2$ is a positive integer.} 
%
%
%
%
\footnote{When defining the function space as Bessel potential space, then no correction terms are needed. In the Hilbert space setting ($p = 2$), the worst-case is given as Bessel-energy. The Bessel kernel on the sphere, however, has a series expansion in spherical harmonics without convenient closed form representation. For general $p>1$ the worst-case error has an integral representation; see \cite{BrSaSl2014arXiv} and also \cite{BCCGST2013}.} 
%
Such a correction term is annihilated when the search for optimal QMC designs for $H^s( \mathbb{S}^d )$, $s \in (d/2 + L, d/2 + 1 + L)$, is restricted to spherical $L$-designs. In that case it suffices to minimise the energy functional (see~\cite{Brauchart_Saff_Sloan+2014:qmc-designs})
\begin{equation*}
\left[ \WCE_{H^s}(X_{N,L}) \right]^2 = \frac{1}{N^2} \sum_{i=1}^{N} \sum_{j=1}^{N} (-1)^{L+1} \left\| \PT{x}_{i,L} - \PT{x}_{j,L} \right\|^{2s-d} - (-1)^{L+1} W_{d-2s}(\mathbb{S}^d)
\end{equation*}
subject to the condition that the node set $X_{N,L} = \{ \PT{x}_{1,L}, \dots,
\PT{x}_{N,L} \} \subset \mathbb{S}^d$ is a spherical $L$-design. Note that the
fixed $L$ (once $s$ is fixed) is ``small'' in the following
sense: 
a QMC method with a more regular (in the sense of spherical designs) node set
is more suitable for integrating functions from a smoother function space $H^s(
\mathbb{S}^d )$. But regularity beyond a critical order will not improve the
worst-case error bound; this should be compared with the optimal order~\eqref{eq:wce2}.
We remark that the Cui and Freeden kernel \cite{Cui_Freeden1997:equidistribution_sphere}
\begin{equation*}
\KERNEL_{\mathrm{CF}}(\PT{x}, \PT{y}) \DEF 2 - 2 \log\Big( 1 + \frac{1}{2} \left\| \PT{x} - \PT{y} \right\| \Big),\qquad \PT{x},\PT{y}\in\mathbb{S}^2,
\end{equation*}
which was used to define a ``generalised discrepancy'' to measure uniform distribution of point set sequences,
can be interpreted as reproducing kernel for $H^{3/2}( \mathbb{S}^2 )$ as observed in \cite{Sloan_Womersley2004:extremal_sphere} and the minimising $\KERNEL_{\mathrm{CF}}$-energy point configurations give rise to a QMC design sequence for $H^{3/2}( \mathbb{S}^2 )$. Recently, C.~Choirat and R.~Seri~\cite{ChSe2013,ChSe2013b} derived the analogous kernel for $d$-spheres. The corresponding minimising configurations then form QMC design sequences for $H^{(d+1)/2}( \mathbb{S}^d )$. A curious observation is that when $H^s( \mathbb{S}^d )$, $d/2 < s < d/2 + 1$, is provided with the reproducing kernel $\KERNEL_{\mathrm{gd}}^{(s)}$ from \eqref{eq:K.gd.s}, a limit process yields that (see~\cite{BrWo2010Manuscript})
\begin{equation*}
\lim_{s \to (d/2)^+} \frac{\left[ \WCE_{H^s}(X_{N}) \right]^2 - \frac{1}{N}}{2s-d} = \frac{1}{N^2} \mathop{\sum_{i=1}^N \sum_{j=1}^N}_{i \neq j} \log \frac{1}{\left\| \PT{x}_i - \PT{x}_j \right\|} -  V_{\mathrm{log}}( \mathbb{S}^d )
\end{equation*} 
for any $N$-point set $X_N = \{ \PT{x}_1, \dots, \PT{x}_N \} \subset \mathbb{S}^d$. This suggests that the function space $H^{d/2}( \mathbb{S}^d )$ (which is not a reproducing kernel Hilbert space) is paired with the logarithmic kernel and the logarithmic energy of an $N$-point set can be understood as a limit of worst-case errors in the above sense. This paring can be extended to $s < d/2$ and the open question is how to define integration and error of integration so that the corresponding Riesz $(2s-d)$-energy is a meaningful measure for this error.

Due to J.~Beck~\cite{Be1984}, the spherical cap discrepancy $D(X_N)$ given in \eqref{eq:discrepancy} of any $X_N \subset \mathbb{S}^d$ is bounded like $\gg N^{-1/2-1/(2d)}$ and, by appealing to a probabilistic argument, 
there are $X_N \subset \mathbb{S}^d$ with
\begin{equation} \label{eq:low-discr.property}
D( X_N ) \ll N^{-1/2-1/(2d)} \, \sqrt{\log N}.
\end{equation}
A sequence of point sets with this property is called a \emph{low-discrepancy sequence on~$\mathbb{S}^d$}.
%
%
%
It is shown in \cite{Brauchart_Saff_Sloan+2014:qmc-designs} that a
low-discrepancy sequence $(X_N^{\mathrm{LD}})$ on $\mathbb{S}^d$ is \emph{almost} a QMC design
sequence for $H^{(d+1)/2}( \mathbb{S}^d )$ in the sense that 
for sufficiently large $N$, 
\begin{equation} \label{eq:almost.QMC.estimate}
c \, N^{-(d+1)/(2d)} \leq \WCE_{H^{(d+1)/2}}( X_N^{\mathrm{LD}} ) \leq C \, N^{-(d+1)/(2d)} \, \sqrt{\log N}.
\end{equation}
One of the deep unresolved questions is if the logarithmic term in
\eqref{eq:low-discr.property} arising from a probabilistic argument can be
removed. It is also unknown how to construct a sequence of $N$-point sets
explicitly with spherical cap discrepancy decaying like $N^{-1/2-1/(2d)} \, \sqrt{\log N}$.
For the $2$-sphere, the construction of A.~Lubotzky, R.~Phillips, and
P.~Sarnak~\cite{LuPhSa1986,LuPhSa1987} satisfies the estimate $D(
X_N^{\mathrm{LPS}} ) \ll ( \log N )^{2/3} \, N^{-1/3}$ with numerical evidence
indicating a convergence rate of $\BigOh( N^{-1/2} )$, whereas the
\emph{spherical Fibonacci lattice point sets} of
\cite{Aistleitner_Brauchart_Dick2012:point_sets_sphere} obey the estimate $D(
X_N^{ABD} ) \leq 44 \sqrt{8} \, N^{-1/2}$ with numerical results showing a
convergence rate of $\BigOh( ( \log N )^c \, N^{-3/4} )$ for some $1/2 \leq c
\leq 1$.
The \emph{typical} spherical cap discrepancy of $N$ i.i.d. uniformly
distributed random points on~$\mathbb{S}^2$ is of exact order $N^{-1/2}$, see \cite{Aistleitner_Brauchart_Dick2012:point_sets_sphere}. 
%
Surprisingly, minimal Coulomb energy points on~$\mathbb{S}^2$ do not have low
spherical cap discrepancy. J.~Korevaar~\cite{Korevaar1996:fekete_points}
conjectured that minimal $(d-1)$-energy configurations on $\mathbb{S}^d$ have
spherical cap discrepancy of order~$N^{-1/d}$. This conjecture was proven by
M.~G{\"o}tz~\cite{Goetz2000:distribution_extremal} up to a logarithmic
factor. He also gave a lower bound of order~$N^{-1/2}$ for $d = 2$. It is open
if Korevaar's conjecture extends to the full potential-theoretic regime. On the
basis of the Poppy-seed Bagel theorem, one could conjecture that minimal
$s$-energy points for hypersingular $s$ should have small spherical cap
discrepancy. The only result obtained so far is the very weak order $\sqrt{\log
  \log N} / \log N$ result
in~\cite{Damelin_Grabner2003:energy_functional_numeric} for the boundary case
$s = d$. The proof employs a smoothened Riesz energy functional.

Stolarsky's invariance principle (see K.~B.~Stolarsky~\cite{Stolarsky1973:sums_distances_ii}) states that the sum of all mutual distances (a Riesz energy with Riesz parameter $-1$) and the spherical cap $L^2$-discrepancy
\begin{equation*}
D_{L^2}(X_{N}) \DEF \left( \int_{0}^{\pi} \int_{\mathbb{S}^{d}} \left| \frac{1}{N}\sum_{\PT{y}\in X_N}\mathbbm{1}_{\PT{y}\in
    C(\PT{z}, \theta)} - \sigma_d(C({\PT z}, \theta)) \right|^{2}\dd{\sigma_d}({\PT z})\, \sin \theta \dd{\theta}\right)^{1/2}
\end{equation*}
is constant regardless of the choice of the node set $X_N$ on $\mathbb{S}^d$; \emph{i.e.}, 
\begin{equation*}
\frac{1}{N^2} \sum_{i=1}^N\sum_{j=1}^{N} \left\| \PT{x}_i - \PT{x}_j \right\| + \frac{1}{C_d} \left[ D_{L^2}(X_{N}) \right]^{2} = \int_{\mathbb{S}^d} \int_{\mathbb{S}^d} \left\| \PT{x}-\PT{y}\right\| \dd \sigma_d(\PT{x}) \,\dd \sigma_d(\PT{y}).
\end{equation*}
This principle connects in a very direct way the three areas optimal energy (maximizing the sum of distances on $\mathbb{S}^d$), uniform distribution (spherical cap $L^2$-discrepancy), and numerical integration with QMC methods for functions on $\mathbb{S}^d$. 
Rearranging terms,
\begin{equation*}
\frac{1}{\sqrt{C_d}} \, D_{L^2}(X_{N})  = \sqrt{ \int_{\mathbb{S}^d} \int_{\mathbb{S}^d} \left\| \PT{x}-\PT{y}\right\| \dd \sigma_d(\PT{x}) \,\dd \sigma_d(\PT{y}) - \frac{1}{N^2} \sum_{i=1}^N\sum_{j=1}^{N} \left\| \PT{x}_i - \PT{x}_j \right\|},
\end{equation*}
one obtains a convenient way of computing the discrepancy $D_{L^2}(X_{N})$. It is shown in \cite{Brauchart_Dick2013:stolarskys_invariance} that the right-hand side represents the worst-case error of a QMC method with node set $\{\PT{x}_1, \dots, \PT{x}_N \} \subset \mathbb{S}^d$ for functions in $H^{s}( \mathbb{S}^d )$ provided with the distance kernel $\KERNEL_{\mathrm{gd}}^{(s)}$ given in \eqref{eq:K.gd.s} with $s = (d+1)/2$.
Moreover, in this setting, the spherical cap $L^2$-discrepancy can be interpreted as worst-case error and \emph{vice versa}.
%
%
A conjecture for the asymptotic expansion (as $N \to \infty$) of the minimal spherical cap $L^2$-discrepancy that is based on the fundamental conjecture for the minimal Riesz $s$-energy and the principle of analytic continuation is proposed in~\cite{Brauchart2011:optimal_riesz}.
%
%
The paper \cite{Brauchart_Dick2013:characterization_sobolev_spaces} gives an extension of Stolarsky's invariance principle to general powers of the distance (raised to the power $2s-d$) involving the generalised spherical $L_2$-discrepancy
\begin{equation*}
\int_{0}^\pi \int_{\mathbb{S}^d} \left| \mathcal{D}_{X_N, \beta}(\PT{z}; \cos \theta) \right|^2 \dd \sigma_d( \PT{z} ) \sin \theta \dd \theta,
\end{equation*}
for the local discrepancy function (with smoothness index $s = \beta + ( d - 1 )/2$)
\begin{equation*}
\mathcal{D}_{X_N, \beta}(\PT{z}; t) \DEF \frac{1}{N} \sum_{\PT{x} \in X_N} (\PT{x} \cdot \PT{z} - t)_+^{\beta-1} - \int_{\mathbb{S}^d} (\PT{y} \cdot \PT{z} - t)_+^{\beta-1} \,\mathrm{d} \sigma_d(\PT{y}).
\end{equation*}
The paper \cite{Brauchart_Dick2014arXiv_spherical_cone} considers the discrepancy with respect to truncated spherical cones that are anchored at infinity and extends Stolarsky's invariance principle to this setting. In ~\cite{Grabner2014:point_minimal_energy} further connections between energy and discrepancy are discussed. We also mention \cite{Le2013} which considers asymptotically uniformly distributed points with an upper bound on the spherical cap discrepancy and a lower bound on the separation. 


\bigskip
\noindent\textbf{Acknowledgement.} The authors are indebted to three anonymous
referees for their valuable remarks and suggestions which 
greatly increased the quality of presentation of this paper.
They are grateful to W\"oden Kusner for his careful proofreading of the
manuscript and his many invaluable comments.

\section*{References}
\addcontentsline{toc}{section}{References}

\bibliographystyle{elsarticle-harv} %
\bibliography{energy}

\begin{thebibliography}{201}
\expandafter\ifx\csname natexlab\endcsname\relax\def\natexlab#1{#1}\fi
\expandafter\ifx\csname url\endcsname\relax
  \def\url#1{\texttt{#1}}\fi
\expandafter\ifx\csname urlprefix\endcsname\relax\def\urlprefix{URL }\fi

\bibitem[{Aistleitner et~al.(2012)Aistleitner, Brauchart, and
  Dick}]{Aistleitner_Brauchart_Dick2012:point_sets_sphere}
Aistleitner, C., Brauchart, J.~S., Dick, J., 2012. Point {S}ets on the {S}phere
  {$\mathbb{S}^2$} with {S}mall {S}pherical {C}ap {D}iscrepancy. Discrete
  Comput. Geom. 48~(4), 990--1024.

\bibitem[{Alexander(1972)}]{Al1972}
Alexander, R., 1972. On the sum of distances between {$n$} points on a sphere.
  Acta Math. Acad. Sci. Hungar. 23, 443--448.

\bibitem[{Alexander(1977)}]{Al1977}
Alexander, R., 1977. On the sum of distances between {$n$} points on a sphere.
  {II}. Acta Math. Acad. Sci. Hungar. 29~(3-4), 317--320.

\bibitem[{Alexander and Stolarsky(1974)}]{AlSt1974}
Alexander, R., Stolarsky, K.~B., 1974. Extremal problems of distance geometry
  related to energy integrals. Trans. Amer. Math. Soc. 193, 1--31.

\bibitem[{Ambrus et~al.(2013)Ambrus, Ball, and Erd{\'e}lyi}]{AmBaEr2013}
Ambrus, G., Ball, K.~M., Erd{\'e}lyi, T., 2013. Chebyshev constants for the
  unit circle. Bull. Lond. Math. Soc. 45~(2), 236--248.

\bibitem[{An et~al.(2010)An, Chen, Sloan, and
  Womersley}]{An_Chen_Sloan+2010:spherical_designs}
An, C., Chen, X., Sloan, I.~H., Womersley, R.~S., 2010. Well conditioned
  spherical designs for integration and interpolation on the two-sphere. SIAM
  J. Numer. Anal. 48~(6), 2135--2157.

\bibitem[{An et~al.(2012)An, Chen, Sloan, and
  Womersley}]{An_Chen_Sloan+2012:regularized_approximations}
An, C., Chen, X., Sloan, I.~H., Womersley, R.~S., 2012. Regularized least
  squares approximations on the sphere using spherical designs. SIAM J. Numer.
  Anal. 50~(3), 1513--1534.

\bibitem[{Apostol(1990)}]{Apostol1990:modular_functions}
Apostol, T.~M., 1990. Modular functions and {D}irichlet series in number
  theory, 2nd Edition. Vol.~41 of Graduate Texts in Mathematics.
  Springer-Verlag, New York.

\bibitem[{Armentano et~al.(2011)Armentano, Beltr{\'a}n, and Shub}]{ArBeSh2011}
Armentano, D., Beltr{\'a}n, C., Shub, M., 2011. Minimizing the discrete
  logarithmic energy on the sphere: the role of random polynomials. Trans.
  Amer. Math. Soc. 363~(6), 2955--2965.

\bibitem[{Bachoc(2005)}]{Bachoc2005:designs_lattices}
Bachoc, C., 2005. Designs, groups and lattices. J. Th{\'e}or. Nombres Bordeaux
  17~(1), 25--44.

\bibitem[{Bachoc and Vallentin(2008)}]{BachocVallentin1}
Bachoc, C., Vallentin, F., 2008. New upper bounds for kissing numbers from
  semidefinite programming. J. Amer. Math. Soc. 21~(3), 909--924.

\bibitem[{Bachoc and Venkov(2001)}]{Bachoc_Venkov2001:modular_forms_designs}
Bachoc, C., Venkov, B., 2001. Modular forms, lattices and spherical designs.
  In: R{\'e}seaux euclidiens, designs sph{\'e}riques et formes modulaires.
  Vol.~37 of Monogr. Enseign. Math. Enseignement Math., Geneva, pp. 87--111.

\bibitem[{Ballinger et~al.(2009)Ballinger, Blekherman, Cohn, Giansiracusa,
  Kelly, and
  Sch{\"u}rmann}]{Ballinger_Blekherman_Cohn+2009:experimental_spheres}
Ballinger, B., Blekherman, G., Cohn, H., Giansiracusa, N., Kelly, E.,
  Sch{\"u}rmann, A., 2009. Experimental study of energy-minimizing point
  configurations on spheres. Experiment. Math. 18~(3), 257--283.

\bibitem[{Bannai(1987)}]{Bannai1987:rigid_designs}
Bannai, E., 1987. Rigid spherical $t$-designs and a theorem of {Y. Hong}. J.
  Fac. Sci. Univ. Tokyo Sect. IA Math. 34, 485--489.

\bibitem[{Bannai and Bannai(2009)}]{Bannai_Bannai2009:survey_designs}
Bannai, E., Bannai, E., 2009. A survey on spherical designs and algebraic
  combinatorics on spheres. European J. Combin. 30~(6), 1392--1425.

\bibitem[{Bannai and Damerell(1979)}]{Bannai_Damerell1979:tight_designs_i}
Bannai, E., Damerell, R.~M., 1979. Tight spherical designs. {I}. J. Math. Soc.
  Japan 31~(1), 199--207.

\bibitem[{Bannai and Damerell(1980)}]{Bannai_Damerell1980:tight_designs_ii}
Bannai, E., Damerell, R.~M., 1980. Tight spherical designs. {II}. J. London
  Math. Soc. (2) 21~(1), 13--30.

\bibitem[{Beck(1984)}]{Be1984}
Beck, J., 1984. Sums of distances between points on a sphere---an application
  of the theory of irregularities of distribution to discrete geometry.
  Mathematika 31~(1), 33--41.

\bibitem[{Beltr{\'a}n(2013{\natexlab{a}})}]{Be2013}
Beltr{\'a}n, C., 2013{\natexlab{a}}. Harmonic properties of the logarithmic
  potential and the computability of elliptic {F}ekete points. Constr. Approx.
  37~(1), 135--165.

\bibitem[{Beltr{\'a}n(2013{\natexlab{b}})}]{Be2013b}
Beltr{\'a}n, C., 2013{\natexlab{b}}. The state of the art in {S}male's 7th
  problem. In: Foundations of computational mathematics, {B}udapest 2011. Vol.
  403 of London Math. Soc. Lecture Note Ser. Cambridge Univ. Press, Cambridge,
  pp. 1--15.

\bibitem[{Beltr{\'a}n(2015)}]{Beltran2015:facility_location_formulation}
Beltr{\'a}n, C., 2015. A facility location formulation for stable polynomials
  and elliptic {F}ekete points. Found. Comput. Math.To appear.

\bibitem[{Bendito et~al.(2009)Bendito, Carmona, Encinas, Gesto, G{\'o}mez,
  Mouri{\~n}o, and S{\'a}nchez}]{BeCaEnGe2009}
Bendito, E., Carmona, A., Encinas, A.~M., Gesto, J.~M., G{\'o}mez, A.,
  Mouri{\~n}o, C., S{\'a}nchez, M.~T., 2009. Computational cost of the {F}ekete
  problem. {I}. {T}he forces method on the 2-sphere. J. Comput. Phys. 228~(9),
  3288--3306.

\bibitem[{Berens et~al.(1968/1969)Berens, Butzer, and
  Pawelke}]{Berens_Butzer_Pawelke1969:limitierungsverfahren_reihen_kugelfunktionen}
Berens, H., Butzer, P.~L., Pawelke, S., 1968/1969. Limitierungsverfahren von
  {R}eihen mehrdimensionaler {K}ugelfunktionen und deren
  {S}aturationsverhalten. Publ. Res. Inst. Math. Sci. Ser. A 4, 201--268.

\bibitem[{Berezin(1986)}]{Be1986}
Berezin, A.~A., 1986. Asymptotics of the maximum number of repulsive particles
  on a spherical surface. J. Math. Phys. 27~(6), 1533--1536.

\bibitem[{B{\'e}termin(2014)}]{Be_arXiv:1404.4485v2}
B{\'e}termin, L., 2014. Renormalized energy and asymptotic expansion of optimal
  logarithmic energy on the sphere, manuscript, \texttt{arXiv:1404.4485v2
  [math.AP]}.

\bibitem[{Bj{\"o}rck(1956)}]{Bj1956}
Bj{\"o}rck, G., 1956. Distributions of positive mass, which maximize a certain
  generalized energy integral. Ark. Mat. 3, 255--269.

\bibitem[{Blum et~al.(1997)Blum, Cucker, Shub, and Smale}]{BCSS1997}
Blum, L., Cucker, F., Shub, M., Smale, S., 1997. Complexity and real
  computation. Foreword by Richard M. Karp. New York, NY: Springer.

\bibitem[{Bondarenko et~al.(2014)Bondarenko, Hardin, and Saff}]{BoHaSa2014}
Bondarenko, A.~V., Hardin, D.~P., Saff, E.~B., 2014. Mesh ratios for
  best-packing and limits of minimal energy configurations. Acta Math. Hung.
  142~(1), 118--131.

\bibitem[{Bondarenko et~al.(2013)Bondarenko, Radchenko, and
  Viazovska}]{Bondarenko_Radchenko_Viazovska2011:optimal_bounds_designs}
Bondarenko, A.~V., Radchenko, D., Viazovska, M.~S., 2013. Optimal asymptotic
  bounds for spherical designs. Ann. of Math. (2) 178~(2), 443--452.

\bibitem[{Bondarenko et~al.(2015)Bondarenko, Radchenko, and
  Viazovska}]{Bondarenko_Radchenko_Viazovska2014:well_separated}
Bondarenko, A.~V., Radchenko, D., Viazovska, M.~S., 2015. Well separated
  spherical designs. Constr. Approx.To appear.

\bibitem[{Borodachov(2012)}]{Borodachov2012:lower_riesz}
Borodachov, S.~V., 2012. Lower order terms of the discrete minimal {R}iesz
  energy on smooth closed curves. Canad. J. Math. 64~(1), 24--43.

\bibitem[{Borodachov and Bosuwan(2014)}]{BoBo2014}
Borodachov, S.~V., Bosuwan, N., 2014. Asymptotics of {D}iscrete {R}iesz
  d-{P}olarization on {S}ubsets of d-{D}imensional {M}anifolds. Potential Anal.
  41~(1), 35--49.

\bibitem[{Borodachov et~al.(2007)Borodachov, Hardin, and
  Saff}]{Borodachov_Hardin_Saff2007:asymptotics_best-packing}
Borodachov, S.~V., Hardin, D.~P., Saff, E.~B., 2007. Asymptotics of
  best-packing on rectifiable sets. Proc. Amer. Math. Soc. 135~(8), 2369--2380
  (electronic).

\bibitem[{Borodachov et~al.(2008)Borodachov, Hardin, and
  Saff}]{Borodachov_Hardin_Saff2008:asymptotics_riesz}
Borodachov, S.~V., Hardin, D.~P., Saff, E.~B., 2008. Asymptotics for discrete
  weighted minimal {R}iesz energy problems on rectifiable sets. Trans. Amer.
  Math. Soc. 360~(3), 1559--1580 (electronic).

\bibitem[{Borodachov et~al.(2014)Borodachov, Hardin, and Saff}]{BoHaSa2014b}
Borodachov, S.~V., Hardin, D.~P., Saff, E.~B., 2014. Low complexity methods for
  discretizing manifolds via riesz energy minimization. Found. Comput. Math.,
  1--36.

\bibitem[{Borodachov et~al.(2015)Borodachov, Hardin, and Saff}]{BoHaSaBook}
Borodachov, S.~V., Hardin, D.~P., Saff, E.~B., 2015. Minimal {D}iscrete
  {E}nergy on the {S}phere and {O}ther {M}anifolds, {S}pringer Verlag, to
  appear.

\bibitem[{Borodin and Serfaty(2013)}]{BoSe2013}
Borodin, A., Serfaty, S., 2013. Renormalized energy concentration in random
  matrices. Comm. Math. Phys. 320~(1), 199--244.

\bibitem[{Bowick et~al.(2002)Bowick, Cacciuto, Nelson, and
  Travesset}]{Bowick_Cacciuto_Nelson+2002:crystalline_order_sphere}
Bowick, M., Cacciuto, A., Nelson, D.~R., Travesset, A., 2002. Crystalline order
  on a sphere and the generalized {T}homson problem. Phys. Rev. Lett. 89,
  185502.

\bibitem[{Bowick et~al.(2014)Bowick, Cecka, Giomi, Middleton, and
  Zielnicki}]{bowick_etal:_thomson_applet}
Bowick, M., Cecka, C., Giomi, L., Middleton, A., Zielnicki, K., 2014. Thomson
  problem @ {S}. {U}.
\newline\urlprefix\url{http://thomson.phy.syr.edu/}

\bibitem[{Bowick et~al.(2006)Bowick, Cacciuto, Nelson, and
  Travesset}]{BoCaNeTr2006}
Bowick, M.~J., Cacciuto, A., Nelson, D.~R., Travesset, A., 2006. Crystalline
  particle packings on a sphere with long-range power-law potentials. Phys.
  Rev. B 73, 024115.

\bibitem[{Bowick and Giomi(2009)}]{Bowick_Giomi2009}
Bowick, M.~J., Giomi, L., 2009. Two-dimensional matter: order, curvature and
  defects. Advances in Physics 58~(5), 449--563.

\bibitem[{Brandolini et~al.(2013 (in press))Brandolini, Choirat, Colzani,
  Gigante, Seri, and Travaglini}]{BCCGST2013}
Brandolini, L., Choirat, C., Colzani, L., Gigante, G., Seri, R., Travaglini,
  G., 2013 (in press). Quadrature rules and distribution of points on
  manifolds. Ann. Sc. Norm. Super. Pisa Cl. Sci., 35.

\bibitem[{Brauchart(2006)}]{Brauchart2006:riesz_energy}
Brauchart, J.~S., 2006. About the second term of the asymptotics for optimal
  {R}iesz energy on the sphere in the potential-theoretical case. Integral
  Transforms Spec. Funct. 17~(5), 321--328.

\bibitem[{Brauchart(2008)}]{Brauchart2008:optimal_logarithmic}
Brauchart, J.~S., 2008. Optimal logarithmic energy points on the unit sphere.
  Math. Comp. 77~(263), 1599--1613.

\bibitem[{Brauchart(2011{\natexlab{a}})}]{Brauchart2011:optimal_riesz}
Brauchart, J.~S., 2011{\natexlab{a}}. Optimal discrete {R}iesz energy and
  discrepancy. Unif. Distrib. Theory 6~(2), 207--220.

\bibitem[{Brauchart(2011{\natexlab{b}})}]{Br2011arXiv}
Brauchart, J.~S., 2011{\natexlab{b}}. A remark on exact formulas for the
  {R}iesz energy of the {$N$th} roots of unity. arXiv:1105.5530v1 [math-ph].

\bibitem[{Brauchart and Dick(2012)}]{Brauchart_Dick2012:QMC_rules}
Brauchart, J.~S., Dick, J., 2012. Quasi-{M}onte {C}arlo rules for numerical
  integration over the unit sphere {$\mathbb{S}^2$}. Numer. Math. 121~(3),
  473--502.

\bibitem[{Brauchart and
  Dick(2013{\natexlab{a}})}]{Brauchart_Dick2013:characterization_sobolev_spaces}
Brauchart, J.~S., Dick, J., 2013{\natexlab{a}}. A {C}haracterization of
  {S}obolev {S}paces on the {S}phere and an {E}xtension of {S}tolarsky's
  {I}nvariance {P}rinciple to {A}rbitrary {S}moothness. Constr. Approx. 38~(3),
  397--445.

\bibitem[{Brauchart and
  Dick(2013{\natexlab{b}})}]{Brauchart_Dick2013:stolarskys_invariance}
Brauchart, J.~S., Dick, J., 2013{\natexlab{b}}. A simple proof of {S}tolarsky's
  invariance principle. Proc. Amer. Math. Soc. 141~(6), 2085--2096.

\bibitem[{Brauchart et~al.(2014{\natexlab{a}})Brauchart, Dick, and
  Fang}]{Brauchart_Dick2014arXiv_spherical_cone}
Brauchart, J.~S., Dick, J., Fang, L., 2014{\natexlab{a}}. Spatial
  low-discrepancy sequences, spherical cone discrepancy, and applications in
  financial modeling, manuscript, \texttt{arXiv:1408.4609}.

\bibitem[{Brauchart et~al.(2014{\natexlab{b}})Brauchart, Dick, Saff, Sloan,
  Wang, and Womersley}]{BrSaSl2014arXiv}
Brauchart, J.~S., Dick, J., Saff, E.~B., Sloan, I.~H., Wang, Y.~G., Womersley,
  R.~S., 2014{\natexlab{b}}. Covering of spheres by spherical caps and
  worst-case error for equal weight cubature in {S}obolev spaces.
  arXiv:1407.8311v1 [math.NA].

\bibitem[{Brauchart et~al.(2009{\natexlab{a}})Brauchart, Dragnev, and
  Saff}]{BrDrSa2009}
Brauchart, J.~S., Dragnev, P.~D., Saff, E.~B., 2009{\natexlab{a}}. Riesz
  extremal measures on the sphere for axis-supported external fields. J. Math.
  Anal. Appl. 356~(2), 769--792.

\bibitem[{Brauchart et~al.(2014{\natexlab{c}})Brauchart, Dragnev, and
  Saff}]{BrDrSa2014}
Brauchart, J.~S., Dragnev, P.~D., Saff, E.~B., 2014{\natexlab{c}}. Riesz
  external field problems on the hypersphere and optimal point separation.
  Potential Analysis, 1--32.

\bibitem[{Brauchart et~al.(2012{\natexlab{a}})Brauchart, Dragnev, Saff, and
  van~de Woestijne}]{BrDrSa2012}
Brauchart, J.~S., Dragnev, P.~D., Saff, E.~B., van~de Woestijne, C.~E.,
  2012{\natexlab{a}}. A fascinating polynomial sequence arising from an
  electrostatics problem on the sphere. Acta Math. Hungar. 137~(1-2), 10--26.

\bibitem[{Brauchart et~al.(2007)Brauchart, Hardin, and
  Saff}]{Brauchart_Hardin_Saff2007:riesz_energy_revolution}
Brauchart, J.~S., Hardin, D.~P., Saff, E.~B., 2007. The support of the limit
  distribution of optimal {R}iesz energy points on sets of revolution in
  {$\mathbb{R}^3$}. J. Math. Phys. 48~(12), 122901, 24.

\bibitem[{Brauchart et~al.(2009{\natexlab{b}})Brauchart, Hardin, and
  Saff}]{Brauchart_Hardin_Saff2009:riesz_energy_revolution}
Brauchart, J.~S., Hardin, D.~P., Saff, E.~B., 2009{\natexlab{b}}. Riesz energy
  and sets of revolution in {$\mathbb{R}^3$}. In: Functional analysis and
  complex analysis. Vol. 481 of Contemp. Math. Amer. Math. Soc., Providence,
  RI, pp. 47--57.

\bibitem[{Brauchart et~al.(2009{\natexlab{c}})Brauchart, Hardin, and
  Saff}]{Brauchart_Hardin_Saff2009:riesz_energy_roots}
Brauchart, J.~S., Hardin, D.~P., Saff, E.~B., 2009{\natexlab{c}}. The {R}iesz
  energy of the {$N$}th roots of unity: an asymptotic expansion for large
  {$N$}. Bull. Lond. Math. Soc. 41~(4), 621--633.

\bibitem[{Brauchart et~al.(2012{\natexlab{b}})Brauchart, Hardin, and
  Saff}]{BrHaSa2012}
Brauchart, J.~S., Hardin, D.~P., Saff, E.~B., 2012{\natexlab{b}}. Discrete
  energy asymptotics on a {R}iemannian circle. Unif. Distrib. Theory 7~(2),
  77--108.

\bibitem[{Brauchart et~al.(2012{\natexlab{c}})Brauchart, Hardin, and
  Saff}]{Brauchart_Hardin_Saff2012:riesz}
Brauchart, J.~S., Hardin, D.~P., Saff, E.~B., 2012{\natexlab{c}}. The
  next-order term for optimal {R}iesz and logarithmic energy asymptotics on the
  sphere. In: Recent advances in orthogonal polynomials, special functions, and
  their applications. Vol. 578 of Contemp. Math. Amer. Math. Soc., Providence,
  RI, pp. 31--61.

\bibitem[{Brauchart and
  Hesse(2007)}]{Brauchart_Hesse2007:numerical_integration}
Brauchart, J.~S., Hesse, K., 2007. Numerical integration over spheres of
  arbitrary dimension. Constr. Approx. 25~(1), 41--71.

\bibitem[{Brauchart et~al.(2014{\natexlab{d}})Brauchart, Saff, Sloan, and
  Womersley}]{Brauchart_Saff_Sloan+2014:qmc-designs}
Brauchart, J.~S., Saff, E.~B., Sloan, I.~H., Womersley, R.~S.,
  2014{\natexlab{d}}. {QMC} designs: optimal order quasi {M}onte {C}arlo
  integration schemes on the sphere. Math. Comp. 83, 2821--2851.

\bibitem[{Brauchart and Womersley(2014)}]{BrWo2010Manuscript}
Brauchart, J.~S., Womersley, R.~S., 2014. Weighted {QMC} designs: numerical
  integration on the unit sphere, $\mathbb{L}_2$ discrepancy and sums of
  distances, in preparation.

\bibitem[{Bresges and Urbanetz(2008)}]{BrUr2008}
Bresges, C., Urbanetz, N.~A., 2008. Determination of the minimum number of
  spacer particles ensuring non-contact between host particles — a new
  approach by numerical modelling. Powder Technology 187~(3), 260--272.

\bibitem[{Calef(2009)}]{Ca2009}
Calef, M.~T., 2009. Theoretical and computational investigations of minimal
  energy problems. Ph.D. thesis, Vanderbilt University.

\bibitem[{Calef et~al.(2013)Calef, Griffiths, Schulz, Fichtl, and
  Hardin}]{CaEtal2013}
Calef, M.~T., Griffiths, W., Schulz, A., Fichtl, C., Hardin, D.~P., 2013.
  Observed asymptotic differences in energies of stable and minimal point
  configurations on {$\mathbb S^2$} and the role of defects. J. Math. Phys.
  54~(10), 101901, 20.

\bibitem[{Calef and Hardin(2009)}]{Calef_Hardin2009:riesz_equilibrium}
Calef, M.~T., Hardin, D.~P., 2009. Riesz {$s$}-equilibrium measures on
  {$d$}-rectifiable sets as {$s$} approaches {$d$}. Potential Anal. 30~(4),
  385--401.

\bibitem[{Chen et~al.(2011)Chen, Frommer, and
  Lang}]{Chen_Frommer_Lang2011:computational_designs}
Chen, X., Frommer, A., Lang, B., 2011. Computational existence proofs for
  spherical {$t$}-designs. Numer. Math. 117~(2), 289--305.

\bibitem[{Chen and Womersley(2006)}]{Chen_Womersley2006:existence_designs}
Chen, X., Womersley, R.~S., 2006. Existence of solutions to systems of
  underdetermined equations and spherical designs. SIAM J. Numer. Anal. 44~(6),
  2326--2341 (electronic).

\bibitem[{Choirat and Seri(2013{\natexlab{a}})}]{ChSe2013b}
Choirat, C., Seri, R., 2013{\natexlab{a}}. Computational aspects of
  {C}ui-{F}reeden statistics for equidistribution on the sphere. Math. Comp.
  82~(284), 2137--2156.

\bibitem[{Choirat and Seri(2013{\natexlab{b}})}]{ChSe2013}
Choirat, C., Seri, R., 2013{\natexlab{b}}. Numerical properties of generalized
  discrepancies on spheres of arbitrary dimension. J. Complexity 29~(2),
  216--235.

\bibitem[{Claeys et~al.(2008)Claeys, Kuijlaars, and Vanlessen}]{ClKuVa2008}
Claeys, T., Kuijlaars, A. B.~J., Vanlessen, M., 2008. Multi-critical unitary
  random matrix ensembles and the general {P}ainlev{\'e} {II} equation. Ann. of
  Math. (2) 168~(2), 601--641.

\bibitem[{Cohn(2002)}]{Cohn2002:bounds_sphere_packings_ii}
Cohn, H., 2002. New upper bounds on sphere packings. {II}. Geom. Topol. 6,
  329--353 (electronic).

\bibitem[{Cohn and Elkies(2003)}]{Cohn_Elkies2003:bounds_sphere_packings_i}
Cohn, H., Elkies, N., 2003. New upper bounds on sphere packings. {I}. Ann. of
  Math. (2) 157~(2), 689--714.

\bibitem[{Cohn and Kumar(2007)}]{Cohn_Kumar2007:universally_optimal_sphere}
Cohn, H., Kumar, A., 2007. Universally optimal distribution of points on
  spheres. J. Amer. Math. Soc. 20~(1), 99--148.

\bibitem[{Cohn and Kumar(2009)}]{Cohn_Kumar2009:optimality_leech}
Cohn, H., Kumar, A., 2009. Optimality and uniqueness of the {L}eech lattice
  among lattices. Ann. of Math. (2) 170~(3), 1003--1050.

\bibitem[{Cohn et~al.(2009)Cohn, Kumar, and Sch{\"u}rmann}]{CohnKumarSchurmann}
Cohn, H., Kumar, A., Sch{\"u}rmann, A., 2009. Ground states and formal duality
  relations in the {G}aussian core model. Phys. Rev. E 80, 061116.

\bibitem[{Cohn and Woo(2012)}]{CoWo2012}
Cohn, H., Woo, J., 2012. Three-point bounds for energy minimization. J. Amer.
  Math. Soc. 25~(4), 929--958.

\bibitem[{Conway and Sloane(1993)}]{Conway_Sloane1993:sphere_packings}
Conway, J.~H., Sloane, N. J.~A., 1993. Sphere packings, lattices and groups.
  Vol. 290 of Grundlehren der Mathematischen Wissenschaften [Fundamental
  Principles of Mathematical Sciences]. Springer-Verlag, New York.

\bibitem[{Coombs et~al.(2009)Coombs, Straube, and Ward}]{CoStWa2009}
Coombs, D., Straube, R., Ward, M., 2009. Diffusion on a sphere with localized
  traps: Mean first passage time, eigenvalue asymptotics, and fekete points.
  SIAM Journal on Applied Mathematics 70~(1), 302--332.

\bibitem[{Coulangeon(2006)}]{Coulangeon2006:spherical}
Coulangeon, R., 2006. Spherical designs and zeta functions of lattices. Int.
  Math. Res. Not., Art. ID 49620, 16.

\bibitem[{Coulangeon and
  Sch{\"u}rmann(2012)}]{Coulangeon_Schuermann2012:energy_minimization}
Coulangeon, R., Sch{\"u}rmann, A., 2012. Energy minimization, periodic sets and
  spherical designs. Int. Math. Res. Not. IMRN 2012~(4), 829--848.

\bibitem[{Cui and Freeden(1997)}]{Cui_Freeden1997:equidistribution_sphere}
Cui, J., Freeden, W., 1997. Equidistribution on the sphere. SIAM J. Sci.
  Comput. 18~(2), 595--609.

\bibitem[{Damelin and
  Grabner(2003)}]{Damelin_Grabner2003:energy_functional_numeric}
Damelin, S.~B., Grabner, P.~J., 2003. Energy functionals, numerical integration
  and asymptotic equidistribution on the sphere. J. Complexity 19, 231--246,
  corrigendum, \textbf{20} (2004), 883--884.

\bibitem[{Delsarte et~al.(1977)Delsarte, Goethals, and
  Seidel}]{Delsarte_Goethals_Seidel1977:spherical_designs}
Delsarte, P., Goethals, J.~M., Seidel, J.~J., 1977. Spherical codes and
  designs. Geometriae Dedicata 6~(3), 363--388.

\bibitem[{Dick and Pillichshammer(2010)}]{DiPi2010}
Dick, J., Pillichshammer, F., 2010. Digital Nets and Sequences. Discrepancy
  Theory and Quasi-Monte Carlo Integration. Cambridge University Press,
  Cambridge.

\bibitem[{Dragnev et~al.(2002)Dragnev, Legg, and Townsend}]{DrLeTo2002}
Dragnev, P.~D., Legg, D.~A., Townsend, D.~W., 2002. Discrete logarithmic energy
  on the sphere. Pacific J. Math. 207~(2), 345--358.

\bibitem[{Dragnev and Saff(2007)}]{Dragnev_Saff2007:riesz_potential_separation}
Dragnev, P.~D., Saff, E.~B., 2007. Riesz spherical potentials with external
  fields and minimal energy points separation. Potential Anal. 26~(2),
  139--162.

\bibitem[{Erber and Hockney(1997)}]{Erber_Hockney1997:complex_systems}
Erber, T., Hockney, G.~M., 1997. Complex systems: equilibrium configurations of
  {$N$} equal charges on a sphere {$(2\leq N\leq 112)$}. In: Advances in
  chemical physics, {V}ol. {XCVIII}. Adv. Chem. Phys., XCVIII. Wiley, New York,
  pp. 495--594.

\bibitem[{Erd{\'e}lyi and Saff(2013)}]{ErSa2013}
Erd{\'e}lyi, T., Saff, E.~B., 2013. Riesz polarization inequalities in higher
  dimensions. J. Approx. Theory 171, 128--147.

\bibitem[{Farkas and Nagy(2008)}]{FaNa2008}
Farkas, B., Nagy, B., 2008. Transfinite diameter, {C}hebyshev constant and
  energy on locally compact spaces. Potential Anal. 28~(3), 241--260.

\bibitem[{Fejes~T{\'o}th(1964)}]{Fe1964}
Fejes~T{\'o}th, L., 1964. Regular figures. A Pergamon Press Book. The Macmillan
  Co., New York.

\bibitem[{Fejes~T{\'o}th(1972)}]{Fe1972}
Fejes~T{\'o}th, L., 1972. Lagerungen in der {E}bene auf der {K}ugel und im
  {R}aum, 2nd Edition. Vol.~65 of Die Grundlehren der mathematischen
  Wissenschaften. Springer-Verlag, Berlin.

\bibitem[{Fekete(1923)}]{Fe1923}
Fekete, M., 1923. {{\"U}ber die Verteilung der Wurzeln bei gewissen
  algebraischen Gleichungen mit ganzzahligen Koeffizienten}. Mathematische
  Zeitschrift 17~(1), 228--249.

\bibitem[{Feng and Zelditch(2013)}]{FeZe2013}
Feng, R., Zelditch, S., 2013. Random {R}iesz energies on compact {K}{\"a}hler
  manifolds. Trans. Amer. Math. Soc. 365~(10), 5579--5604.

\bibitem[{Forrester(2010)}]{Fo2010}
Forrester, P.~J., 2010. Log-gases and random matrices. Vol.~34 of London
  Mathematical Society Monographs Series. Princeton University Press,
  Princeton, NJ.

\bibitem[{Fuselier and Wright(2009)}]{FuWr2009}
Fuselier, E.~J., Wright, G.~B., 2009. Stability and error estimates for vector
  field interpolation and decomposition on the sphere with {RBF}s. SIAM J.
  Numer. Anal. 47~(5), 3213--3239.

\bibitem[{G{\'o}rski et~al.(2005)G{\'o}rski, Hivon, Banday, Wandelt, Hansen,
  Reinecke, and Bartelmann}]{HEALPix2005}
G{\'o}rski, K.~M., Hivon, E., Banday, A.~J., Wandelt, B.~D., Hansen, F.~K.,
  Reinecke, M., Bartelmann, M., 2005. {HEALP}ix: A framework for
  high-resolution discretization and fast analysis of data distributed on the
  sphere. The Astrophysical Journal 622~(2), 759.

\bibitem[{G{\"o}tz(2000)}]{Goetz2000:distribution_extremal}
G{\"o}tz, M., 2000. On the distribution of weighted extremal points on a
  surface in {${\bf R}^d,\ d\ge3$}. Potential Anal. 13~(4), 345--359.

\bibitem[{G{\"o}tz(2003)}]{Go2003}
G{\"o}tz, M., 2003. On the {R}iesz energy of measures. J. Approx. Theory
  122~(1), 62--78.

\bibitem[{G{\"o}tz and Saff(2001)}]{GoSa2001}
G{\"o}tz, M., Saff, E.~B., 2001. Note on {$d$}-extremal configurations for the
  sphere in {$\mathbb{R}^{d+1}$}. In: Recent progress in multivariate
  approximation ({W}itten-{B}ommerholz, 2000). Vol. 137 of Internat. Ser.
  Numer. Math. Birkh{\"a}user, Basel, pp. 159--162.

\bibitem[{Gourary and Adrian(1960)}]{GoAd1960}
Gourary, B.~S., Adrian, F.~J., 1960. Wave functions for electron-excess color
  centers in alkali halide crystals. In: Seitz, F., Turnball, D. (Eds.), Solid
  {S}tate {P}hysics: Advances in Research and Applications. Vol.~10. Academic
  Press, New York and London, pp. 127--247.

\bibitem[{Grabner(1991)}]{Grabner1991:erdoes_turan_type}
Grabner, P.~J., 1991. Erd{\H{o}}s-{T}ur\'an type discrepancy bounds. Monatsh.
  Math. 111, 127--135.

\bibitem[{Grabner(2014)}]{Grabner2014:point_minimal_energy}
Grabner, P.~J., 2014. Point sets of minimal energy. In: Larcher, G.,
  Pillichshammer, F., Winterhof, A., Xing, C. (Eds.), Applications of Algebra
  and Number Theory. Cambridge University Press, pp. 104--117, essays in Honour
  of Harald Niederreiter.

\bibitem[{Grabner and
  Tichy(1993)}]{Grabner_Tichy1993:spherical_designs_discrepancy}
Grabner, P.~J., Tichy, R.~F., 1993. Spherical designs, discrepancy and
  numerical integration. Math. Comp. 60, 327--336.

\bibitem[{Gr{\"a}f(2013)}]{Graef2013:efficient_algorithms_computation}
Gr{\"a}f, M., 2013. Efficient algorithms for the computation of optimal
  quadrature points on riemannian manifolds. Ph.D. thesis, Technische
  Universit{\"a}t Chemnitz.

\bibitem[{Gr{\"a}f and Potts(2011)}]{Graef_Potts2011:fourier}
Gr{\"a}f, M., Potts, D., 2011. On the computation of spherical designs by a new
  optimization approach based on fast spherical {F}ourier transforms. Numer.
  Math. 119~(4), 699--724.

\bibitem[{Gr{\"a}f and Potts(2013)}]{Graef_Potts2013:table}
Gr{\"a}f, M., Potts, D., 2013. Table of spherical designs. Website.
\newline\urlprefix\url{http://www-user.tu-chemnitz.de/~potts/workgroup/graef/quadrature}

\bibitem[{Gr{\"a}f et~al.(2012)Gr{\"a}f, Potts, and Steidl}]{GrPoSt2012}
Gr{\"a}f, M., Potts, D., Steidl, G., 2012. Quadrature errors, discrepancies,
  and their relations to halftoning on the torus and the sphere. SIAM Journal
  on Scientific Computing 34~(5), A2760--A2791.

\bibitem[{Habicht and van~der
  Waerden(1951)}]{Habicht_Waerden1951:lagerung_von_punkten}
Habicht, W., van~der Waerden, B.~L., 1951. Lagerung von {P}unkten auf der
  {K}ugel. Math. Ann. 123, 223--234.

\bibitem[{Hales(2005)}]{Hales2005:kepler_conjecture}
Hales, T.~C., 2005. A proof of the {K}epler conjecture. Ann. of Math. (2)
  162~(3), 1065--1185.

\bibitem[{Harbrecht et~al.(2012)Harbrecht, Wendland, and
  Zori{\u\i}}]{HaWeZo2012}
Harbrecht, H., Wendland, W.~L., Zori{\u\i}, N.~V., 2012. On {R}iesz minimal
  energy problems. J. Math. Anal. Appl. 393~(2), 397--412.

\bibitem[{Harbrecht et~al.(2014{\natexlab{a}})Harbrecht, Wendland, and
  Zori{\u\i}}]{Harbrecht_Wendland_Zorii2014:rapid}
Harbrecht, H., Wendland, W.~L., Zori{\u\i}, N.~V., 2014{\natexlab{a}}. Rapid
  solution of minimal {R}iesz energy problems. Preprint 2014-09, Mathematisches
  Institut, Universit{\"a}t Basel, Switzerland.

\bibitem[{Harbrecht et~al.(2014{\natexlab{b}})Harbrecht, Wendland, and
  Zori{\u\i}}]{HaWeZo2014}
Harbrecht, H., Wendland, W.~L., Zori{\u\i}, N.~V., 2014{\natexlab{b}}. Riesz
  minimal energy problems on {$C^{k-1,1}$}-manifolds. Math. Nachr. 287~(1),
  48--69.

\bibitem[{Hardin et~al.(2013)Hardin, Kendall, and Saff}]{HaKeSa2013}
Hardin, D.~P., Kendall, A.~P., Saff, E.~B., 2013. Polarization optimality of
  equally spaced points on the circle for discrete potentials. Discrete Comput.
  Geom. 50~(1), 236--243.

\bibitem[{Hardin and Saff(2004)}]{Hardin_Saff2004:discretizing_manifolds}
Hardin, D.~P., Saff, E.~B., 2004. Discretizing manifolds via minimum energy
  points. Notices Amer. Math. Soc. 51~(10), 1186--1194.

\bibitem[{Hardin and Saff(2005)}]{Hardin_Saff2005:minimal_riesz}
Hardin, D.~P., Saff, E.~B., 2005. Minimal {R}iesz energy point configurations
  for rectifiable {$d$}-dimensional manifolds. Adv. Math. 193~(1), 174--204.

\bibitem[{Hardin et~al.(2014)Hardin, Saff, and Simanek}]{HaSaSi2014arXiv}
Hardin, D.~P., Saff, E.~B., Simanek, B., 2014. Periodic discrete energy for
  long-range potentials, manuscript, \texttt{arXiv:1403.7505v1 [math-ph]}.

\bibitem[{Hardin et~al.(2007)Hardin, Saff, and
  Stahl}]{Hardin_Saff_Stahl2007:support_logarithmic}
Hardin, D.~P., Saff, E.~B., Stahl, H., 2007. Support of the logarithmic
  equilibrium measure on sets of revolution in {$\mathbb{R}^3$}. J. Math. Phys.
  48~(2), 022901, 14.

\bibitem[{Hardin et~al.(2012)Hardin, Saff, and Whitehouse}]{HaSaWh2012}
Hardin, D.~P., Saff, E.~B., Whitehouse, J.~T., 2012. Quasi-uniformity of
  minimal weighted energy points on compact metric spaces. Journal of
  Complexity 28~(2), 177--191.

\bibitem[{Hardin and Sloane(1993)}]{Hardin_Sloane1993:optimal_designs}
Hardin, R.~H., Sloane, N. J.~A., 1993. A new approach to the construction of
  optimal designs. J. Statist. Plann. Inference 37~(3), 339--369.

\bibitem[{Hardin and Sloane(1995)}]{Hardin_Sloane1995:codes_designs}
Hardin, R.~H., Sloane, N. J.~A., 1995. Codes (spherical) and designs
  (experimental). In: Different aspects of coding theory ({S}an {F}rancisco,
  {CA}, 1995). Vol.~50 of Proc. Sympos. Appl. Math. Amer. Math. Soc.,
  Providence, RI, pp. 179--206.

\bibitem[{Hardin and Sloane(1996)}]{Hardin_Sloane1996:mclarens_snub_cube}
Hardin, R.~H., Sloane, N. J.~A., 1996. Mc{L}aren's improved snub cube and other
  new spherical designs in three dimensions. Discrete Comput. Geom. 15~(4),
  429--441.

\bibitem[{Hardin and Sloane(2002)}]{Hardin_Sloane2002:table}
Hardin, R.~H., Sloane, N. J.~A., 2002. Table of spherical designs. website,
  \texttt{http://neilsloane.com/sphdesigns/dim3/}.

\bibitem[{Hardin et~al.(1997)Hardin, Sloane, and
  Smith}]{Hardin_Sloane_Smith1997:table}
Hardin, R.~H., Sloane, N. J.~A., Smith, W.~D., 1997. Minimal energy
  arrangements of points on a sphere. website,
  \texttt{http://neilsloane.com/electrons/}.

\bibitem[{Harman(1982)}]{Ha1982}
Harman, G., 1982. Sums of distances between points of a sphere. Internat. J.
  Math. Math. Sci. 5~(4), 707--714.

\bibitem[{Hesse(2006)}]{Hesse2006:lower_worst-case}
Hesse, K., 2006. A lower bound for the worst-case cubature error on spheres of
  arbitrary dimension. Numer. Math. 103~(3), 413--433.

\bibitem[{Hesse and Sloan(2005{\natexlab{a}})}]{Hesse_Sloan2005:optimal_lower}
Hesse, K., Sloan, I.~H., 2005{\natexlab{a}}. Optimal lower bounds for cubature
  error on the sphere {$S^2$}. J. Complexity 21~(6), 790--803.

\bibitem[{Hesse and
  Sloan(2005{\natexlab{b}})}]{Hesse_Sloan2005:worst_case_sobolev}
Hesse, K., Sloan, I.~H., 2005{\natexlab{b}}. Worst-case errors in a {S}obolev
  space setting for cubature over the sphere {$S^2$}. Bull. Austral. Math. Soc.
  71~(1), 81--105.

\bibitem[{Hesse and Sloan(2006)}]{Hesse_Sloan2006:cubature_s_sobolev}
Hesse, K., Sloan, I.~H., 2006. Cubature over the sphere {$S^2$} in {S}obolev
  spaces of arbitrary order. J. Approx. Theory 141~(2), 118--133.

\bibitem[{Holho{\c{s}} and Ro{\c{s}}ca(2014)}]{HoRo2014}
Holho{\c{s}}, A., Ro{\c{s}}ca, D., 2014. An octahedral equal area partition of
  the sphere and near optimal configurations of points. Comput. Math. Appl.
  67~(5), 1092--1107.

\bibitem[{Hou and Shao(2011)}]{HouSh2011}
Hou, X., Shao, J., 2011. Spherical distribution of 5 points with maximal
  distance sum. Discrete Comput. Geom. 46~(1), 156--174.

\bibitem[{Korevaar(1996)}]{Korevaar1996:fekete_points}
Korevaar, J., 1996. Fekete extreme points and related problems. In:
  Approximation theory and function series ({B}udapest, 1995). Vol.~5 of Bolyai
  Soc. Math. Stud. J{\'a}nos Bolyai Math. Soc., Budapest, pp. 35--62.

\bibitem[{Korevaar and
  Meyers(1993)}]{Korevaar_Meyers1993:spherical_faraday_chebyshev}
Korevaar, J., Meyers, J. L.~H., 1993. Spherical {F}araday cage for the case of
  equal point charges and {C}hebyshev-type quadrature on the sphere. Integral
  Transform. Spec. Funct. 1~(2), 105--117.

\bibitem[{Korevaar and Meyers(1994)}]{Korevaar_Meyers1994:chebyshev_quadrature}
Korevaar, J., Meyers, J. L.~H., 1994. Chebyshev-type quadrature on
  multidimensional domains. J. Approx. Theory 79~(1), 144--164.

\bibitem[{Korobov(1959)}]{Korobov1959:approximate_integrals}
Korobov, N.~M., 1959. Approximate evaluation of repeated integrals. Dokl. Akad.
  Nauk SSSR 124, 1207--1210.

\bibitem[{Kuijlaars and
  Saff(1998)}]{Kuijlaars_Saff1998:asymptotics_minimal_energy}
Kuijlaars, A. B.~J., Saff, E.~B., 1998. Asymptotics for minimal discrete energy
  on the sphere. Trans. Amer. Math. Soc. 350~(2), 523--538.

\bibitem[{Kuipers and
  Niederreiter(1974)}]{Kuipers_Niederreiter1974:uniform_distribution_sequences}
Kuipers, L., Niederreiter, H., 1974. Uniform distribution of sequences.
  Wiley-Interscience, New York.

\bibitem[{LaFave~Jr.(2013)}]{LaF2013}
LaFave~Jr., T., 2013. Correspondences between the classical electrostatic
  {T}homson problem and atomic electronic structure. Journal of Electrostatics
  71~(6), 1029--1035.

\bibitem[{LaFave~Jr.(2014)}]{LaF2014}
LaFave~Jr., T., 2014. Discrete transformations in the {T}homson problem.
  Journal of Electrostatics 72~(1), 39--43.

\bibitem[{Landkof(1972)}]{Landkof1972:potential_theory}
Landkof, N.~S., 1972. Foundations of modern potential theory. Springer-Verlag,
  New York, translated from the Russian by A. P. Doohovskoy, Die Grundlehren
  der mathematischen Wissenschaften, Band 180.

\bibitem[{Le~Gia et~al.(2010)Le~Gia, Sloan, and Wendland}]{LeGSlWe2010}
Le~Gia, Q.~T., Sloan, I.~H., Wendland, H., 2010. Multiscale analysis in
  {S}obolev spaces on the sphere. SIAM J. Numer. Anal. 48~(6), 2065--2090.

\bibitem[{Leopardi(2006)}]{Le2006}
Leopardi, P., 2006. A partition of the unit sphere into regions of equal area
  and small diameter. Electron. Trans. Numer. Anal. 25, 309--327 (electronic).

\bibitem[{Leopardi(2013)}]{Le2013}
Leopardi, P., 2013. Discrepancy, separation and riesz energy of finite point
  sets on the unit sphere. Adv. Comput. Math. 39~(1), 27--43.

\bibitem[{Li and Vaaler(1999)}]{Li_Vaaler1999:trigonometric_extremal}
Li, X.-J., Vaaler, J.~D., 1999. Some trigonometric extremal functions and the
  {E}rd{\H o}s-{T}ur{\'a}n type inequalities. Indiana Univ. Math. J. 48~(1),
  183--236.

\bibitem[{L{\'o}pez~Garc{\'{\i}}a and
  Saff(2010)}]{Lopez_Saff2010:asymptotics_greedy}
L{\'o}pez~Garc{\'{\i}}a, A., Saff, E.~B., 2010. Asymptotics of greedy energy
  points. Math. Comp. 79~(272), 2287--2316.

\bibitem[{Lubotzky et~al.(1986)Lubotzky, Phillips, and Sarnak}]{LuPhSa1986}
Lubotzky, A., Phillips, R., Sarnak, P., 1986. Hecke operators and distributing
  points on the sphere. {I}. Comm. Pure Appl. Math. 39~(S, suppl.), S149--S186,
  frontiers of the mathematical sciences: 1985 (New York, 1985).

\bibitem[{Lubotzky et~al.(1987)Lubotzky, Phillips, and Sarnak}]{LuPhSa1987}
Lubotzky, A., Phillips, R., Sarnak, P., 1987. Hecke operators and distributing
  points on {$S^2$}. {II}. Comm. Pure Appl. Math. 40~(4), 401--420.

\bibitem[{Lyubich and
  Vaserstein(1993)}]{Lyubich_Vaserstein1993:isometric_banach}
Lyubich, Y.~I., Vaserstein, L.~N., 1993. Isometric embeddings between classical
  {B}anach spaces, cubature formulas, and spherical designs. Geom. Dedicata
  47~(3), 327--362.

\bibitem[{Magnus et~al.(1966)Magnus, Oberhettinger, and
  Soni}]{Magnus_Oberhettinger_Soni1966:formulas_theorems}
Magnus, W., Oberhettinger, F., Soni, R.~P., 1966. Formulas and theorems for the
  special functions of mathematical physics. Vol.~52 of Grundlehren der
  mathematischen Wissenschaften. Springer-Verlag, third enlarged edition.

\bibitem[{Mart{\'{\i}}nez-Finkelshtein
  et~al.(2004)Mart{\'{\i}}nez-Finkelshtein, Maymeskul, Rakhmanov, and
  Saff}]{MaMaRaSa2004}
Mart{\'{\i}}nez-Finkelshtein, A., Maymeskul, V., Rakhmanov, E.~A., Saff, E.~B.,
  2004. Asymptotics for minimal discrete {R}iesz energy on curves in
  {$\mathbb{R}^d$}. Canad. J. Math. 56~(3), 529--552.

\bibitem[{Marzo and
  Ortega-Cerd{\`a}(2010)}]{Marzo_Ortega-Cerda2010:equidistribution_fekete}
Marzo, J., Ortega-Cerd{\`a}, J., 2010. Equidistribution of {F}ekete points on
  the sphere. Constr. Approx. 32~(3), 513--521.

\bibitem[{Mays(2013)}]{Mays2013}
Mays, A., 2013. A real quaternion spherical ensemble of random matrices. J.
  Stat. Phys. 153~(1), 48--69.

\bibitem[{Melnyk et~al.(1977)Melnyk, Knop, and Smith}]{MeKnSm1977}
Melnyk, T.~W., Knop, O., Smith, W.~R., 1977. Extremal arrangements of points
  and unit charges on a sphere: equilibrium configurations revisited. Canad. J.
  Chem. 55~(10), 1745--1761.

\bibitem[{Mhaskar and Saff(1985)}]{MhSa1985}
Mhaskar, H.~N., Saff, E.~B., 1985. Where does the sup norm of a weighted
  polynomial live? ({A} generalization of incomplete polynomials). Constr.
  Approx. 1~(1), 71--91.

\bibitem[{Montgomery(1988)}]{Montgomery1988:minimal_theta}
Montgomery, H.~L., 1988. Minimal theta functions. Glasgow Math. J. 30~(1),
  75--85.

\bibitem[{M{\"u}ller(1966)}]{Mueller1966:spherical_harmonics}
M{\"u}ller, C., 1966. Spherical harmonics. Vol.~17 of Lecture Notes in
  Mathematics. Springer-Verlag, Berlin.

\bibitem[{Musin(2008)}]{Musin}
Musin, O.~R., 2008. The kissing number in four dimensions. Ann. of Math. (2)
  168~(1), 1--32.

\bibitem[{Nebe(2013)}]{Nebe2013:venkov_lattices_designs}
Nebe, G., 2013. Boris {V}enkov's theory of lattices and spherical designs. In:
  Diophantine methods, lattices, and arithmetic theory of quadratic forms. Vol.
  587 of Contemp. Math. Amer. Math. Soc., Providence, RI, pp. 1--19.

\bibitem[{Nebe and Venkov(2009)}]{NebeVenkov1}
Nebe, G., Venkov, B., 2009. On lattices whose minimal vectors form a 6-design.
  European J. Combin. 30~(3), 716--724.

\bibitem[{Nerattini et~al.(2014)Nerattini, Brauchart, and
  Kiessling}]{NeBrKie2013}
Nerattini, R., Brauchart, J., Kiessling, M.-H., 2014. {O}ptimal {$N$}-{P}oint
  {C}onfigurations on the {S}phere: "{M}agic" {N}umbers and {S}male’s 7th
  {P}roblem. Journal of Statistical Physics, 1--69.

\bibitem[{Nodari and Serfaty(2014)}]{NoSe2014}
Nodari, S.~R., Serfaty, S., 2014. Renormalized energy equidistribution and
  local charge balance in 2d coulomb systems. International Mathematics
  Research Notices.

\bibitem[{Novak and
  Wo{\'z}niakowski(2008)}]{Novak_Wozniakowski2008:tractability_vol1}
Novak, E., Wo{\'z}niakowski, H., 2008. Tractability of multivariate problems.
  {V}ol. 1: {L}inear information. Vol.~6 of EMS Tracts in Mathematics. European
  Mathematical Society (EMS), Z\"urich.

\bibitem[{Of et~al.(2010)Of, Wendland, and Zori{\u\i}}]{OfWeZo2010}
Of, G., Wendland, W.~L., Zori{\u\i}, N.~V., 2010. On the numerical solution of
  minimal energy problems. Complex Var. Elliptic Equ. 55~(11), 991--1012.

\bibitem[{P{\'o}lya and Szeg{\"o}(1931)}]{Polya_Szego:transfiniten_durchmesser}
P{\'o}lya, G., Szeg{\"o}, G., 1931. {\"U}ber den transfiniten {D}urchmesser
  ({K}apazit\"atskonstante) von ebenen und r{\"a}umlichen {P}unktmengen. J.
  Reine Angew. Math. 165, 4--49.

\bibitem[{Pritsker(2011)}]{Pr2011}
Pritsker, I.~E., 2011. Distribution of point charges with small discrete
  energy. Proc. Amer. Math. Soc. 139~(10), 3461--3473.

\bibitem[{Pritsker et~al.(2014)Pritsker, Saff, and Wise}]{PrSaWi2014}
Pritsker, I.~E., Saff, E.~B., Wise, W., 2014. Reverse triangle inequalities for
  {R}iesz potentials and connections with polarization. J. Math. Anal. Appl.
  410~(2), 868--881.

\bibitem[{Rakhmanov et~al.(1994)Rakhmanov, Saff, and
  Zhou}]{Rakhmanov_Saff_Zhou1994:minimal_discrete_energy}
Rakhmanov, E.~A., Saff, E.~B., Zhou, Y.~M., 1994. Minimal discrete energy on
  the sphere. Math. Res. Lett. 1~(6), 647--662.

\bibitem[{Reimer(2000)}]{Reimer2000:hyperinterpolation_sphere}
Reimer, M., 2000. Hyperinterpolation on the sphere at the minimal projection
  order. J. Approx. Theory 104~(2), 272--286.

\bibitem[{Reimer(2001)}]{Reimer2001:geometry_nodes}
Reimer, M., 2001. The geometry of nodes in a positive quadrature on the sphere.
  In: Recent progress in multivariate approximation ({W}itten-{B}ommerholz,
  2000). Vol. 137 of Internat. Ser. Numer. Math. Birkh{\"a}user, Basel, pp.
  245--248.

\bibitem[{Reimer(2003)}]{Reimer2003:multivariate_approximation}
Reimer, M., 2003. Multivariate polynomial approximation. Vol. 144 of
  International Series of Numerical Mathematics. Birkh{\"a}user Verlag, Basel.

\bibitem[{Rougerie and Serfaty(2013)}]{RoSe2013arXiv}
Rougerie, N., Serfaty, S., 2013. Higher dimensional {C}oulomb gases and
  renormalized energy functionals, manuscript, \texttt{arXiv:1307.2805v3
  [math-ph]}.

\bibitem[{Sadoc and Mosseri(1999)}]{SaMo1999}
Sadoc, J.-F., Mosseri, R., 1999. Geometrical frustration. Collection
  Al{\'e}a-Saclay: Monographs and Texts in Statistical Physics. Cambridge
  University Press, Cambridge.

\bibitem[{Saff and Kuijlaars(1997)}]{Saff_Kuijlaars1997:distributing_many}
Saff, E.~B., Kuijlaars, A. B.~J., 1997. Distributing many points on a sphere.
  Math. Intelligencer 19~(1), 5--11.

\bibitem[{Saff and Totik(1997)}]{Saff_Totik1997:logarithmic_potentials}
Saff, E.~B., Totik, V., 1997. Logarithmic potentials with external fields. Vol.
  316 of Grundlehren der Mathematischen Wissenschaften [Fundamental Principles
  of Mathematical Sciences]. Springer-Verlag, Berlin, appendix B by Thomas
  Bloom.

\bibitem[{Sandier and Serfaty(2012)}]{SaSe2012}
Sandier, E., Serfaty, S., 2012. From the {G}inzburg-{L}andau model to vortex
  lattice problems. Comm. Math. Phys. 313~(3), 635--743.

\bibitem[{Sarnak and
  Str{\"o}mbergsson(2006)}]{Sarnak_Stroembergsson2006:minima_epsteins}
Sarnak, P., Str{\"o}mbergsson, A., 2006. Minima of {E}pstein's zeta function
  and heights of flat tori. Invent. Math. 165~(1), 115--151.

\bibitem[{Schaback(1995)}]{Sch1995}
Schaback, R., 1995. Error estimates and condition numbers for radial basis
  function interpolation. Adv. Comput. Math. 3~(3), 251--264.

\bibitem[{Schoenberg(1938)}]{Schoenberg1938:positive_definite}
Schoenberg, I.~J., 1938. Metric spaces and positive definite functions. Trans.
  Amer. Math. Soc. 44~(3), 522--536.

\bibitem[{Schwartz(2013)}]{Sch2013}
Schwartz, R.~E., 2013. The five-electron case of {T}homson’s problem.
  Experimental Mathematics 22~(2), 157--186.

\bibitem[{Serfaty(2014)}]{Se2014}
Serfaty, S., 2014. Ginzburg-{L}andau {V}ortices, {C}oulomb {G}ases, and
  {R}enormalized {E}nergies. J. Stat. Phys. 154~(3), 660--680.

\bibitem[{Seymour and
  Zaslavsky(1984)}]{Seymour_Zaslavsky1984:averaging_designs}
Seymour, P.~D., Zaslavsky, T., 1984. Averaging sets: a generalization of mean
  values and spherical designs. Adv. in Math. 52~(3), 213--240.

\bibitem[{Shub and Smale(1993{\natexlab{a}})}]{ShSm1993I}
Shub, M., Smale, S., 1993{\natexlab{a}}. Complexity of {B}ezout's theorem. {I}:
  Geometric aspects. J. Am. Math. Soc. 6~(2), 459--501.

\bibitem[{Shub and Smale(1993{\natexlab{b}})}]{ShSm1993III}
Shub, M., Smale, S., 1993{\natexlab{b}}. Complexity of {B}ezout's theorem.
  {III}. {C}ondition number and packing. J. Complexity 9~(1), 4--14,
  festschrift for Joseph F. Traub, Part I.

\bibitem[{Sloan and Womersley(2004)}]{Sloan_Womersley2004:extremal_sphere}
Sloan, I.~H., Womersley, R.~S., 2004. Extremal systems of points and numerical
  integration on the sphere. Adv. Comput. Math. 21~(1-2), 107--125.

\bibitem[{Sloan and Womersley(2009)}]{Sloan_Womersley2009:variational_designs}
Sloan, I.~H., Womersley, R.~S., 2009. A variational characterisation of
  spherical designs. J. Approx. Theory 159~(2), 308--318.

\bibitem[{Smale(1998)}]{Sm1998}
Smale, S., 1998. Mathematical problems for the next century. Math.
  Intelligencer 20~(2), 7--15.

\bibitem[{Stolarsky(1972)}]{Stolarsky1972:sums_distances_i}
Stolarsky, K.~B., 1972. Sums of distances between points on a sphere. Proc.
  Amer. Math. Soc. 35, 547--549.

\bibitem[{Stolarsky(1973)}]{Stolarsky1973:sums_distances_ii}
Stolarsky, K.~B., 1973. Sums of distances between points on a sphere. {II}.
  Proc. Amer. Math. Soc. 41, 575--582.

\bibitem[{Tammes(1930)}]{Tammes1930:pollen_grains}
Tammes, P. M.~L., 1930. On the origin of number and arrangement of the places
  of exit on the surface of pollen grains. Recueil des travaux botaniques
  n{\'e}erlandais 27, 1--84.

\bibitem[{Teuber et~al.(2011)Teuber, Steidl, Gwosdek, Schmaltz, and
  Weickert}]{TeStGw2011}
Teuber, T., Steidl, G., Gwosdek, P., Schmaltz, C., Weickert, J., 2011.
  Dithering by differences of convex functions. SIAM J. Imaging Sci. 4~(1),
  79--108.

\bibitem[{Thomson(1904)}]{thomson:1904}
Thomson, J.~J., 1904. On the structure of the atom: an investigation of the
  stability and periods of oscillation of a number of corpuscles arranged at
  equal intervals around the circumference of a circle; with application of the
  results to the theory of atomic structure. Philos. Mag. 7~(39), 237--265.

\bibitem[{Thue(1910)}]{Thue1910}
Thue, A., 1910. {\"U}ber die dichteste {Z}usammenstellung von kongruenten
  {K}reisen in einer {E}bene. Christiania Vid.-Selsk. Skr. 1, 9p.

\bibitem[{Torquato and Stillinger(2006)}]{ToSt2006}
Torquato, S., Stillinger, F.~H., 2006. New conjectural lower bounds on the
  optimal density of sphere packings. Experiment. Math. 15~(3), 307--331.

\bibitem[{Tumanov(2013)}]{Tu2013}
Tumanov, A., 2013. Minimal biquadratic energy of five particles on a 2-sphere.
  Indiana Univ. Math. J. 62~(6), 1717--1731.

\bibitem[{Venkov(1984)}]{Venkov1984:even_unimodular_lattices}
Venkov, B.~B., 1984. Even unimodular extremal lattices. Trudy Mat. Inst.
  Steklov. 165, 43--48, algebraic geometry and its applications.

\bibitem[{Wagner(1990)}]{Wagner1990:means_distances_lower}
Wagner, G., 1990. On means of distances on the surface of a sphere (lower
  bounds). Pacific J. Math. 144~(2), 389--398.

\bibitem[{Wagner(1992)}]{Wagner1992:means_distances_upper}
Wagner, G., 1992. On means of distances on the surface of a sphere. {II}.
  {U}pper bounds. Pacific J. Math. 154~(2), 381--396.

\bibitem[{Wales et~al.(2014)Wales, Doye, Dullweber, Hodges, Naumkin, Calvo,
  Hern{\'a}ndez-Rojas, and Middleton}]{CCD2014}
Wales, D.~J., Doye, J. P.~K., Dullweber, A., Hodges, M.~P., Naumkin, F.~Y.,
  Calvo, F., Hern{\'a}ndez-Rojas, J., Middleton, T.~F., 2014. The {C}ambridge
  {C}luster {D}atabase.
\newline\urlprefix\url{http://www-wales.ch.cam.ac.uk/CCD.html}

\bibitem[{Yudin(1997)}]{Yudin1997:lower_design}
Yudin, V.~A., 1997. Lower bounds for spherical designs. Izv. Ross. Akad. Nauk
  Ser. Mat. 61~(3), 213--223.

\bibitem[{Zori{\u\i}(2003)}]{Zo2003}
Zori{\u\i}, N.~V., 2003. Equilibrium potentials with external fields.
  Ukra{\"\i}n. Mat. Zh. 55~(9), 1178--1195.

\bibitem[{Zori{\u\i}(2004)}]{Zo2004}
Zori{\u\i}, N.~V., 2004. Potential theory with respect to consistent kernels: a
  completeness theorem, and sequences of potentials. Ukra{\"\i}n. Mat. Zh.
  56~(11), 1513--1526.

\end{thebibliography}
\end{document}